\documentclass[a4paper,fleqn,usenatbib]{mnras}

\usepackage{newtxtext,newtxmath}

\usepackage[T1]{fontenc}
\usepackage{ae,aecompl}

\usepackage{tikz}

\usepackage{graphicx}	
\usepackage{amsmath}	
\usepackage{amssymb}	

\newcommand{\msun}{~\mathrm{M}_{\odot}}

\newcommand{\bs}[1]{\boldsymbol{#1}}
\newcommand{\M}{\mathcal{M}}

\title[The physics of multiphase gas flows]{The physics of multiphase gas flows: fragmentation of a radiatively cooling gas cloud in a hot wind}

\author[Sparre, Pfrommer \& Vogelsberger]{
Martin Sparre$^{1,2,3}$\thanks{E-mail: sparre@uni-potsdam.de}, Christoph Pfrommer$^{2,1}$ and Mark Vogelsberger$^{3}$\\
\\
$^{1}$Institut f\"ur Physik und Astronomie, Universit\"at Potsdam, Karl-Liebknecht-Str.\,24/25, 14476 Golm, Germany\\
$^{2}$Leibniz-Institut f\"ur Astrophysik Potsdam (AIP), An der Sternwarte 16, 14482 Potsdam, Germany\\
$^{3}$Department of Physics, Kavli Institute for Astrophysics and Space Research, Massachusetts Institute of Technology,\\$\phantom{aaaaaaaaaaaaaaaaaaaaaaaaaaaaaaaaaaaaaaaaaaaaaaaaaaaaaaaaaaaaaaaaa}$ Cambridge, MA 02139, USA
}

\pubyear{2018}

\begin{document}
\label{firstpage}
\pagerange{\pageref{firstpage}--\pageref{lastpage}}
\maketitle

\begin{abstract}
Galactic winds exhibit a multiphase structure that consists of hot-diffuse and cold-dense phases. Here we present high-resolution idealised simulations of the interaction of a hot supersonic wind with a cold cloud with the moving-mesh code {\sc arepo} in setups with and without radiative cooling. We demonstrate that cooling causes clouds with sizes larger than the cooling length to fragment in two- and three-dimensional simulations (2D and 3D). We confirm earlier 2D simulations by \citet{2018MNRAS.473.5407M} and highlight differences of the shattering processes of 3D clouds that are exposed to a hot wind. The fragmentation process is quantified with a friends-of-friends analysis of shattered cloudlets and density power spectra. Those show that radiative cooling causes the power spectral index to gradually increase when the initial cloud radius is larger than the cooling length and with increasing time until the cloud is fully dissolved in the hot wind. A resolution of around 1 pc is required to reveal the effect of cooling-induced fragmentation of a 100 pc outflowing cloud. Thus, state-of-the-art cosmological zoom simulations of the circumgalactic medium (CGM) fall short by orders of magnitudes from resolving this fragmentation process. This physics is, however, necessary to reliably model observed column densities and covering fractions of Lyman-$\alpha$ haloes, high-velocity clouds, and broad-line regions of active galactic nuclei.
\end{abstract}

\begin{keywords}
galaxies: formation -- methods: numerical -- ISM: jets and outflows
\end{keywords}


\section{Introduction}\label{Intro}

Strong feedback is required to prevent galaxies from forming more stars than observed particularly in low-mass galaxies \citep{2012RAA....12..917S}. The observed gas outflows in starbursts and normal galaxies \citep{1990ApJS...74..833H,1996ApJ...462..651L,2005ApJS..160..115R, 2005ARA&A..43..769V,2012MNRAS.426..801B, 2014ApJ...794..156R,2017ApJ...846..151H} provide more direct evidence for the existence of feedback processes. These winds are \emph{multiphase} because they contain hot diffuse and cold dense gas \citep{2009ApJ...697.2030S,2013ApJ...768...75R}. Some observations have even found evidence of molecular gas \citep{2010A&A...518L.155F,2011ApJ...733L..16S} and star formation \citep{2017Natur.544..202M} in outflows from galaxies with an active galactic nucleus (AGN).

Even the highest resolution cosmological galaxy formation simulations \citep{2014MNRAS.437.1750M,2015ApJ...804...18A,2015MNRAS.454...83W,2017MNRAS.467..179G,2017arXiv170206148H} are unable to resolve the sub-parsec-scale multiphase structure of winds, and instead subgrid-treatments of winds are required \citep{2006MNRAS.373.1265O,2008MNRAS.387.1431D,2012MNRAS.426..140D,2013MNRAS.428.2966P,2016MNRAS.462.3265D}. An example of a scheme that models such winds has been proposed by \citet{2003MNRAS.339..289S}, where winds are launched from star-forming gas cells, and the winds are decoupled from hydrodynamical interactions until they are outside the star-forming region.

A hot diffuse wind naturally arises in analytical and numerical modelling of energy and mass injections into starburst galaxies \citep{1985Natur.317...44C,2018arXiv180301005S}. The \emph{cloud crushing problem} studies how such a wind affects a cold gas cloud. Important analytical modelling of the problem was done by \citet{1994ApJ...420..213K}. They derived the \emph{cloud crushing time-scale} defined as the time it takes for the initial shock to propagate through the cloud:
\begin{align}
t_\text{cc} \equiv \frac{R_\text{cloud}}{\varv_\text{wind}} \sqrt{\frac{\rho_\text{cloud}}{\rho_\text{wind}}}. \label{tcc}
\end{align}
Here $\rho$ is the gas density, $\varv_\text{wind}$ is the wind velocity and $R_\text{cloud}$ is the cloud radius. In addition to being dynamically perturbed by a shock the cloud is also affected by instabilities, because the time-scales of the Kelvin-Helmholtz (KH) and Rayleigh-Taylor (RT) instabilities are comparable and proportional to $t_\text{cc}$. It is therefore theoretically well-motivated that a cloud gets evaporated on a time-scale proportional to $t_\text{cc}$, and several simulations have confirmed such a picture for non-radiative clouds \citep{1992ApJ...390L..17S,1995ApJ...454..172X,2006ApJS..164..477N}. Simulations that adopt different hydrodynamical methods to treat turbulence have established that instabilities become very important after the initial shock compression stage \citep{2007MNRAS.380..963A,2010MNRAS.406.2289H,2012MNRAS.424.2999S}.

\citet{2009ApJ...703..330C} show that radiative cooling delays fluid-instabilities, and as a result, clouds survive longer than in the non-radiative case. \citet{2015ApJ...805..158S} quantified the lifetimes of clouds influenced by radiative cooling with realistic parameters of a starburst wind. They show that 100 pc clouds typically survive travel lengths of $40 R_\text{cloud}$ before they are completely evaporated, and they furthermore parameterise the lifetime of clouds by a function of the form, $\alpha t_\text{cc} \sqrt{1+\mathcal{M}_\text{hot}}$. Here $\mathcal{M}_\text{hot}$ is the Mach number of the hot wind, and the constant of proportionality, $\alpha$, is $1.75, 2.5, 4$ and $6$ for the time when $90$, $75$, $50$, and $25$\% of the original cold cloud mass is still not evaporated. They hence find clouds to survive for substantially longer times than predicted by Eq.~\ref{tcc}, but still not long time enough to be transported to the outskirts of galaxies.

An important step for the study of multiphase gas flows was done by \citet{2018MNRAS.473.5407M}, who suggested that radiative cooling causes gas structures to \emph{shatter} into \emph{cloudlets}, which assume sizes comparable to the cooling length, which for a temperature around $10^4$ K is given by
\begin{align}
l_\text{cloudlet}=c_\text{s} t_\text{cool} = 0.1 \;\text{pc} \times \left( \frac{n}{\text{cm}^{-3}} \right)^{-1},\label{CoolingScale}
\end{align}
implying that gas with a characteristic density of $n \simeq 0.1-1 \; \text{cm}^{-3}$, corresponding to the gas in the outskirts of several observed haloes \citep{1999ApJ...515..500R,2002ApJ...565..743R,2009ApJ...690.1558P,2015Sci...348..779H}, should fragment to $\simeq 0.1-1$ pc structures. The physical motivation for the shattering process derives from considering a gas perturbation with a temperature $\gg 10^4$ K well below the surroundings, and with a size ($\mathcal{R}$) that greatly exceeds the cooling length, $\mathcal{R}\gg c_\text{s} t_\text{cool}$. The density of the cloud and surroundings are assumed to be similar. Such a cloud cools faster than the timescale for maintaining a pressure equilibrium with the surroundings, and it is thus very unstable. A possible evolutionary path is that the cloud maintains its geometrical shape (the process is isochoric, i.e., at constant volume) and hence evolves at constant density. After the cloud has cooled to $\simeq 10^4$ K, where radiative cooling is no longer efficient, a pressure equilibrium with the surroundings is established after a sound crossing time across $\mathcal{R}$. A different and faster path to equilibrium can be realised if the unstable cloud \emph{shatters} into smaller cloudlets of size, $c_\text{s} t_\text{cool}$, which can obtain pressure equilibrium with the surroundings much faster in comparison to a cloud with the initial cloud radius, accelerating the process by a factor $\mathcal{R}/l_\text{cloudlet}$. A system consisting of cloudlets shattered to these high densities is potentially able to produce a large area covering fraction, but at the same time only occupies a small volume filling fraction. To study the physics of shattering with simulations, we require a very high resolution. For this reason \citet{2018MNRAS.473.5407M} was limited to examine shattering in 2D simulations.

For the first time, we here present 3D simulations with high enough resolution to resolve the shattering of $n=0.1$ cm$^{-3}$ gas structures. We focus on the above-mentioned problem, where a cold spherical gas cloud is accelerated and disrupted by a hot supersonic wind. In Section~\ref{SimulationOverview} we present our simulations and in Section~\ref{2D3D} we reproduce the 2D results of \citet{2018MNRAS.473.5407M} and examine our 3D simulations. We furthermore quantify the effect of shattering by measuring the density power spectrum and by characterising the gas distribution with a friends-of-friends cloudlet finder. We discuss our results in Section~\ref{Sec:Discussion}, and conclude in Section~\ref{Sec:Conclusion}.

\section{Simulation overview}\label{SimulationOverview}

We have performed a set of simulations, which enable us to test whether dense outflows accelerated by a hot diffuse wind undergo \emph{shattering}. The hot wind initially has a temperature of $T_\text{hot}=10^7$ K and the cold cloud has $T_\text{cloud}=10^4$ K. The number densities are $n_\text{cold}=0.1$ cm$^{-3}$ and $n_\text{hot}=10^{-4}$ cm$^{-3}$, respectively, and the hot wind has a Mach number of $\mathcal{M}_\text{hot} = 1.5$. These initial conditions are selected to be identical to \citet{2018MNRAS.473.5407M}, which enables a direct comparison.

In 2D we perform radiative cooling simulations for clouds with an initial radius of $R_\text{cloud} = $ 1, 10 and 100 pc.  We also perform a non-radiative simulation (without radiative cooling) of a 1 pc cloud. The non-radiative simulation is self-similar, so it is straightforward to rescale this simulation to any given initial cloud radius. Each cloud is simulated at four different resolution levels with the number of cells per cloud radius, $R_\text{cloud}/\Delta x$, ranging from 76 to 607. The former corresponds to the resolution typically used in modern 3D cloud crushing simulations \citep{2015ApJ...805..158S,2017ApJ...834..144S} and the latter is identical to the high resolution used in the 2D simulations of \citet{2018MNRAS.473.5407M}. See Table~\ref{Table:SimulationOverview} for an overview of our simulations.

In the 3D simulations we limit ourselves to simulations with $80\leq R_\text{cloud}/\Delta x\leq 160$, because 3D simulations are computationally more expensive than the 2D cases. We adopt identical cloud radii in 2D and 3D simulations.

To study the differences between instabilities in 2D and 3D simulations we also perform simulations of a set of 3D cylindrical clouds, where the cylinder height is equal to the height of the simulations box. Due to the periodic boxes used in our simulations this corresponds to infinite cylinders. In all cases the symmetry axis of the cylinder is along the $z$-axis, so it is perpendicular to the flow of the hot wind, which is along the $y$-axis. Hence, the initial conditions of cold gas phase is either in the form of a 2D sphere, 3D sphere or 3D cylinder. With a $N$D sphere we refer to the domain encapsulated by a sphere of a given radius, $R_\text{cloud}$, in $N$-dimensional space.\footnote{We note, that this nomenclature differs from the mathematical notation, where a sphere in $N$D space is referred to as a $N-1$-sphere.}

\begin{table*}
\centering
\begin{minipage}{.53\textwidth}
\centering
{\bf 2D sphere simulations} 	\\
\begin{tabular}{l|crr}
\hline\hline
Name  & Cooling & $R_{\text{cloud}}/$pc & $R_{\text{cloud}}/\Delta x$  \\
\hline
{\tt 2D-1pc-607-NonRad}$\phantom{Cylin.}$ &  no & 1 & 607 \\
{\tt 2D-1pc-607} & yes  & 1 & 607 \\
{\tt 2D-10pc-607} & yes & 10 & 607 \\
{\tt 2D-100pc-607} & yes & 100 & 607 \\
\hline
{\tt 2D-1pc-304-NonRad} &  no & 1 & 304 \\
{\tt 2D-1pc-304} & yes  & 1 & 304 \\
{\tt 2D-10pc-304} & yes & 10 & 304 \\
{\tt 2D-100pc-304} & yes & 100 & 304 \\
\hline
{\tt 2D-1pc-152-NonRad} &  no & 1 & 152 \\
{\tt 2D-1pc-152} & yes  & 1 & 152 \\
{\tt 2D-10pc-152} & yes & 10 & 152 \\
{\tt 2D-100pc-152} & yes & 100 & 152 \\
\hline
{\tt 2D-1pc-76-NonRad} &  no & 1 & 76 \\
{\tt 2D-1pc-76} & yes  & 1 & 76 \\
{\tt 2D-10pc-76} & yes & 10 & 76 \\
{\tt 2D-100pc-76} & yes & 100 & 76 \\
\hline\hline
\end{tabular}\\
{\bf 3D sphere simulations}\\
\begin{tabular}{l|crr}
\hline\hline
Name  & Cooling & $R_{\text{cloud}}/$pc & $R_{\text{cloud}}/\Delta x$   \\
\hline
{\tt 3D-1pc-160-NonRad}$\phantom{Cylin.}$ &  no & 1 & 160 \\
{\tt 3D-1pc-160} & yes  & 1 & 160 \\
{\tt 3D-10pc-160} & yes & 10 & 160\\
{\tt 3D-100pc-160} & yes & 100 & 160\\
\hline
{\tt 3D-1pc-80-NonRad} &  no & 1 & 80 \\
{\tt 3D-1pc-80} & yes  & 1 & 80 \\
{\tt 3D-10pc-80} & yes & 10 & 80\\
{\tt 3D-100pc-80} & yes & 100 & 80\\
\hline\hline
\end{tabular}\\
{\bf 3D cylinder simulations}\\
\begin{tabular}{l|crrr}
\hline\hline
Name  & Cooling & $R_{\text{cloud}}/$pc & $R_{\text{cloud}}/\Delta x$   \\
\hline
{\tt 3D-Cylinder-NonRad}$\phantom{aaaa}$ &  no & 1 & 80 \\
{\tt 3D-Cylinder-1pc} &  yes & 1 & 80 \\
{\tt 3D-Cylinder-10pc} &  yes & 10 & 80  \\
{\tt 3D-Cylinder-100pc} &  yes & 100 & 80  \\
\hline\hline
\end{tabular}
    \end{minipage}%
    \begin{minipage}{.46\textwidth}
        \centering
		\includegraphics[width=0.5\textwidth]{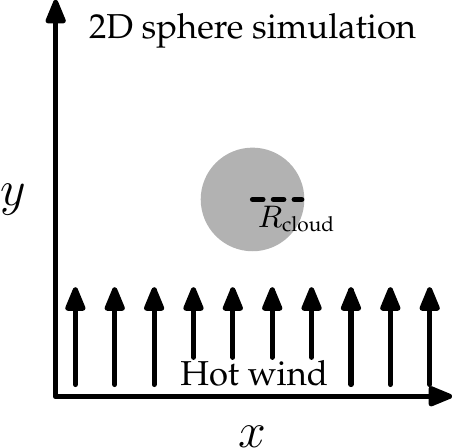}\vspace*{2.5cm}
		\includegraphics[width=1.0\textwidth]{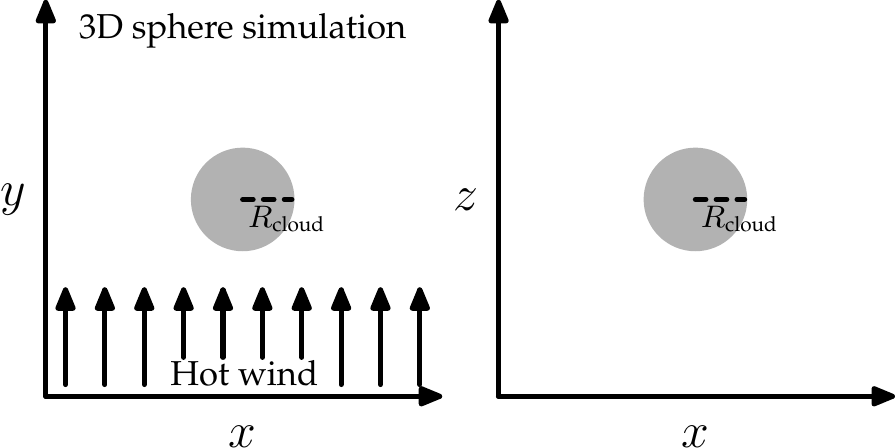}\vspace*{0.4cm}
		\includegraphics[width=1.0\textwidth]{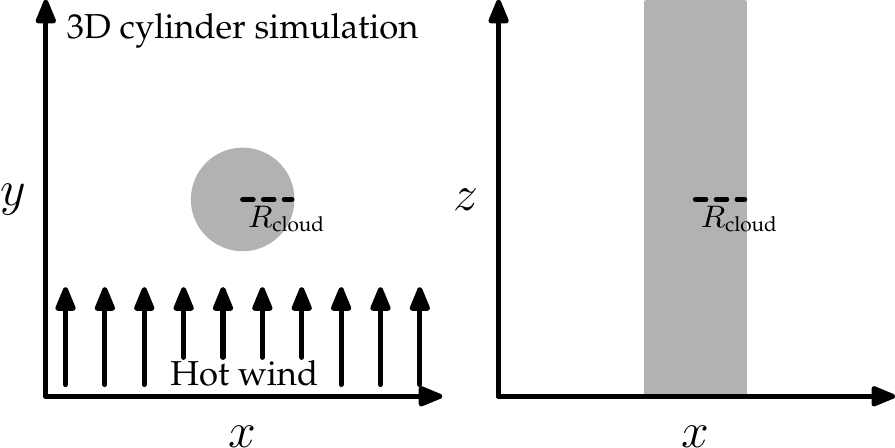}
    \end{minipage}%

\caption{An overview of the simulations presented in this paper. The setups span 2D spheres, 3D spheres and 3D cylinders. The sketches to the right show these initial configurations, where the dark colours show the location of the cold phases seen in planes through the cloud centre. The sphere simulations are named according to the scheme $N$D-$R_\text{cloud}$-$R_\text{cloud}/\Delta x$, where $N$ is the number of dimensions. A suffix, -NonRad, is added for the \emph{non-radiative} simulations without radiative cooling. The initial radius of the cold cloud and the number of cells per cloud radius are noted in the columns. Our high-resolution 2D sphere simulations with $R_\text{cloud}=1,10$ and 100 pc have a spatial resolution of $\Delta x = 0.00165, 0.0165$ and $0.165$ pc, respectively. For the high-resolution 3D spheres we have  $\Delta x = 0.00625, 0.0625$ and $0.625$ pc, respectively.}
\label{Table:SimulationOverview}
\end{table*}

\subsection{Temperature floor and cooling function}

We use a temperature floor at $T=5000$ K, implying that heat is added to gas cells so they cannot cool below this value. This slightly reduces the amount of small-scale structure compared to a run without such a temperature floor. The motivation for this is to ease comparison to \citealt{2018MNRAS.473.5407M} (who use a temperature floor of $10^4$ K), and also to avoid very dense clouds, which are numerically expansive to resolve.

All gas cells initially have a solar composition of elements. For each cell the cooling function is calculated by summing up the contributions from primordial species \citep[using rates from][]{1992ApJS...78..341C,1996ApJS..105...19K}, metal line cooling and Compton cooling from interaction with CMB photons. The metal line cooling rates, which are based on the density and metallicity of the gas cells, are calculated with {\sc cloudy} \citep{1998PASP..110..761F,2013RMxAA..49..137F}. We assume the ionizing radiation background from \citet{2009ApJ...703.1416F}. This method for calculating the cooling function is identical to what is used in the Illustris simulation project \citep{2014Natur.509..177V,2014MNRAS.444.1518V,2014MNRAS.445..175G,2015MNRAS.452..575S}, see Section 2.4 of \citet{2013MNRAS.436.3031V} for a more detailed description.

\begin{figure*}
    \centering
    \begin{minipage}{.32\textwidth}
        \centering
        \includegraphics[width=\linewidth]{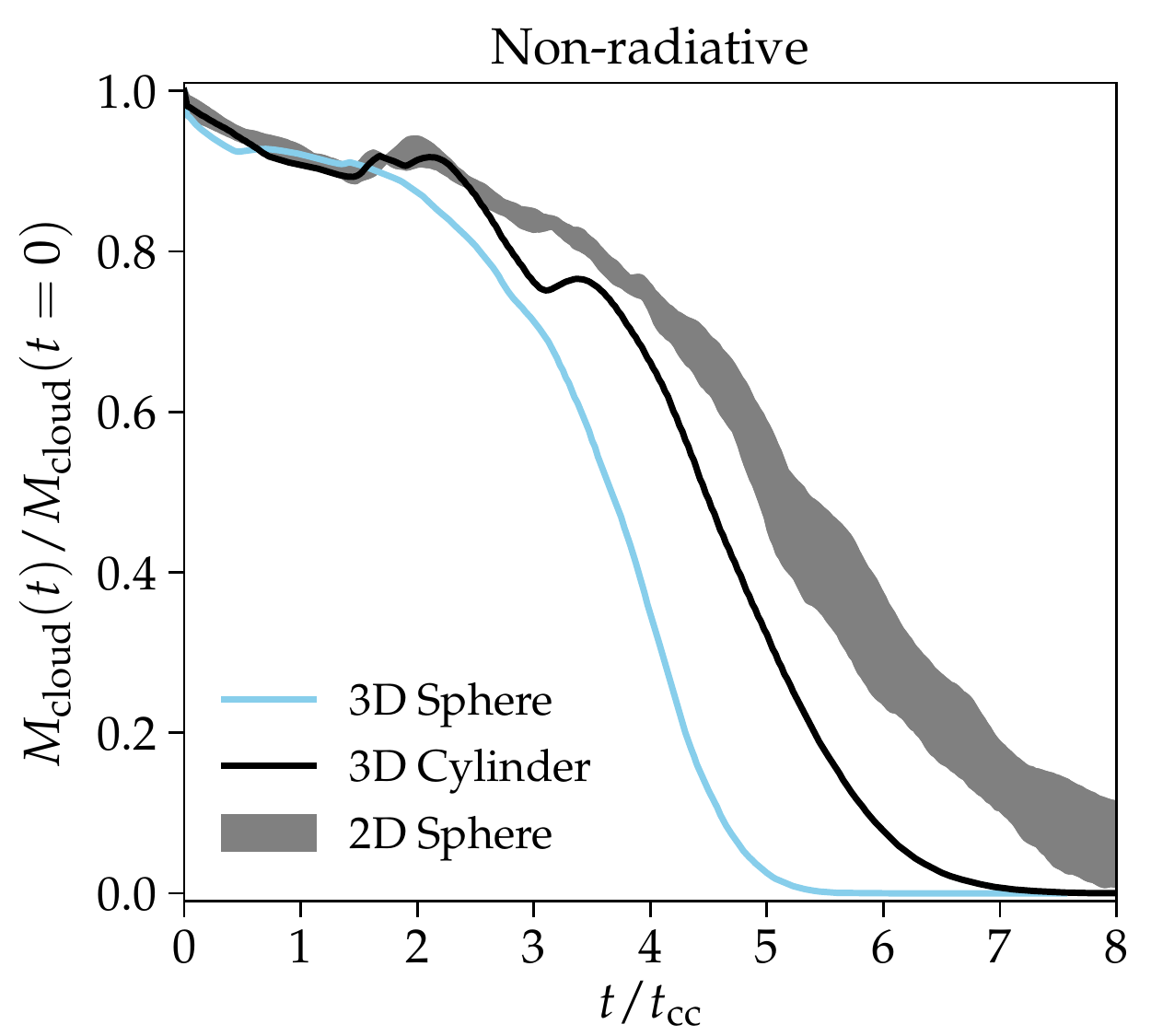}
    \end{minipage}%
    \begin{minipage}{.32\textwidth}
        \centering
        \includegraphics[width=\linewidth]{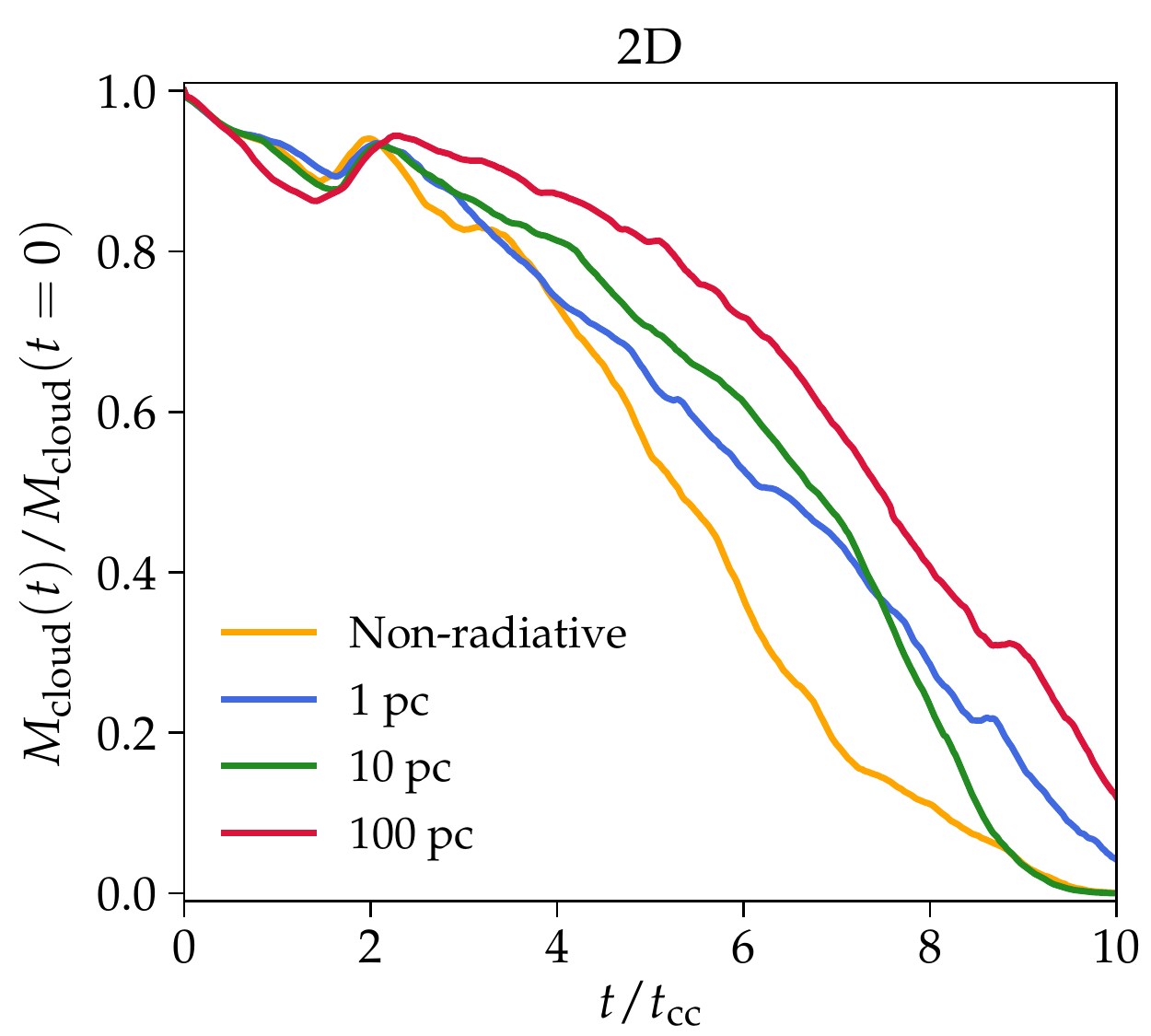}
    \end{minipage}%
    \begin{minipage}{.32\textwidth}
        \centering
        \includegraphics[width=\linewidth]{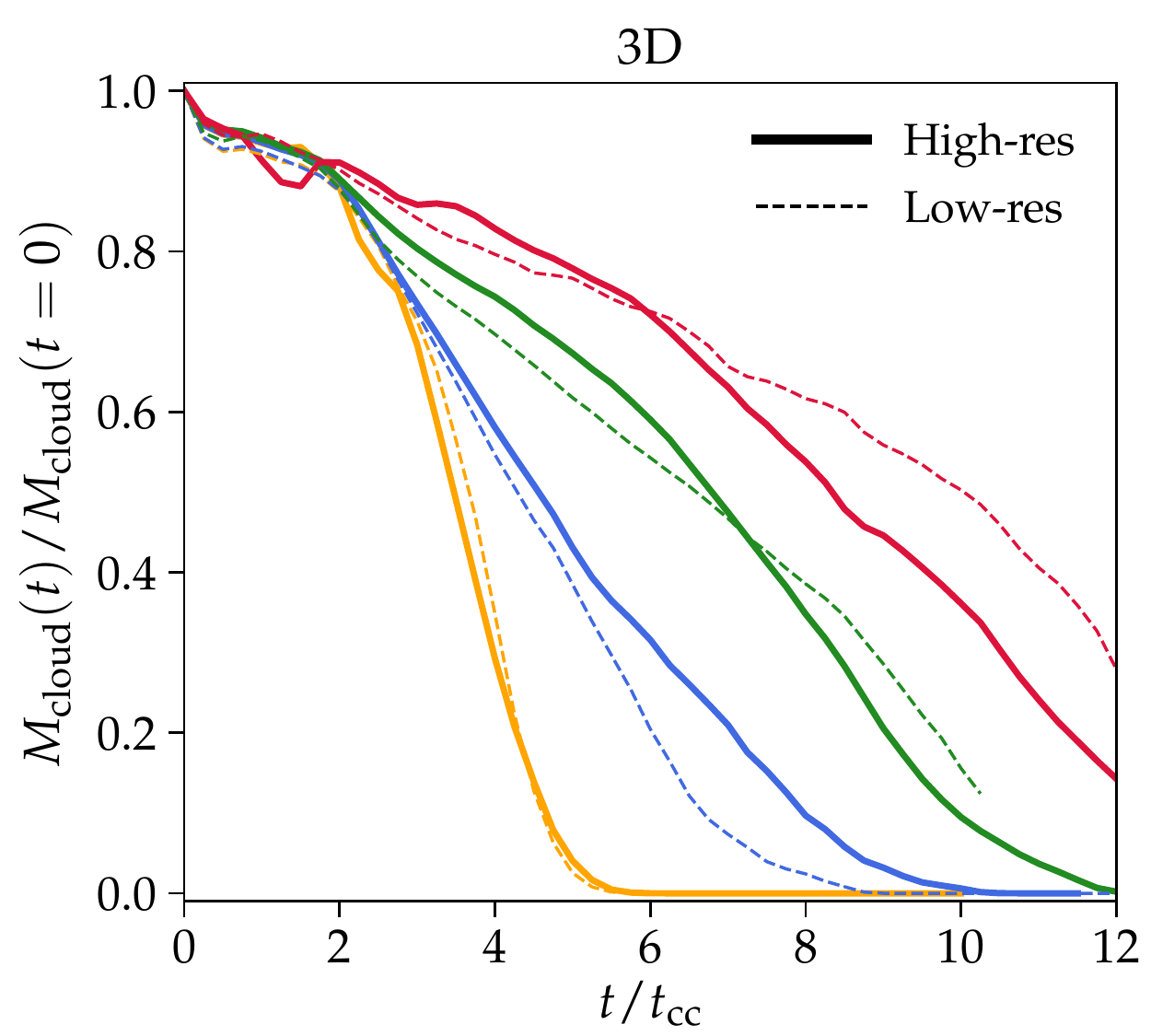}
    \end{minipage}%
        \caption{The survival mass fraction of gas with $n\geq \frac{1}{3}n_\text{cloud}$. Left panel: we show the non-radiative simulation of a 2D sphere, a 3D sphere and a 3D cylinder. For the 2D sphere we show a band spanned by the well-resolved 2D simulations (see text for details). For $t\lesssim 2.5 t_\text{cc}$ the evolution of the 2D sphere and the 3D cylinder is remarkably similar, which is expected from geometrical correspondences. At later times instabilities in the $z$-direction cause the 3D cylinder to evaporate faster than the 2D sphere. The 3D sphere is evaporated faster than the 3D cylinder and the 2D sphere. Central panel: the 2D high-resolution simulations with different cloud sizes. Cooling extends the lifetime of clouds, and in clouds with large radii cooling plays the largest role in extending the lifetime. Right panel: we show the 3D simulations. The thick lines show high-resolution simulations with a resolution of $ R_\text{cloud}/\Delta x =160$ and the thin lines show the counterparts with $R_\text{cloud}/\Delta x =80$. As also seen in 2D a larger value of the cloud radius implies a larger survival time. The relative difference between the different cloud sizes are, however, larger in 3D than in 2D. This can be explained by the interaction of instabilities and radiative cooling in 3D (see Fig.~\ref{Fig02Instability}).}
\label{Fig1CloudSurvival}
\end{figure*}

\subsection{Simulation code and refinement scheme}\label{Refinement}

We use the moving-mesh code {\sc arepo} \citep{2010MNRAS.401..791S,2016MNRAS.455.1134P}. A gas cell is de-refined if the mass is more than two times smaller than the target mass, $m_\text{target}$, of a simulation. For each 3D simulation the target mass is defined as $m_\text{target}\equiv  (\Delta x)^3 n_\text{cold} \mu m_\text{p}$, where $\mu=0.599$ is the mean particle mass (we assume fully ionized gas, and solar metallicity following \citealt{2009ARA&A..47..481A}) in units of the proton mass, $m_\text{p}$, and $\Delta x$ is the average cells size inside the cold cloud, which can be calculated as the ratio of the last two columns in Table~\ref{Table:SimulationOverview}. In 2D we define $m_\text{target}\equiv  (\Delta x)^2 n_\text{cold} \mu m_\text{p}\cdot 1 \text{ kpc}$, since a 1 kpc height is assumed in the $z$-direction in the 2D calculations. Gas cells with masses exceeding $2 m_\text{target}$ are refined.

Adopting the same mass resolution throughout the simulation box implies a higher spatial resolution in dense regions in comparison to diffuse regions. We obtain a factor of $\sqrt{n_\text{cold}/n_\text{hot}}\simeq 31.6$ higher spatial resolution inside the cold cloud in comparison to the hot wind in the initial conditions of the 2D simulations. For the 3D setup this factor is $(n_\text{cold}/n_\text{hot})^{1/3}=10$. The hydrodynamical processes affecting the clouds are, however, mostly happening near the contact boundary between the hot and cold gas. To better resolve this boundary, we adopt a \emph{volume refinement criterion}, which ensures that the volume of a cell does not exceed a factor of $\beta$ in comparison to its neighbour cells. We choose $\beta = 4$ in 2D simulations and $\beta = 8$ for the 3D cases. With these values of $\beta$ the volume refinement criterion is analogous to requiring neighbouring cells in an adaptive-mesh-refinement code to differ at most by one refinement level. To speed up the simulations adaptive time-steps are used. Overall, this setup ensures that the majority of the computational power is spent on simulating the cold clouds and their immediate surroundings.

\subsection{Detailed simulation setup}

All simulations are carried out in a rectangular box domain. For the 2D simulations the box dimensions in the $x$- and $y$-directions are $40 R_\text{cloud}$ and $80 R_\text{cloud}$, respectively. For the simulations with 607 cells per $R_\text{cloud}$ the hot wind is initialised on a uniform grid with $768 \times 1536 $ cells. For the simulation with 304 cells per $R_\text{cloud}$ a grid of $384 \times 768 $ cells is used instead (this scaling continues down to lower resolution). The centre of the cold cloud is defined to be $(x,y) = (20 R_\text{cloud}, 10 R_\text{cloud})$. All hot gas cells inside the cloud radius are then removed and replaced with mesh-generating points with a density corresponding to $n_\text{cloud}$.

Starting from $y=0$ the first two rows of cells are static and have a density, temperature, volume and metallicity that is equal to the wind properties. The rest of the cells in the simulations move with the Lagrangian flow. The wind in our windtunnel setup is blowing in the $\hat{y}$-direction with a speed, $\varv_\text{inject}$. Cells near the $x$-edges have properties fixed to the windtunnel injection properties of the wind to prevent the bow shock from propagating into the direction of the cloud, which would happen otherwise, if we adopted periodic boundaries for the $x$-dimension (and for the $z$-dimension, in the 3D simulations).

In 3D a smaller ratio of box-size to cloud-size is chosen to limit the number of cells, and thus make the simulations less expensive. We choose $L_x = 8\times R_\text{cloud}$ and $L_y / L_x=4$. The hot wind is initially distributed on a grid with $N_x\times N_y \times N_z = 128\times 512 \times 128$ cells for the 3D-simulations with 160 cells per cloud radius (for the low-resolution simulation we use a grid of size $64\times 256 \times 64$). The initial position of the cloud is $(x,y,z) = (4 R_\text{cloud}, 8 R_\text{cloud},4 R_\text{cloud})$. The behaviour of the injection region and the sampling of cloud cells are all done similarly to the 2D setup.

For the cylindrical simulations we select a box with $L_x=L_z=16R_\text{cloud}$ and $L_y/L_x=2$, and the cloud is initiated at $(x,y,z) = (8 R_\text{cloud}, 10 R_\text{cloud},8 R_\text{cloud})$. The hot wind is initialised with $N_x\times N_y \times N_z = 128\times 256 \times 128$ cells. Cells near the $y=0$ boundary have the density, temperature and velocity fixed to the injection values.

We run all simulations for a time of $10 t_\text{cc}$, except for {\tt 3D-10pc-160} and {\tt 3D-100pc-160}, which we run for $13 t_\text{cc}$, because the clouds here have longer survival times. If a cloud in any of the simulations gets near the upper $y$-boundary, we make a spatial translation in the $y$-direction, so the cloud is in the centre of the box.

\begin{figure*}
    \centering
    \begin{minipage}{.24\textwidth}
        \centering
        \includegraphics[width=\linewidth]{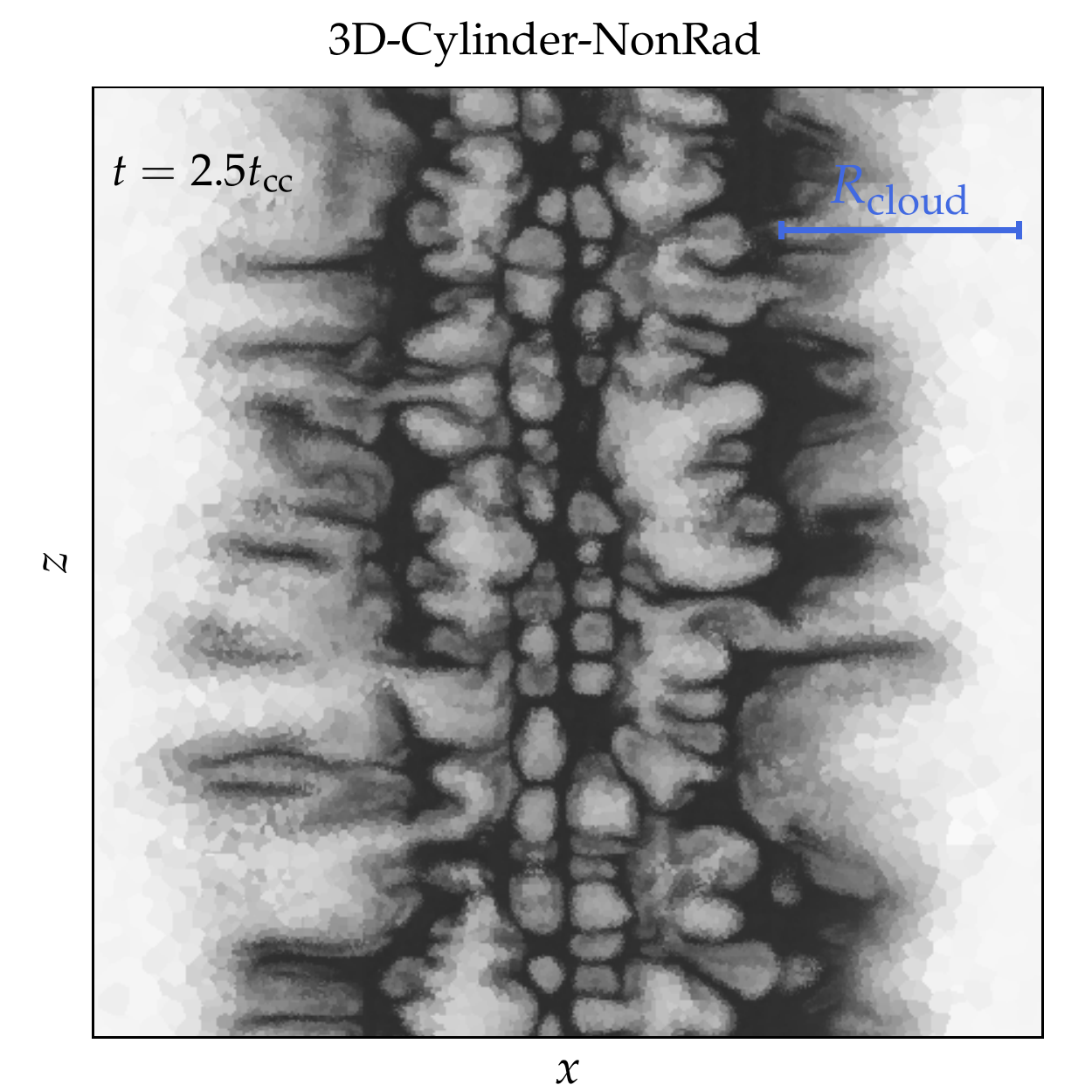}
    \end{minipage}%
    \begin{minipage}{.24\textwidth}
        \centering
        \includegraphics[width=\linewidth]{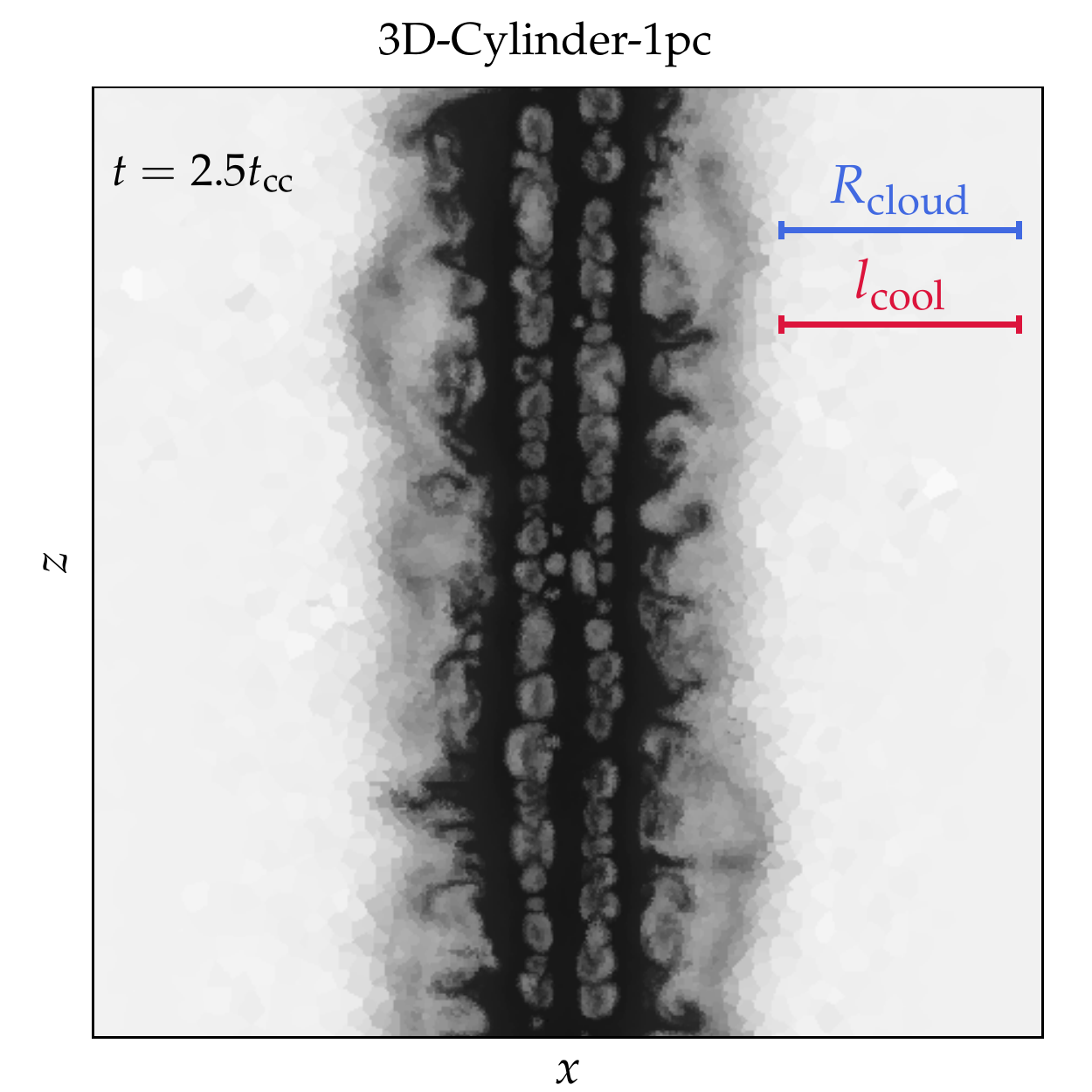}
    \end{minipage}
    \begin{minipage}{.24\textwidth}
        \centering
        \includegraphics[width=\linewidth]{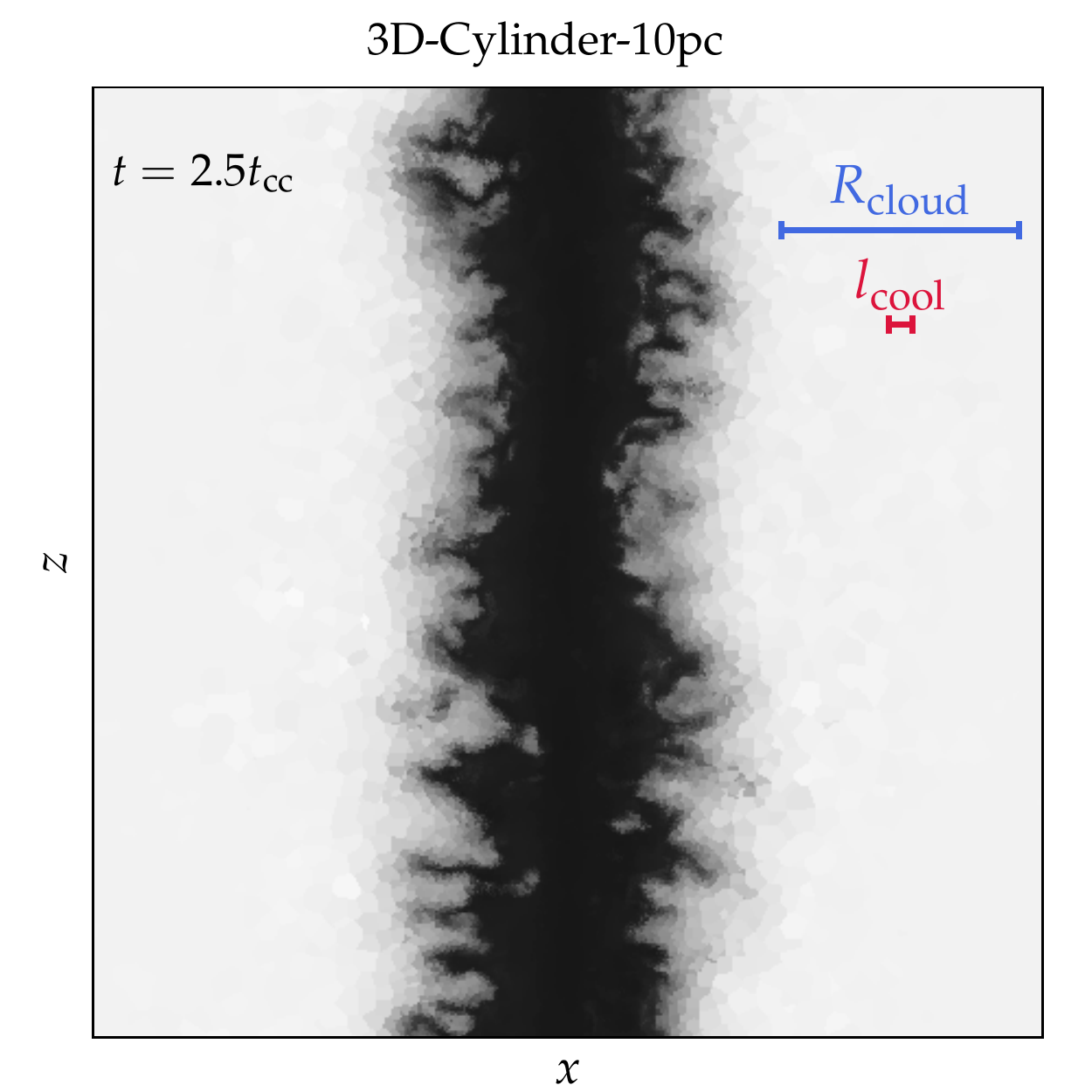}
    \end{minipage}
    \begin{minipage}{.24\textwidth}
        \centering
        \includegraphics[width=\linewidth]{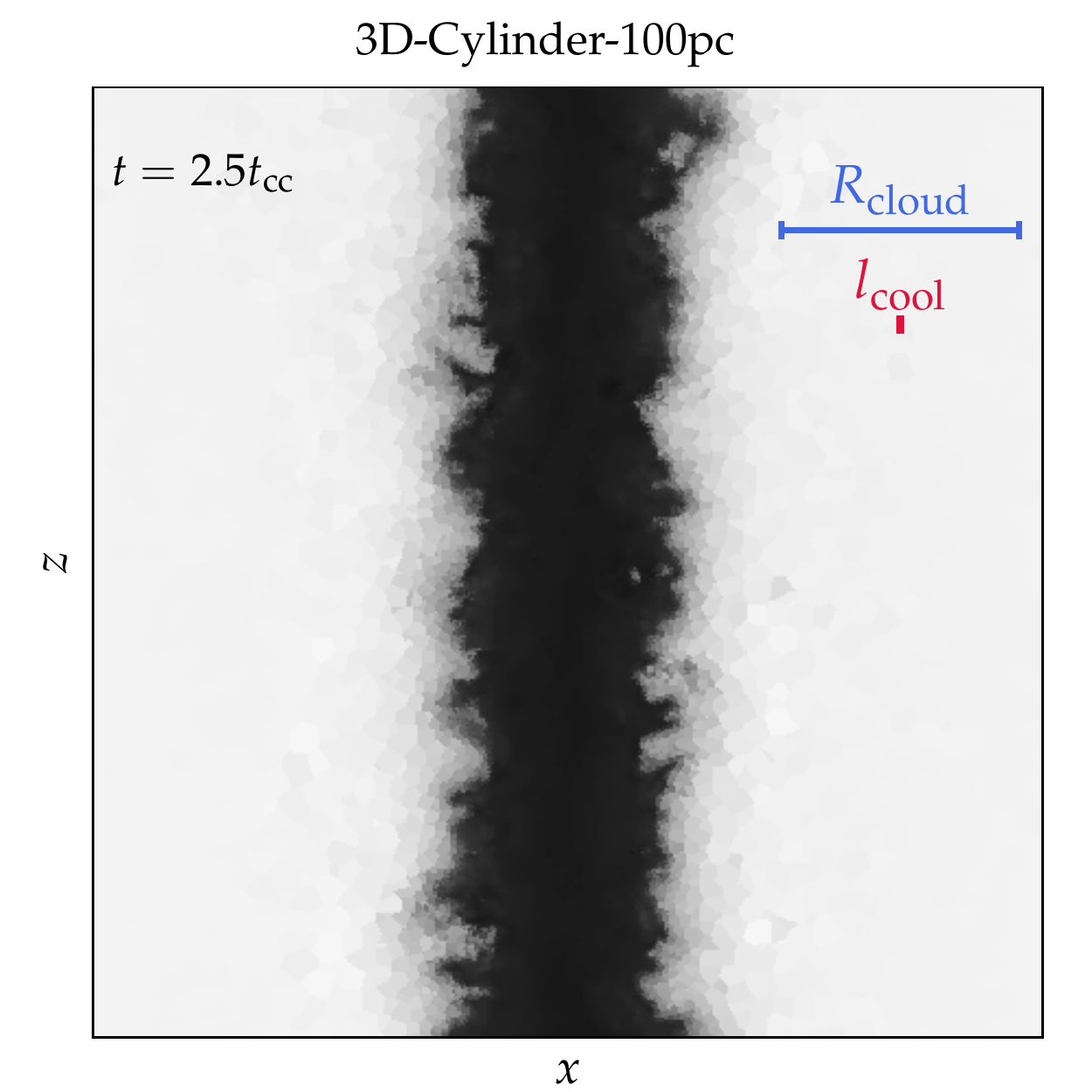}
    \end{minipage}\\
    \centering
    \begin{minipage}{.24\textwidth}
        \centering
        \includegraphics[width=\linewidth]{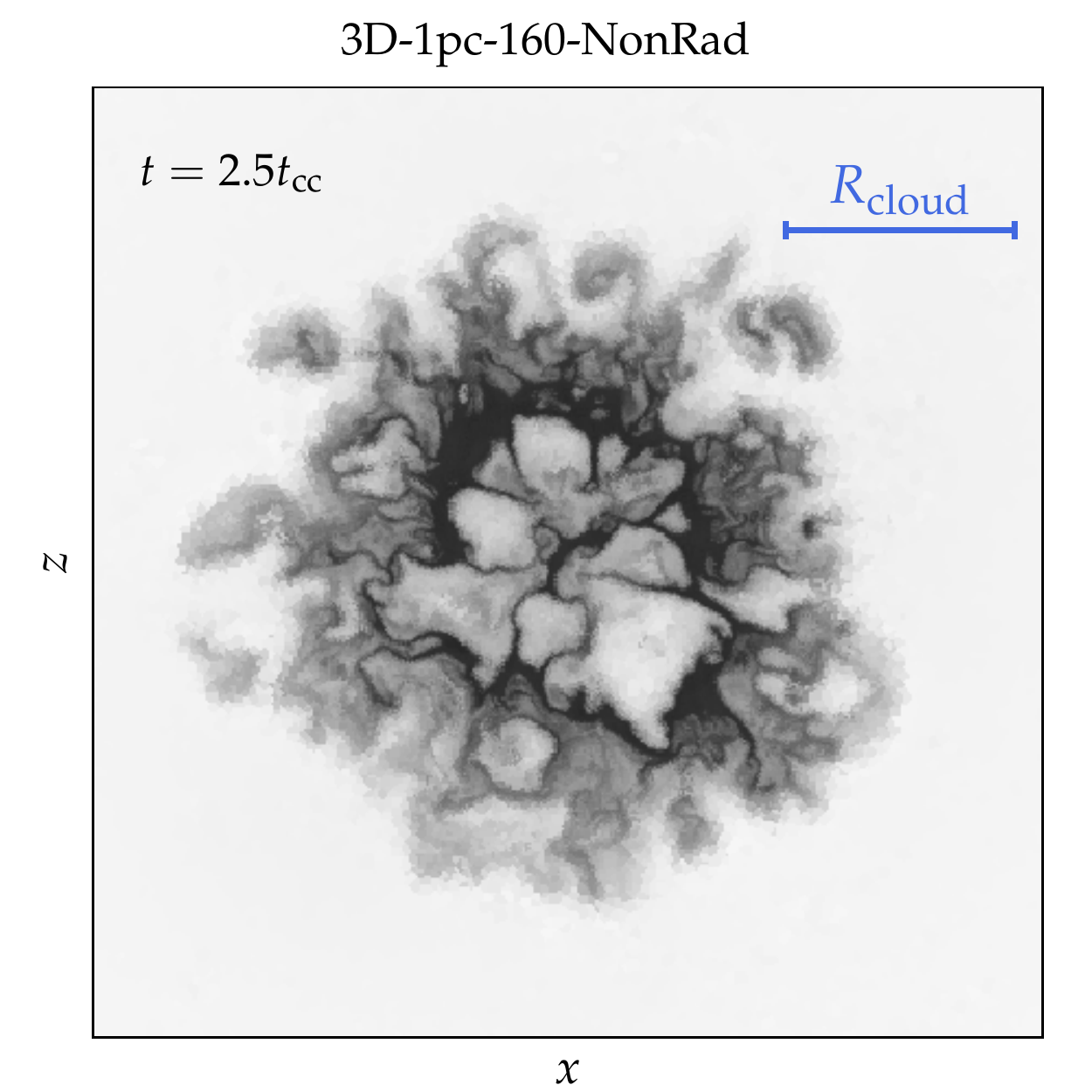}
    \end{minipage}%
    \begin{minipage}{.24\textwidth}
        \centering
        \includegraphics[width=\linewidth]{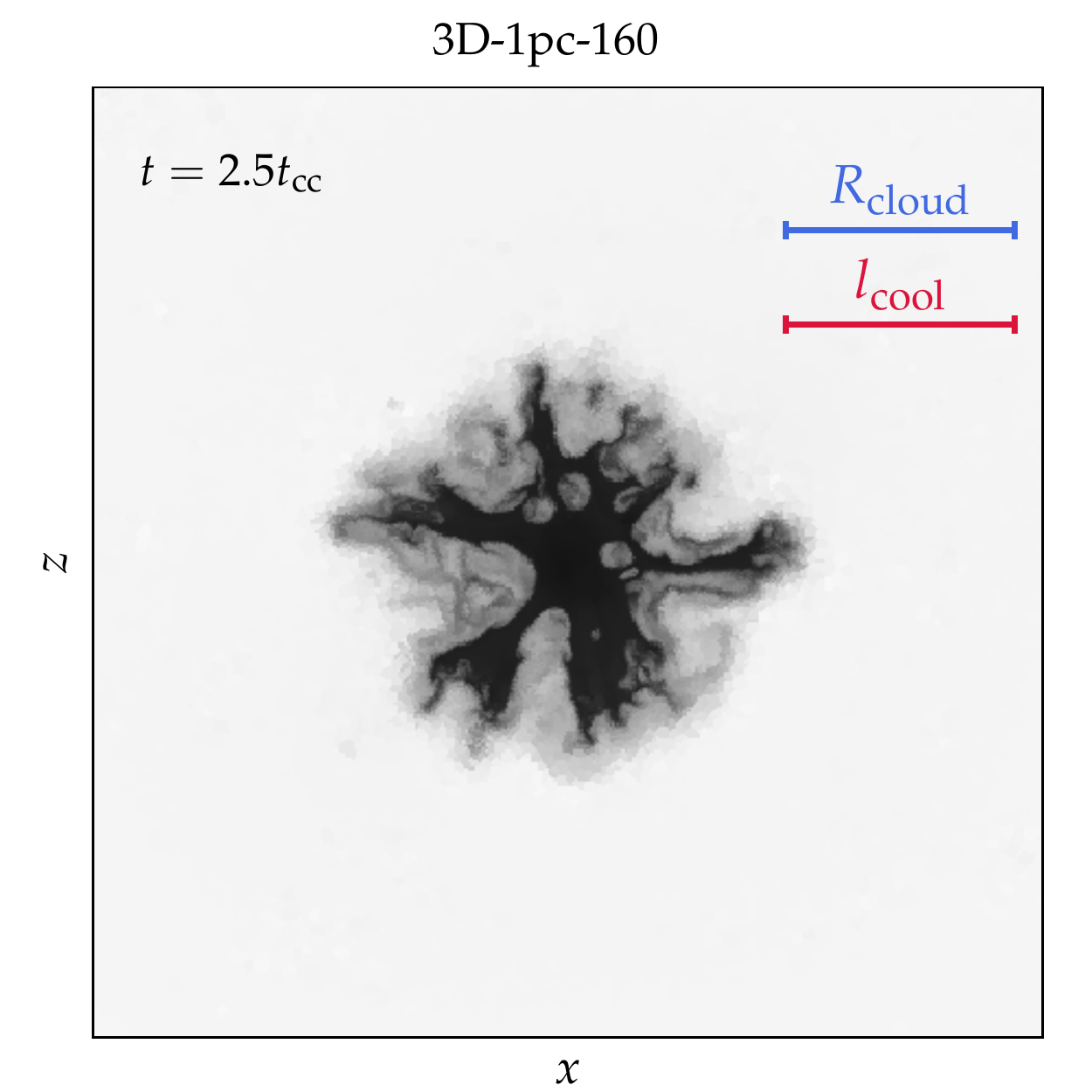}
    \end{minipage}%
    \begin{minipage}{.24\textwidth}
        \centering
        \includegraphics[width=\linewidth]{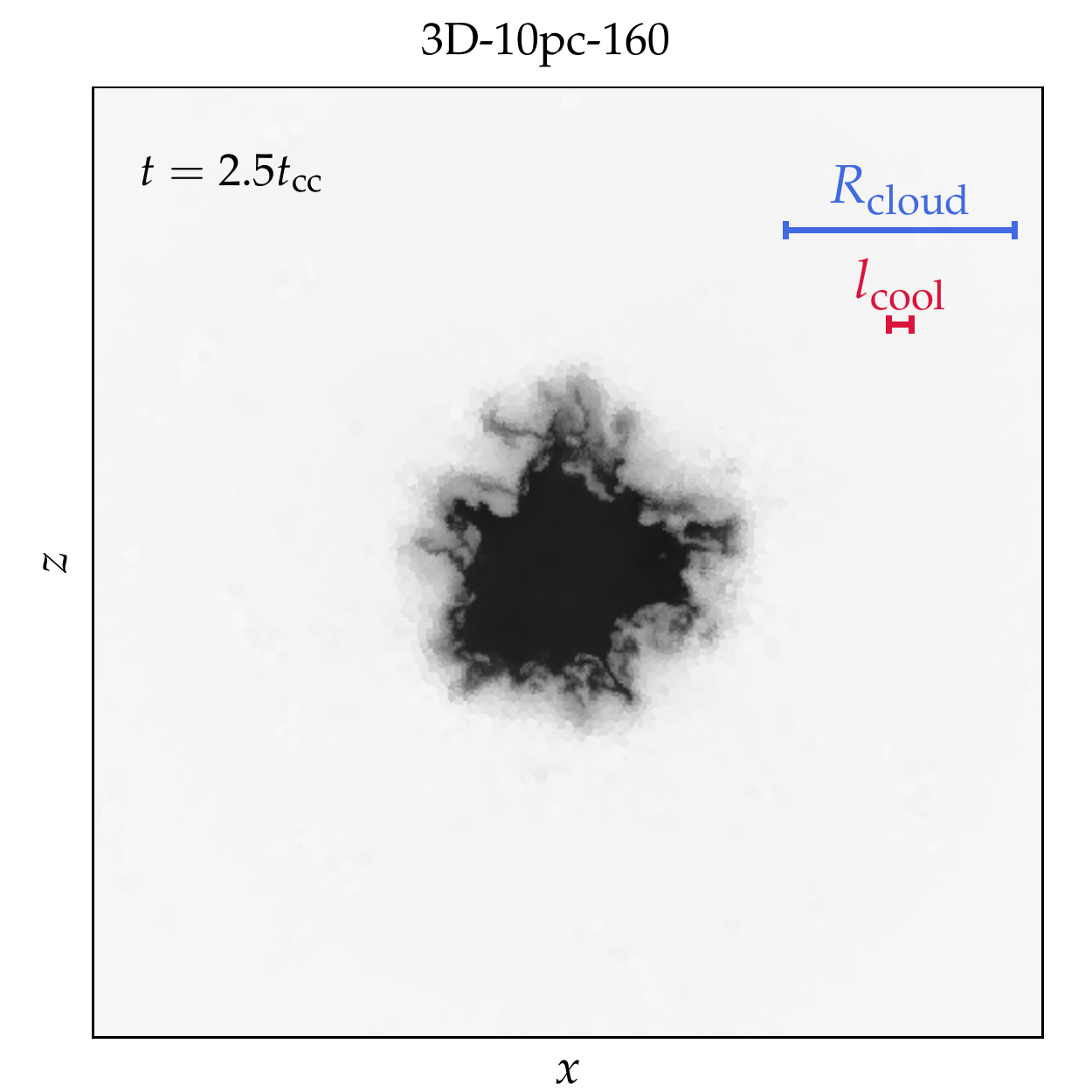}
    \end{minipage}%
    \begin{minipage}{.24\textwidth}
        \centering
        \includegraphics[width=\linewidth]{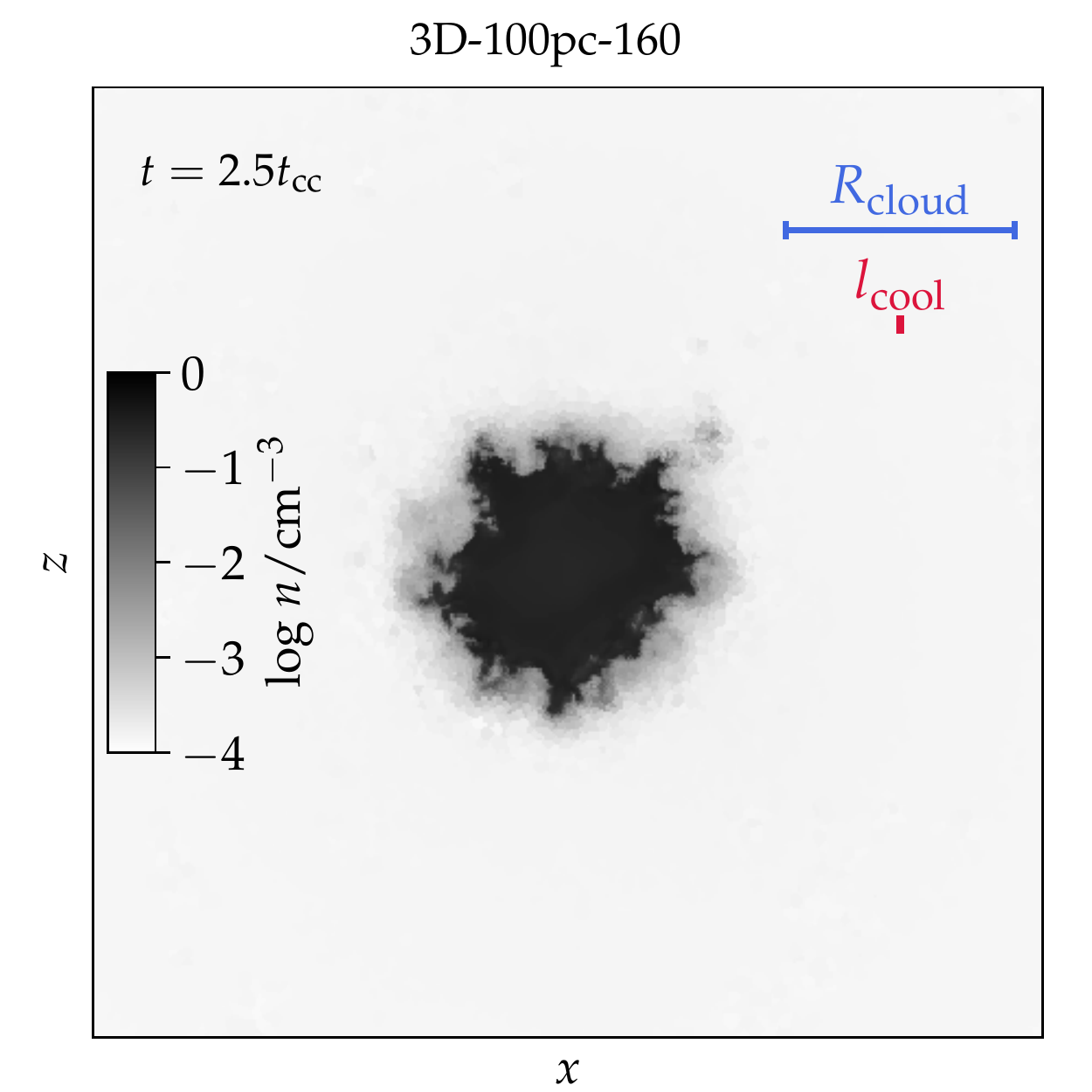}
    \end{minipage}%

        \caption{Upper panels: a slice through the cylinder at $y=y_\text{centre}-R_\text{cloud}/2$ showing the centre of the simulation box with $|x-L_x/2|\leq 2 R_\text{cloud}$. The hot wind is blowing into the figure in the projection shown. In the non-radiative and 1 pc simulation an instability creates cavities in the cylinders. For the large cylinder with a 100 pc radius radiative cooling suppresses the growth of this instability. Lower panels: the same but for the high-resolution 3D sphere simulations. Here the instability is also only efficient for the non-radiative and the 1 pc simulations with cooling; for the 10 pc and 100 pc spheres the instability is suppressed. In 2D such instabilities cannot develop, because a $z$-axis is not included. This explains the large spread in the lifetime of the 3D spheres compared to 2D spheres (from Fig.~\ref{Fig1CloudSurvival}).}
\label{Fig02Instability}
\end{figure*}

\begin{figure*}
 	  \centering
    \begin{minipage}{.47\textwidth}
        \centering
        \includegraphics[width=\linewidth]{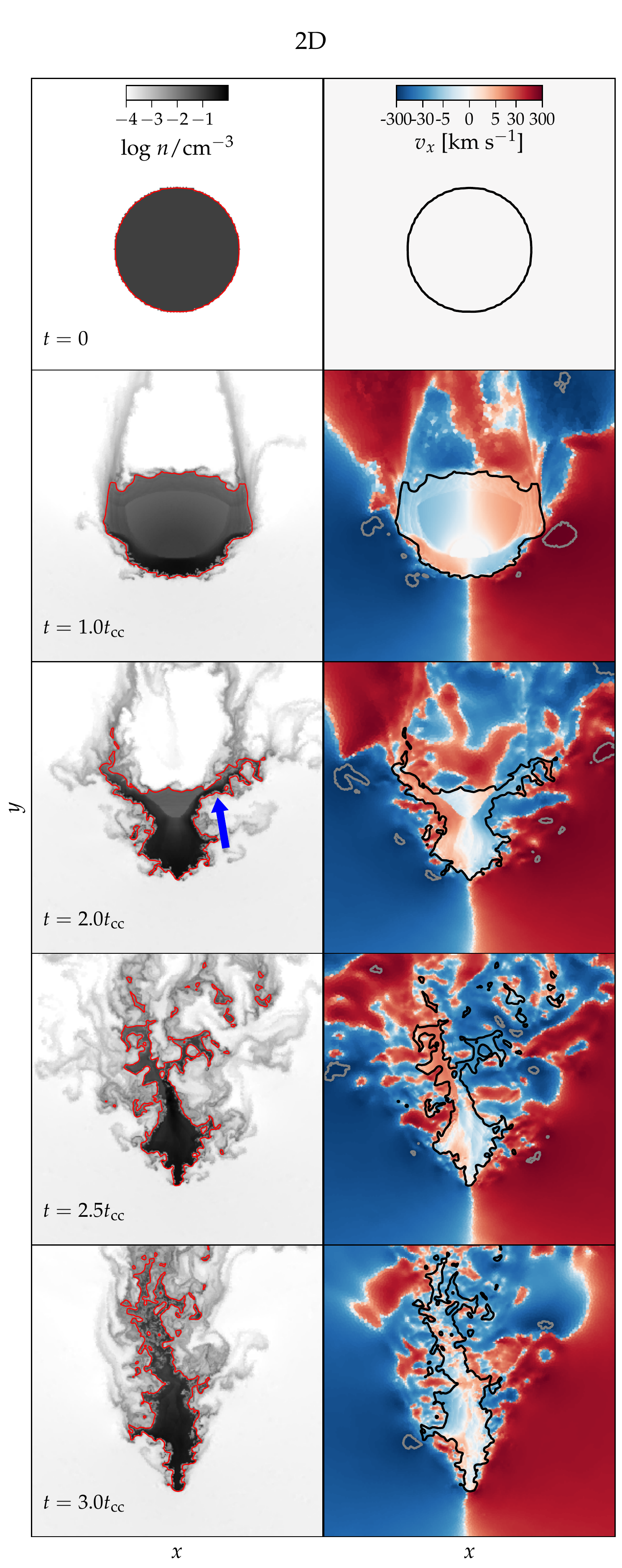}
    \end{minipage}%
    \begin{minipage}{.47\textwidth}
        \centering
        \includegraphics[width=\linewidth]{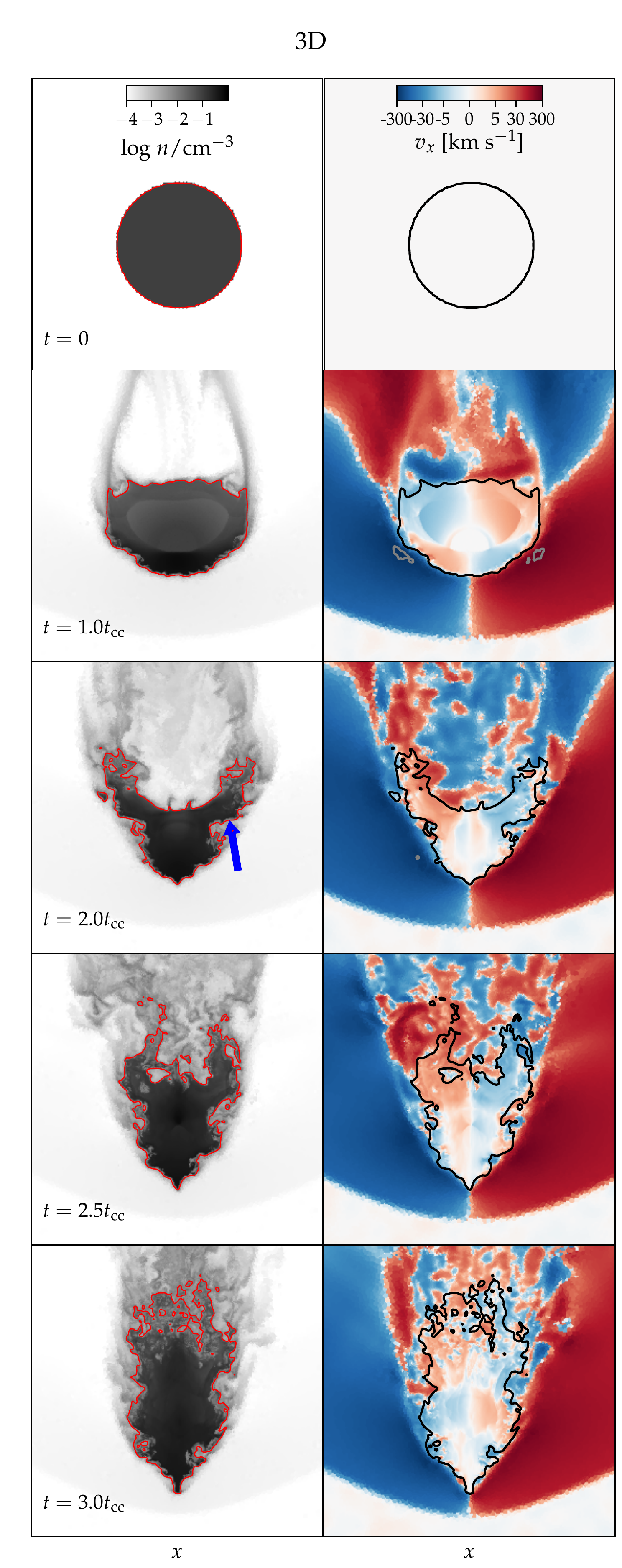}
    \end{minipage}%
    \caption{The early evolution of the density and $\varv_x$ for the 10 pc 2D and 3D spheres ({\tt 2D-10pc-607} and {\tt 3D-10pc-160}, respectively). The 3D simulations are visualised by a slice in the $z=L_z/2$ plane. We also show a density contour of $n=10^{-2}~\rmn{cm}^{-3}$ in the density (red line) and velocity panels (black line). From $t=0$ to $t=2t_\text{cc}$ the 2D and 3D spheres evolve almost identically, both in terms of the density field and the $\varv_x$-value inside the clouds. The peak value of $\varv_x$ is slightly larger in 2D than in 3D; this is revealed by a larger size of velocity peaks indicated by the grey contours showing regions with $|\varv_x|\geq 265$ km s$^{-1}$. The downstream \emph{tails} (indicated by the arrows at $t=2t_\text{cc}$), however, evolve very differently in 2D and 3D. At $t=2.5 t_\text{cc}$ these tails have been stripped and transformed into downstream cloudlets/fragments in 2D, but in 3D their material remains attached to the main-cloud. }
\label{Fig03EarlyEvolution}
\end{figure*}

\section{Results}\label{2D3D}

\subsection{Cloud survival mass fractions}

From a theoretical point of view cloud crushing simulations in 2D and 3D are expected to behave differently. A first insight can be obtained by analysing the flow around a 2D sphere, a 3D sphere and a 3D cylinder. The flow around a cylinder initially corresponds to the flow around a 2D sphere. This correspondence remains true until fluid instabilities break the cylindrical symmetry of the gaseous 3D cylinder.

These differences and correspondences are revealed in the left panel of Fig.~\ref{Fig1CloudSurvival}, which shows the time-evolution of the survival mass fraction of gas with $n\geq n_\text{cloud}/3$ for the non-radiative simulations of the 2D-sphere, the 3D-sphere and the 3D cylinder. The time is normalised to $t_\text{cc}$ and the dense gas mass is shown in units of the initial cloud mass. With these normalisations the non-radiative simulations are self-similar. The same metric is also used by \citet{2015ApJ...805..158S} and \citet{2017ApJ...834..144S}. For the cleanest possible comparison the simulations of the 3D sphere and 3D cylinder have the exact same resolution; the simulations are {\tt 3D-Cylinder-NonRad} and {\tt 3D-1pc-80-NonRad}, which have $R_\text{cloud}/\Delta x =80$ inside the cold cloud. For the 2D simulations it is impossible with our mass-criterion for refinement to match the resolution to the 3D simulations both inside and outside the cloud. Instead we show a grey band bounded by the non-radiative 2D simulations ({\tt 2D-1pc-607-NonRad}, {\tt 2D-1pc-304-NonRad}, {\tt 2D-1pc-152-NonRad}), which we determine to be well converged in Appendix~\ref{Sec:ShatteringAndConvergence}.

The geometrical correspondence between the flow around a 2D sphere and a 3D cylinder is reflected by their almost identical evolution in Fig.~\ref{Fig1CloudSurvival} (left panel) until $2.5 t_\text{cc}$. After this time the correspondence breaks down, because a 3D instability starts to dissolve the 3D cylinder. This instability is visualised for the non-radiative cylinder in the upper left panel of Fig.~\ref{Fig02Instability}. Since the 2D sphere corresponds to a 3D cylinder simulation, where the symmetry along the $z$-axis is strictly enforced, it is not surprising that the 2D sphere survives longer than the 3D simulations (as we saw in Fig.~\ref{Fig1CloudSurvival}, left panel).

At early times, $t\lesssim 2.5 t_\text{cc}$, the 3D-sphere simulation behaves slightly (but still significantly) different than the 2D-sphere and the 3D-cylinder. At later time the dense gas in the 3D sphere is also evaporated much faster than for the two other simulation geometries. For a non-radiative simulation setup it is evidently easier for a hot wind to penetrate the 3D sphere in comparison to the 2D sphere or a 3D cylinder.

The high-resolution 2D sphere simulations ({\tt 2D-1pc-607-NonRad}, {\tt 2D-1pc-607}, {\tt 2D-10pc-607} and {\tt 2D-100pc-607}) are shown in the central panel of Fig.~\ref{Fig1CloudSurvival}. The presence of radiative cooling in a simulation extends the cloud lifetime, in comparison to the non-radiative simulation. This can for example be concluded by the time it takes to evaporate half of the original cold-cloud mass. The same conclusion is reached for the 2D simulations in \citet{2017MNRAS.470..114A}. It is expected that cooling is more efficient in larger than in smaller clouds. The cooling time is mainly a function of temperature, whereas the cloud crushing time-scale is proportional to the radius of a cloud (see Eq. \ref{tcc}). In large clouds, the relative importance of cooling is therefore larger, which explains why large clouds survive longer than small clouds.

The right panel of Fig.~\ref{Fig1CloudSurvival} shows the same quantity for the 3D sphere simulations. The trend that large clouds survive longer in the presence of cooling is also seen in 3D. \citet{2015ApJ...805..158S} used 3D cloud crushing simulations with $R_\text{cloud}=100$ pc to provide fitting functions for the survival time of clouds in a starburst wind following \citet{1985Natur.317...44C}. When considering clouds with different $R_\text{cloud}$-values, we note that their estimate of cloud survival time (which we have summarised in our Section~\ref{Intro}) may change by a factor of a few.

Remarkably, we find that the relative difference of survival time-scales for clouds with different radii is larger in 3D than in 2D. This relative difference can be estimated by comparing the time, at which half of the cloud mass is evaporated, e.g., for the non-radiative and 10 pc simulations. In 2D the 10 pc simulation has a 25--30 per cent longer survival time, whereas the increase is around 100 per cent in 3D. This larger difference in 3D can be explained by instabilities, which grow differently in 3D if they are not suppressed by cooling (see also Section~\ref{sec:early} for an in-depth discussion of this point).

Fig.~\ref{Fig02Instability} (upper panels) demonstrate the presence of such instabilities in the four cylindrical simulations. Each panel shows a slice in the $x$--$z$-plane, and the $y$-coordinate of each slice is chosen to be downstream from the centre of the cloud at $y_\text{centre}-R_\text{cloud}/2$, where $y_\text{centre}$ marks the centre of the cylinder, which has been determined by fitting a normal distribution to a mass-histogram along the $y$-axis. For the non-radiative simulation and the 1 pc simulation with radiative cooling an instability creates holes in the cylinders. For the 10 pc and 100 pc simulations the instability is suppressed by radiative cooling. A similar instability is present in the 3D sphere simulations seen in the lower panels of Fig.~\ref{Fig02Instability}. From left to right cooling plays a gradually larger role, which leads to a stronger suppression of the instability.

A 2D simulation formally corresponds to a 3D simulation, where symmetry is strictly enforced in the $z$-direction. It is therefore impossible for the 2D simulations to take the growth of these instabilities into account. In 3D these instabilities are suppressed by cooling for the 10 and 100 pc radiative simulations. This provides the explanation for the larger relative difference in cloud survival times for small and big clouds in 3D in comparison to the 2D case.

\begin{figure*}
\centering
\includegraphics[width=1.0\linewidth]{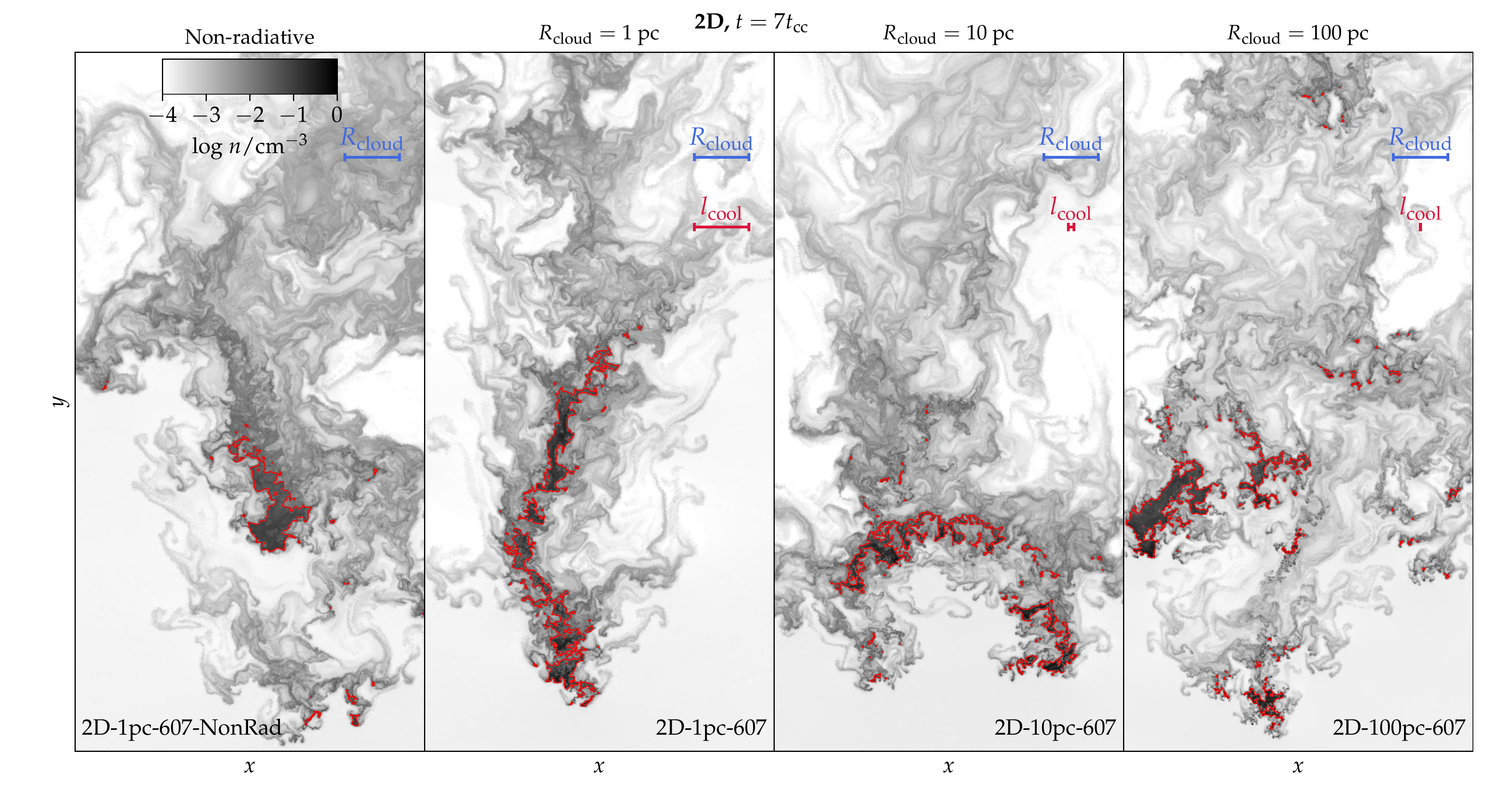}
\includegraphics[width=1.0\linewidth]{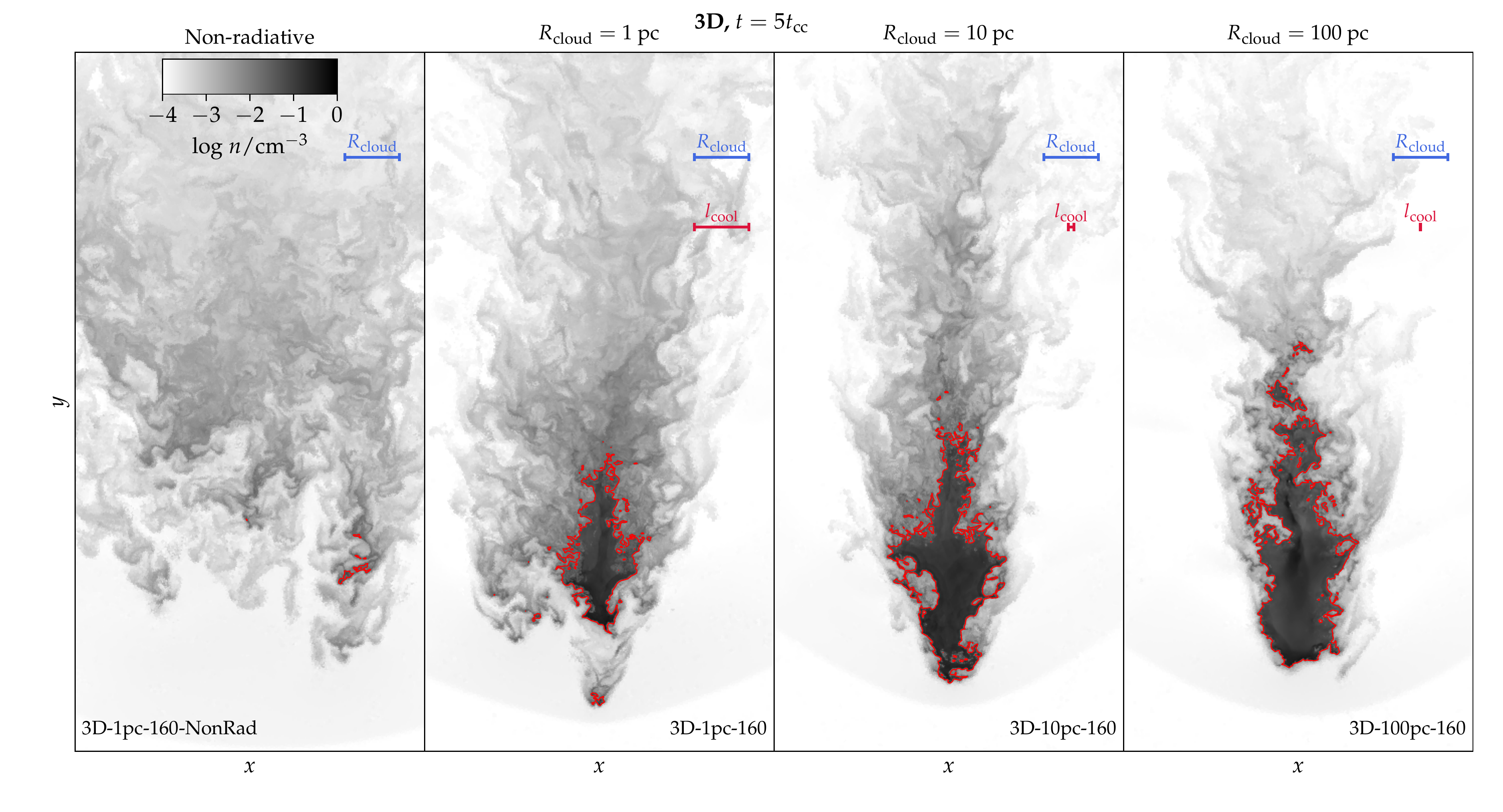}
\caption{Upper panels: the density in the high-resolution 2D simulations at $t=7t_\text{cc}$. From left to right we show the non-radiative simulation and the simulations with radiative cooling for $R_\text{cloud}=1$ pc, $10$ pc and $100$ pc. In each panel the initial cloud radius is shown as well as the cooling length for the initial cloud density (only for the simulations with radiative cooling enabled). To visualise the presence of dense, cold cloudlets/clouds we show red contours encompassing volumes with $n\geq n_\text{cold,init}/4=0.025~\rmn{cm}^{-3}$. Sub-$R_\text{cloud}$ cloudlets are most abundant in the 10 pc and 100 pc simulations, where the cooling length is smaller than the initial cloud radius. This is consistent with the 2D simulations of \citet{2018MNRAS.473.5407M}. Lower panels: slices of the density field in the high-resolution 3D simulations. At $t=5t_\text{cc}$ the cloud of the non-radiative simulation is almost dissolved in the hot wind. The 1 pc, 10 pc and 100 pc simulations with cooling have a gradually larger covering fraction of dense, $n\geq 0.1$ cm$^{-3}$, gas, which is consistent with the longer cloud survival time in Fig.~\ref{Fig1CloudSurvival}. Sub-$R_\text{cloud}$ cloudlets appear slightly more abundant in the simulations of the large clouds, but their presence seems to be less pronounced in comparison to the 2D case.}
\label{Fig04Density}
\end{figure*}

In the right panel of Fig.~\ref{Fig1CloudSurvival} we perform a resolution test of the 3D simulations. The non-radiative, 1 pc and 10 pc clouds exhibit a good agreement between the high- and low-resolution simulations, so convergence is achieved here. The late-time evolution ($t\gtrsim 7t_\text{cc}$) of {\tt 3D-100pc-160} and {\tt 3D-100pc-80} hints that our 100 pc simulations in 3D are not well converged. For {\tt 3D-100pc-160} the cooling length, which is 1 pc for our initial conditions, is resolved with less than 2 cells, so this is a plausible reason for the lack of convergence. When interpreting the results of the {\tt 3D-100pc-160} simulation we note they may not be fully converged.

In Appendix~\ref{Sec:ShatteringAndConvergence} we study convergence of the 2D simulations, and demonstrate that the high-resolution 2D simulations are well converged.

\subsection{The early evolution of clouds in 2D and 3D}
\label{sec:early}

From a visual point of view the 2D- and 3D-spheres behave similarly at the early stages of the simulations. This is for example revealed in the 10 pc simulations with radiative cooling at $t=0,1$ and $2$ $t_\text{cc}$, where the hot wind strips material from the cloud surface (see Fig.~\ref{Fig03EarlyEvolution}). This similarity is easily revealed in the visualisation of the density field, and in addition the $\varv_x$-velocity also reveals that the velocity flow is similar inside and outside the main cloud at these early times.

The first remarkable differences between the 2D- and 3D-spheres arise at $2t_\text{cc}\leq t\leq2.5t_\text{cc}$, where the \emph{wings} of the clouds (marked with a blue arrow in the $t=2t_\text{cc}$ panels of Fig.~\ref{Fig03EarlyEvolution}) are affected differently by the hot wind. In the 2D simulations the wings are accelerated downstream and transformed into several smaller cloudlets shortly after $t=2t_\text{cc}$, whereas the wings in the 3D simulation remain connected to the main cloud. The reason for these differences is that the flow can go above and below the cloud (i.e., in the positive and negative $z$-direction, respectively) in the 3D simulations. Due to the dimensional constraints in 2D all the flow momentum is deposited into the cloud, which causes it to break into small cloudlets via the action of instabilities. This difference is important, because it implies that shattering is less likely to happen in 3D than in 2D at least for the initial stages ($t\leq 3t_\text{cc}$) of our simulations.

\subsubsection{Instabilities}

Overall, Figs.~\ref{Fig02Instability}~and~\ref{Fig03EarlyEvolution} reveal that instabilities grow differently in 2D and 3D simulations. First, instabilities along the $z$-axis are explicitly suppressed in 2D simulations, and secondly, instabilities in the $x$--$y$-plane can grow faster in 2D, because the hot wind cannot use the $z$-direction to flow around the cold--dense gas. Due to these differences it is advisable only to use 3D simulations to provide reliable predictions for the CGM. 2D simulations may still be useful for studying gaseous systems at a very high resolution (at a relatively low computational cost), but their results should be considered with caution.

After the initial shock has propagated through the dense cloud, the main driver of evaporation of the dense cloud is the KH-instability \citep{2007MNRAS.380..963A,1994ApJ...420..213K}. This instability occurs when there is a velocity shear across a boundary, which is the case near the head of the cloud. Another instability is the Richtmyer--Meshkov instability \citep[RM instability; see][]{doi:10.1002/cpa.3160130207,Meshkov1969}, which comes into play when fluids of different densities are accelerated. This occurs when the hot gas interacts with the wings of the cloud, as marked with the blue arrows in Fig.~\ref{Fig03EarlyEvolution}. It follows from the evolution of the wings that the RM instability is able to efficiently trigger the formation of cloudlets in the initial stages of the 2D simulation, but not in 3D. Clouds in 2D and 3D may therefore be affected very differently by the RM instability.

The RT instability that occurs at the interface between dense and dilute gas phases also affects the cloud. The growth of small perturbations from 1 to 2$t_\text{cc}$ into elongated structures (\emph{RT fingers}) at the rear of the 2D and 3D clouds is mainly due to the RT instability. At later times the RT instability likely continues to play a role at the rear of the cloud, where the shear velocity difference between the hot and cold phase is smaller in comparison to the front of the cloud.

\begin{figure*}
\centering
\includegraphics[width = 0.9\linewidth]{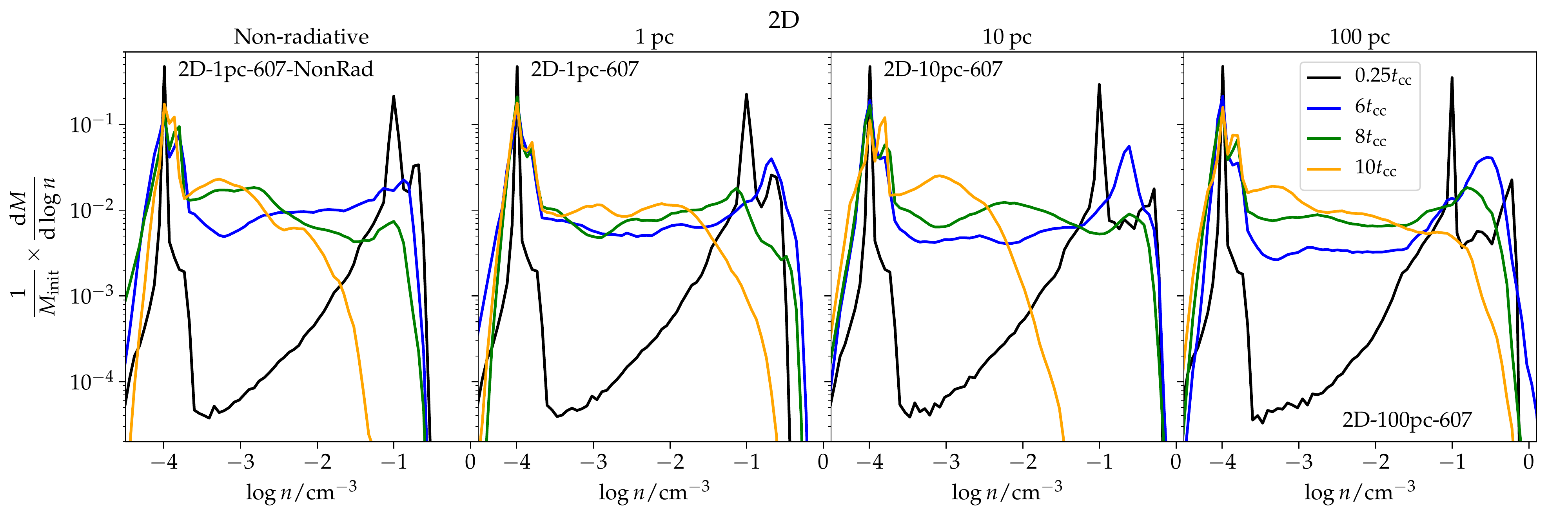}
\includegraphics[width = 0.9\linewidth]{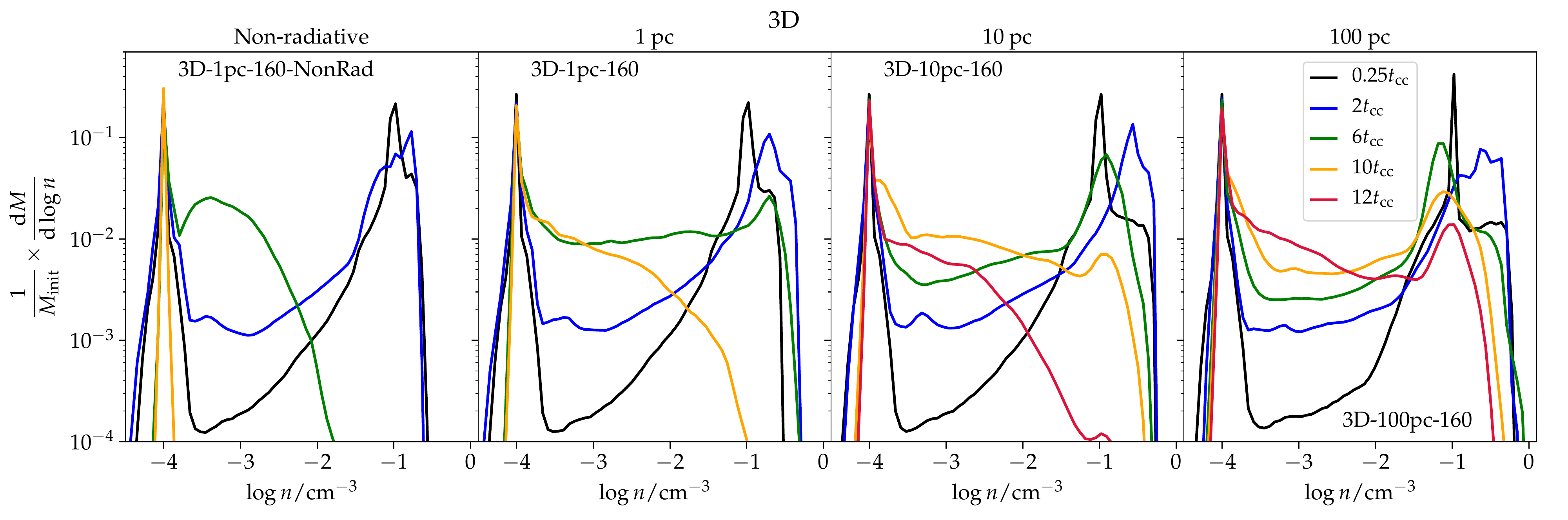}
\caption{Mass-weighted histograms of gas density. The histograms are normalised to the initial mass, $M_\text{init}$, in each simulation. 2D sphere and 3D sphere simulations are shown in upper and lower panels, respectively. The initial distribution of gas is bimodal, with densities $n= 10^{-1}$ and $10^{-4}$~cm$^{-3}$. However, already after $t=0.25t_\text{cc}$ the gas at the cloud-wind interface has started to dissolve, which populates the range of intermediate densities. A significant amount of gas is compressed to higher densities via shocks and radiative cooling (if present). As time evolves the dense gas with $n\geq 0.1$ cm$^{-3}$ is gradually evaporated (see also Fig.~\ref{Fig1CloudSurvival}).}\label{Fig4ADensityHist}
\end{figure*}

\subsubsection{The stand-off distance and potential flow solutions}

Figure~\ref{Fig03EarlyEvolution} indicates that the \emph{stand-off-distance}, which is the distance from the front of the cold cloud to the head of the bow shock, is larger in 2D than in 3D. Indeed, the head of the bow-shock is not even visible in any of the 2D panels. To validate our simulations, we compare the stand-off-distance to the analytical theory of \citet{Moeckel49approximatemethod} in Appendix~\ref{standoffdistance}, and find good agreement. The differences are a direct consequence of the geometrical differences between 2D and 3D spheres.

A better understanding of the differences between the flow around 2D and 3D spheres can be obtained by examining analytical solutions for the flow around them. The flow around a static sphere or cylinder can be obtained analytically under the assumption that the fluid is incompressible and follows a \emph{potential flow}. Such flows satisfy the conditions, $\bs{\nabla\cdot\varv}=0$ and $\bs{\nabla\times\varv}=\bs{0}$. The solutions are derived in Appendix~\ref{Sec:PotentialFlow}.

In these cases, the peak values of the gas velocity around 2D and 3D spheres are $2 \varv_\text{inject}$ and $3 \varv_\text{inject}/2$, respectively. The peak occurs just outside the sphere at $(x,y)=(x_\text{centre}\pm R_\text{cloud},y_\text{centre})$. Such a difference in velocity directly affects the evolution of the cloud, because the time-scale of the KH instability is proportional to the velocity of the gas flow. Hence, in the initial stages, the KH instability grows faster in 2D, which leads to larger density amplitudes and more ablated material (see second-row panels in Fig.~\ref{Fig03EarlyEvolution} at $t=t_\rmn{cc}$). This is exemplified with grey contours that encapsulate regions with $|\varv_x|\geq 265$ km s$^{-1}$ in Fig.~\ref{Fig03EarlyEvolution}. The abundance and sizes of such regions are indeed larger in 2D in comparison with 3D, as predicted by the potential flow solutions.

However, this initial behaviour does not carry over to the non-linear evolution of the clouds in different dimensions. In 3D, instabilities are additionally able to develop along the $z$-axis and eventually lead to a faster disruption of a non-radiative cloud. \citet{2018arXiv180605677M} study the KH instability in a slightly different setup, where the wind flows along the symmetry axis of the cylinder in 3D and along a slab in 2D. Despite of their geometrical setup being different from ours, we can still draw useful analogies between our findings. E.g., they find that instabilities develop faster in 3D in comparison to 2D, which is perfectly consistent with our non-radiative simulations of 2D and 3D spheres. \citet{2018arXiv180605677M} furthermore establish that body- and surface-modes are present in their cylinders, which is again consistent with Fig.~\ref{Fig02Instability}.

\begin{figure*}
    \centering
    \begin{minipage}{.44\textwidth}
        \centering
        \includegraphics[width=\linewidth]{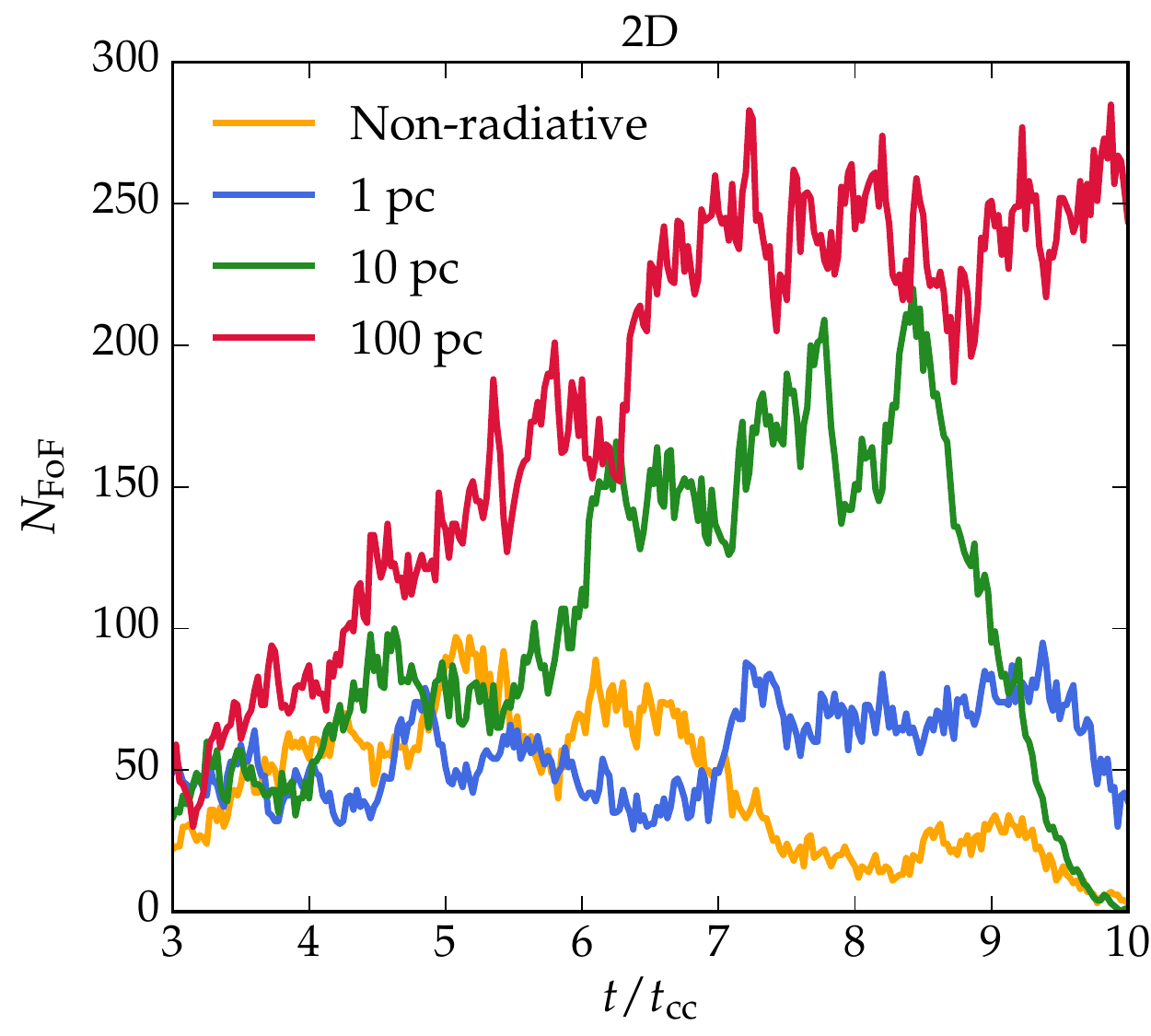}
    \end{minipage}%
    \begin{minipage}{0.44\textwidth}
        \centering
        \includegraphics[width=\linewidth]{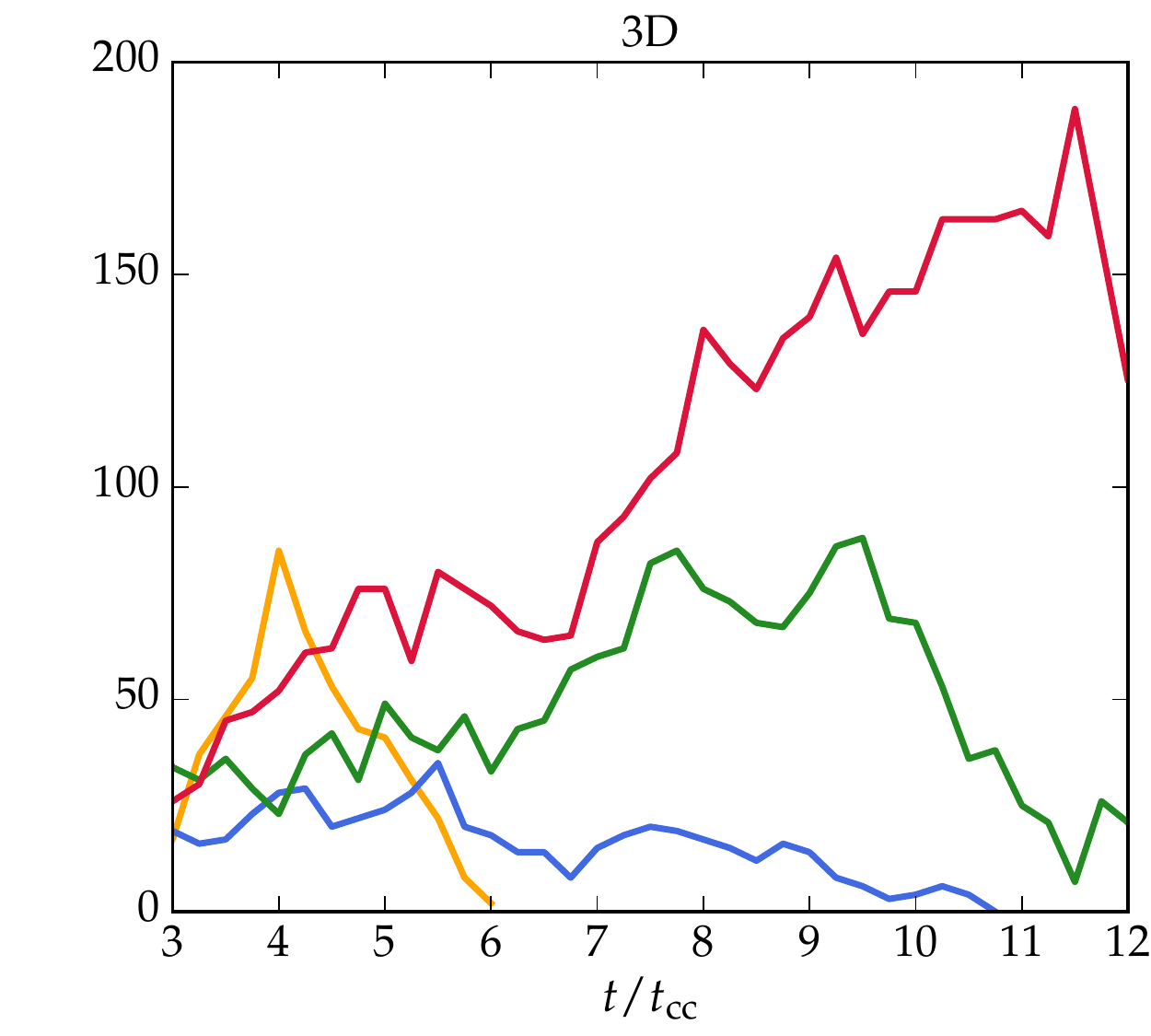}
    \end{minipage}
        \caption{The time-evolution of the number of friends-of-friends groups, $N_\text{FoF}$, which consist of connected regions with cell densities above $n\geq 0.25 n_\text{cloud}$. $N_\text{FoF}$ is a measure of the level of fragmentation of a simulation. Left panel: the simulations of the 2D spheres. As also revealed by the visual analysis, the non-radiative and 1 pc cooling simulations have a similar level of fragmentation, and the 10 pc and 100 pc are more fragmented. Right panel: the 3D sphere simulations. For the simulations with 1 pc, 10 pc and 100 pc the clouds become gradually more fragmented, especially after $6t_\text{cc}$.}
\label{Fig5FoF}
\end{figure*}

\subsection{The density distribution}

As in \citet{2018MNRAS.473.5407M} we first inspect the density distribution of the 2D sphere simulations at $t=7t_\text{cc}$ in the upper panels of Fig.~\ref{Fig04Density}. In the 100 pc simulation many dense cloudlets with  $n\gtrsim 0.1 $ cm$^{-3}$ are visible, and the 10 pc simulation also reveals structures on a much smaller scale than the initial cloud radius. Such sub-$R_\text{cloud}$ structures are essentially absent in the 1 pc and the non-radiative simulations. Overall, our 2D simulations are in very good agreement with \citet{2018MNRAS.473.5407M}, who predicts fragmentation to occur when $R_\text{cloud}\gg l_\text{cool}$.

The lower panels of Fig.~\ref{Fig04Density} show the visual appearance of the 3D spheres. We have shifted the analysis time to $t=5t_\text{cc}$, so that the 1 pc simulation with radiative cooling has a mass survival fraction that corresponds to the value at $t=7t_\text{cc}$ for the 2D sphere simulation (see Fig.~\ref{Fig1CloudSurvival}). The presence of cloudlets is visible downstream from the main cloud in the 100 pc simulation. In the other simulations such cloudlets are not easily visible. The non-radiative simulation is almost evaporated, which precludes the survival of long-lived cloudlets. Based on the visual inspection of the density it is unclear when fragmentation becomes important in 3D. To conclusively address this issue we will now introduce quantitative measures of the degree of fragmentation in the simulations.

First, we show the evolution of the density field in form of mass-weighted density-histograms in Fig.~\ref{Fig4ADensityHist}. Each histogram is normalised by the mass at $t=0$, $M_\text{init}$, and the bins are distributed evenly in $\log n$. We show a selection of times chosen to highlight the evaporation of differently-sized clouds. Note, that different times are chosen for the 2D and 3D simulations, respectively.

At $t=0.25t_\text{cc}$ the 2D simulations exhibit two peaks near the initial cold-cloud- and hot-wind-densities at $n= 10^{-1}$ and $10^{-4}$~cm$^{-3}$, respectively. At this early time a small fraction of gas has already been shock-compressed (and cooled) to exceed densities of $n \simeq 0.25$~cm$^{-3}$ in all simulations. By comparing the density histograms at different times it can be seen how the dense phase with $n\geq 0.1$ cm$^{-3}$ is gradually dissolved into gas with intermediate densities of $n= 10^{-4}$ to $10^{-1}$~cm$^{-3}$. At $t=10 t_\text{cc}$ most of the dense gas of the initial cold cloud is dissolved in all simulations, except for the case of $R_\text{cloud}=100$ pc case, which has a longer lifetime as already shown in Fig.~\ref{Fig1CloudSurvival}.

As in 2D, the 3D simulations also show a bimodal density-distribution at $t=0.25 t_\text{cc}$ with an imprint of the initial conditions. We have already established that there is a larger spread in the survival time of dense gas in 3D in comparison to 2D. This is also reflected by the density histograms. The 1 pc and 10 pc simulations with radiative cooling are almost depleted of dense gas with $n\geq 0.1$ cm$^{-3}$ at 10 and 12$t_\text{cc}$, respectively, which is consistent with Fig.~\ref{Fig1CloudSurvival}. As the dense gas dissolves we again see that intermediate densities are present; a behaviour, which is consistent with the study of multiphase gas in \citet{2017ApJ...834..144S}.

\begin{figure*}
\centering
\includegraphics[width = 1.0\linewidth]{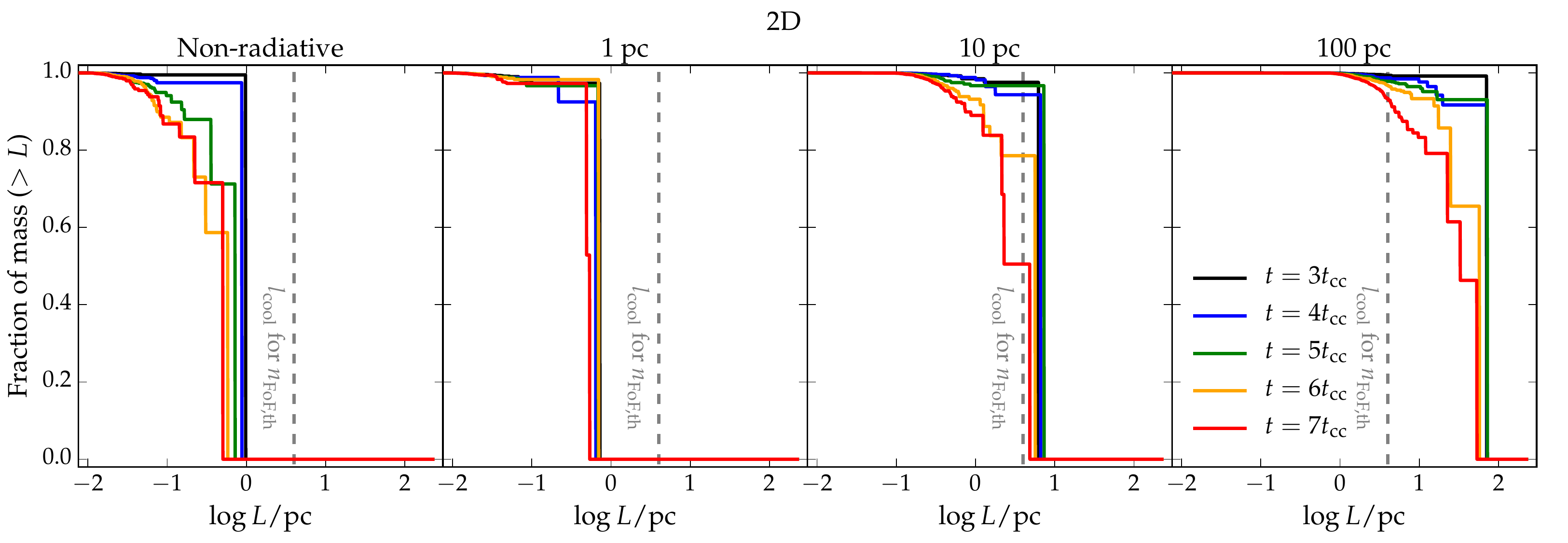}
\centering
\includegraphics[width=1.0\linewidth]{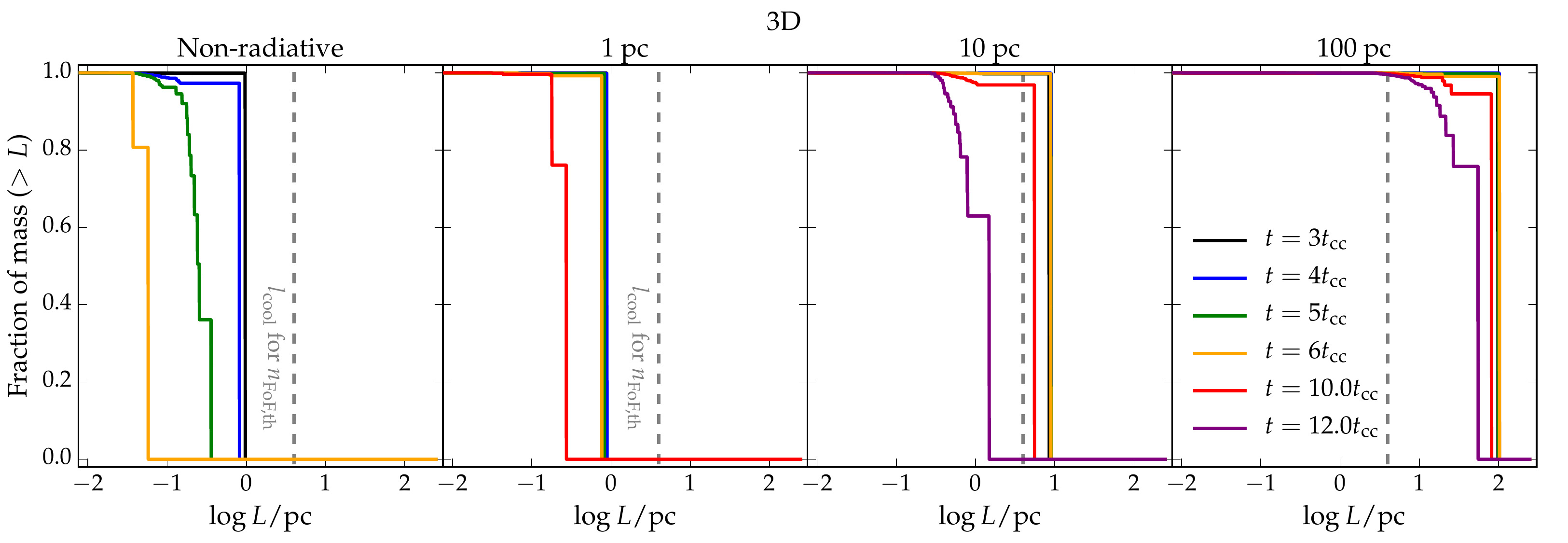}
\caption{Upper panels: the evolution of the cloud length in the 2D simulations ({\tt 2D-1pc-607-NonRad}, {\tt 2D-1pc-607}, {\tt 2D-10pc-607} and {\tt 2D-100pc-607}, from left to right). The cooling length corresponding to $n_\text{FoF,th}=0.25n_\text{cloud}$ is marked in each panel. When radiative cooling is enabled clouds with sizes larger than this cooling length fragment. This fragmentation is accompanied by an overall increased mass of dense gas in the simulation (Fig.~\ref{Fig1CloudSurvival}). The non-radiative simulation also fragments, but here the dense gas survival fraction is smaller than in the radiative cooling simulations. Here, the fragmentation is a signature of the cloud being close to final disruption. Lower panels: the evolution of the FoF-mass function in 3D simulations ({\tt 3D-1pc-160-NonRad}, {\tt 3D-1pc-160}, {\tt 3D-10pc-160} and {\tt 3D-100pc-160}). The very different time-evolution of the dense gas survival fraction in Fig.~\ref{Fig1CloudSurvival} complicates interpretation of the evolution of the 3D FoF mass functions in the context of the shattering hypothesis of \citet{2018MNRAS.473.5407M}.}
\label{Fig017A}
\end{figure*}

\subsection{Quantifying fragmentation with a friends-of-friends analysis}\label{FoFSubsec}

To quantify the presence of small-scale structures in the simulations, we construct friends-of-friends (FoF) groups consisting of connected cells above a density threshold, $n_\text{FoF,th}$, which is a free parameter in the analysis. A \emph{linking length} is defined based on the maximum linear cell size allowed by the refinement criterion for a cell with this threshold density. First, we assess the evolution of FoF groups in our simulations and then study the evolution of the FoF mass function.

\subsubsection{Evolution of friends-of-friends groups}

The fiducial value of $n_\text{FoF,th}$ is selected to be $0.25n_\text{cloud}$. In Appendix~\ref{Sec:FoFMass} we check that our results are insensitive to the exact value of this parameter. In order to only include relatively dense and resolved clouds in the FoF-catalogues, we require the peak density of a group to be above $0.5 n_\text{cloud}$. Furthermore, a cloud should contain at least 20 cells. $N_\text{FoF}$ refers to the number of FoF-groups full-filling these criteria.

Fig.~\ref{Fig5FoF} (left panel) shows the number of FoF groups as a function of time in the high-resolution 2D simulations. This confirms the visual appearance of the density maps; in simulations with radiative cooling fragmentation becomes gradually more important when the cloud size becomes larger than the cooling length. The 100 pc simulation has many more FoF-groups than the other simulations already at $t\gtrsim 4 t_\text{cc}$. The 1 pc and 10 pc simulations with cooling behave relatively similar until $t\simeq 6 t_\text{cc}$, but then more FoF groups appear in the latter simulation, implying more fragmentation. The non-radiative and the 1 pc cooling simulations have a similar number of FoF-groups throughout the simulations.

The FoF-analysis for the 3D simulations are shown in the right panel of Fig.~\ref{Fig5FoF}. The main result from the 2D simulations -- that cooling causes fragmentation of clouds with $R_\text{cloud}\gg l_\text{cool}$ -- is also obtained in 3D. This is the first time this has been established in 3D simulations.

To further quantify the analysis, we examine mass-functions of the FoF-groups in Section~\ref{SecFoFMassFunction}. For the 2D simulations these reveal evidence for shattering in clouds with $R_\text{cloud}\gg l_\text{cool}$, but in the 3D simulations, different effects shape the mass functions, which are very sensitive to the different survival times seen for the dense gas in 3D.

\subsubsection{Friends-of-friends mass function}\label{SecFoFMassFunction}

With our FoF-analysis it is possible to obtain additional information about the evolution of FoF-groups. As a measure of the cloud-size distribution we generate a mass-weighed cumulative histogram of the distribution of cloud size, $L$, which for the 2D simulations is defined as $L\equiv \sqrt{A/\pi}$, where $A$ is the area of a FoF group. Figure~\ref{Fig017A} shows the time-evolution of this distribution for the 2D simulations. At $t=3 t_\text{cc}$ all simulations contain one large cloud with a linear scale comparable to the initial cloud radius.

At $t=7t_\text{cc}$ the non-radiative simulation is more fragmented than the 1 pc simulation with cooling. This is because the dense gas is almost dissolved in the non-radiative simulation with only 15 per cent of its original mass remaining, whereas the 1 pc radiative cooling simulation has a two times higher dense gas mass (see Fig.~\ref{Fig1CloudSurvival}). This is also consistent with the non-radiative simulations having more FoF groups than the simulation with cooling enabled at this time (Fig.~\ref{Fig5FoF}).

At the same time the survival fraction of cold material is similar for the 1, 10 and 100 pc simulations with radiative cooling (40-50 per cent), and this makes it straightforward to interpret their evolution in Fig.~\ref{Fig017A}. The 1 pc simulation remains stable against fragmentation, and for the 10 and 100 pc cloud simulations fragmentation is gradually more important. The mass fraction of structure in sub-$R_\text{cloud}$ cloudlets is relatively small, but if small clouds are abundant in number (but not necessarily in mass) they may still have a high area covering fraction.

In Fig.~\ref{Fig017A} the cooling length corresponding to the $n_\text{FoF,th}$-value is marked by a dashed vertical line. Given that the boundary of a FoF-group has a density similar to $n_\text{FoF,th}$, this is the relevant cooling length for a FoF group. Following the hypothesis of \citet{2018MNRAS.473.5407M} clouds with sizes above this length will fragment, and smaller clouds should be more stable to fragmentation. Of the simulations with radiative cooling, only the 1 pc cloud is able to resist excessive fragmentation until $7t_\text{cc}$, and this is also the only simulation with an initial cloud size $\lesssim$ the cooling length. The 2D simulations are thus in good agreement with the shattering hypothesis from \citet{2018MNRAS.473.5407M}.

For the 3D simulations we define the cloud size as $L\equiv [3V/4\pi]^{1/3}$, where $V$ is the cell volume. The lower panels of Fig.~\ref{Fig017A} show that the interpretation of the FoF-group masses is complicated by the very different survival times of the 3D clouds revealed in Fig.~\ref{Fig1CloudSurvival}. {\tt 3D-1pc-160-NonRad} has a rich structure of sub-$R_\text{cloud}$ cloudlets at $t=5t_\text{cc}$, but these are not long-lived, since the mass in the dense phase reaches a fraction of zero shortly after this time. A comparison between Fig.~\ref{Fig1CloudSurvival} and Fig.~\ref{Fig017A} reveals that the evolution of the cloud-mass-function for the 3D simulations is mainly driven by the difference in dense gas survival fraction, rather than probing shattering. To establish shattering it is thus more appropriate to study the power spectrum evolution, or the actual peak number of FoF-groups for the 3D simulations than the FoF-mass-distribution.

\begin{figure*}
\centering
\includegraphics[width = 1.0\linewidth]{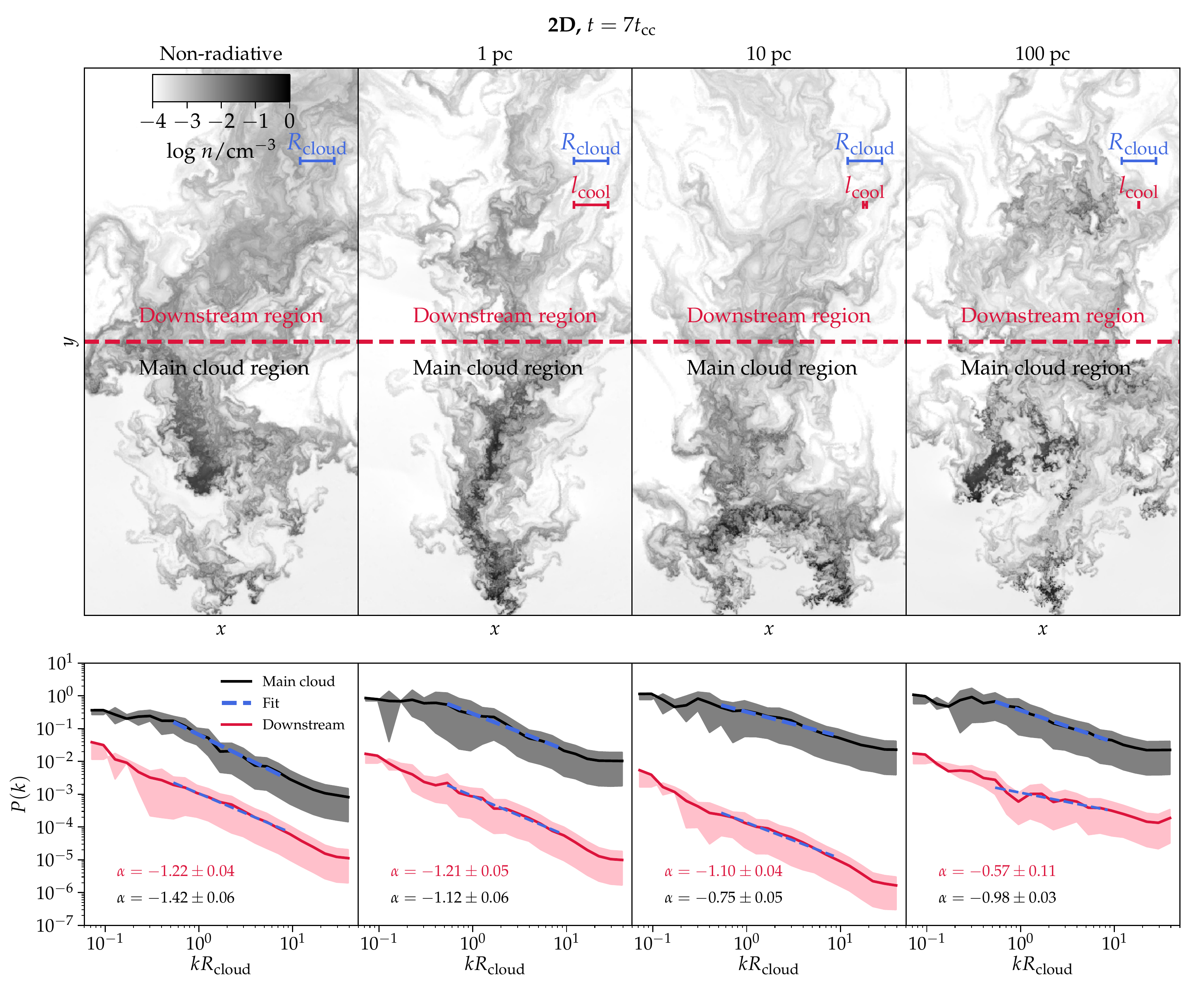}
\caption{Upper panels: the density field of the 2D simulations is split into two regions, the \emph{main cloud region} and the \emph{downstream region}, divided by the dashed horizontal line. The power spectrum for each region is shown in the lower panels. The mean power spectrum is shown with solid lines, and the $16-84\%$ percentiles are indicated by the shaded areas. Power law fits are performed (thick dashed lines), and the resulting slopes, $\alpha$, are shown in the lower left corner.}
\label{Fig6PowerSpectrum}
\end{figure*}

\begin{figure*}
\centering
\includegraphics[width = 1.0\linewidth]{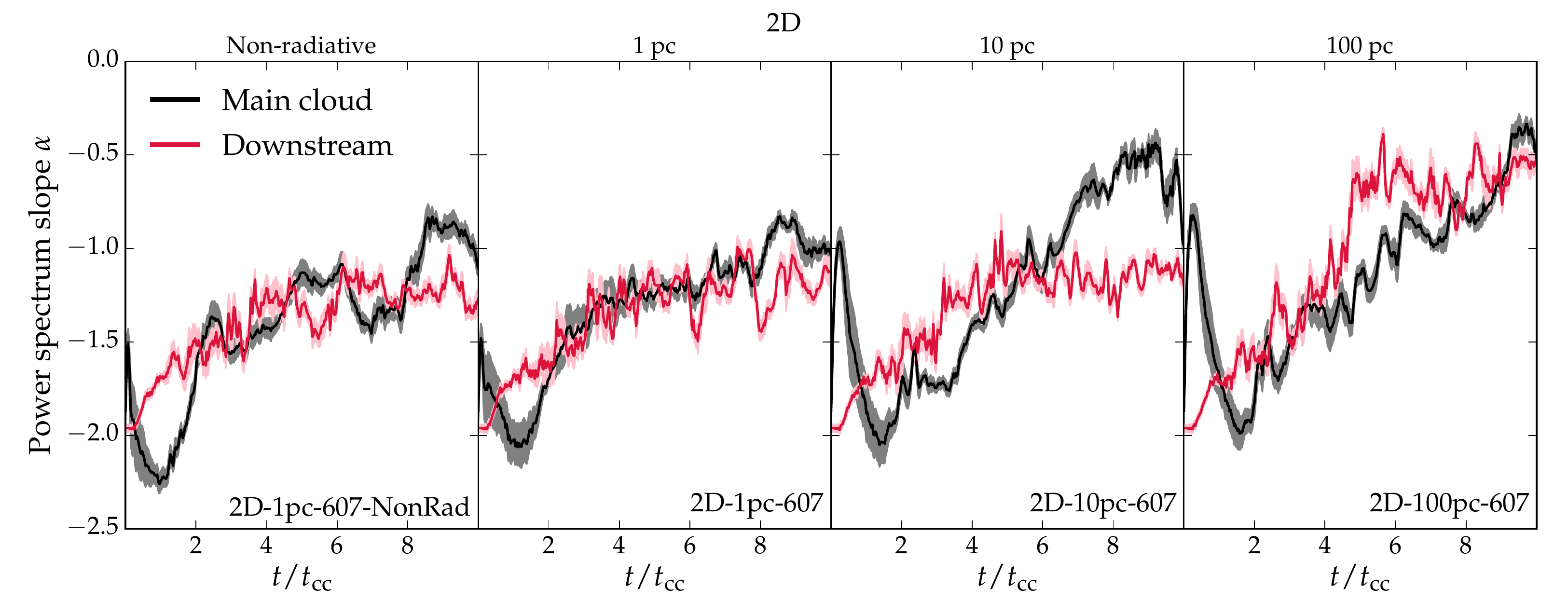}
\includegraphics[width = 1.0\linewidth]{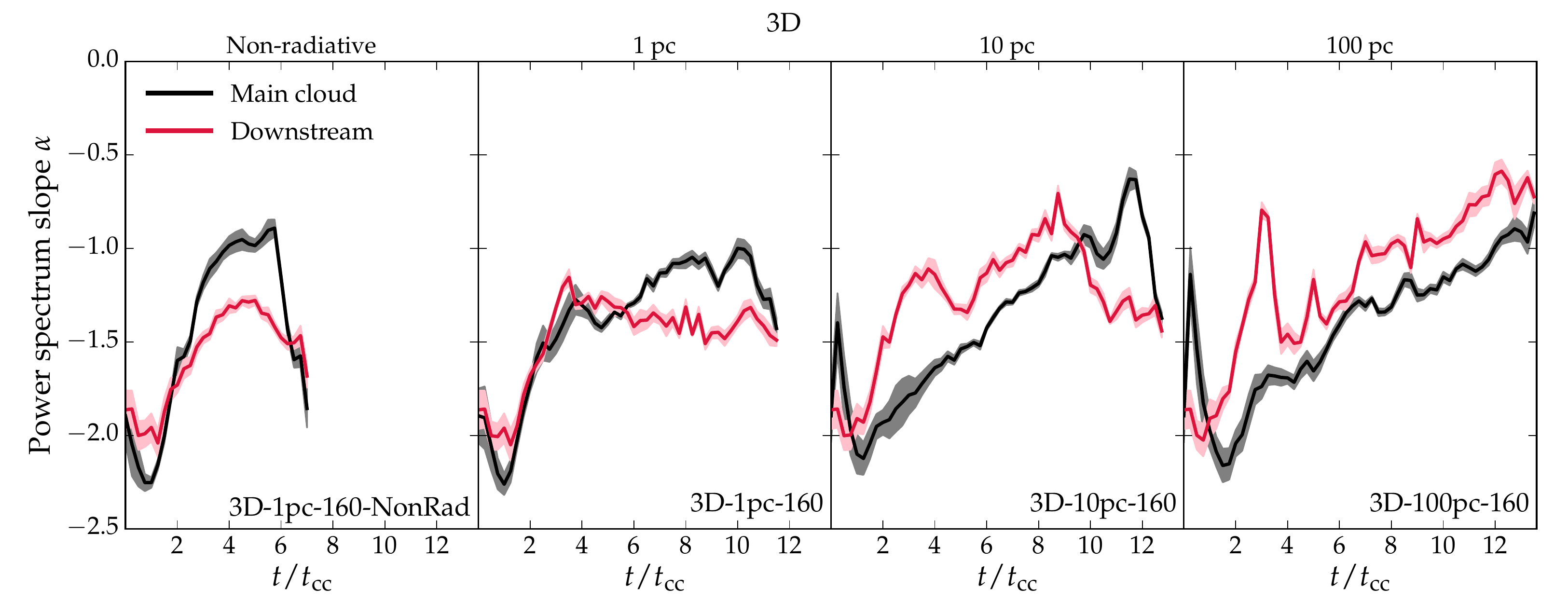}
\caption{Upper panels: the evolution of the spectral slope for the 2D simulations. For a 1 pc cloud the evolution of the spectral index is comparable to the non-radiative simulation. Near the end of the 100 pc simulation the fragmentation caused by cooling makes the spectral slope shallower than in the other simulations. In the main cloud region of the 10 pc simulation the same is seen. Lower panels: the same for the high-resolution 3D sphere simulations. The peak value of the slope is slightly larger for the 10 pc and 100 pc simulation than for the 1 pc (radiative) simulation. This is a consequence of the fragmentation of clouds with $R_\text{cloud}\gg l_\text{cool}$, as predicted by \citet{2018MNRAS.473.5407M}.}
\label{Fig7SpectralSlope}
\end{figure*}

\subsection{Power spectrum analysis}

As a complementary method of characterising the gas structure in the simulations, we perform a power spectrum analysis of the density field.

\subsubsection{Calculation of the 2D power spectrum}

For each 2D simulation two quadratic regions are defined, as shown in Fig.~\ref{Fig6PowerSpectrum}. One region contains the main cloud of the simulation and the other is located downstream. The dimension of each region is $8 R_\text{cloud}\times 8 R_\text{cloud}$. The head of the main cloud, $y_\text{head}$, is determined to be the 1\% percentile of the $y$-coordinates of all gas cells with $n\geq 10n_\text{inject}$. The centre in the $x$-direction, $x_\text{centre}$, is defined to be the median $x$-coordinate of the same gas cells. The main cloud region is then defined as having $|x-x_\text{centre}|\leq 4 R_\text{cloud}$ and $ y_\text{head}\leq y \leq y_\text{head}+8 R_\text{cloud}$. The downstream region has the same $x$-coordinates, and $ y_\text{head}+8 R_\text{cloud}\leq y \leq y_\text{head}+16 R_\text{cloud}$.

For both regions the density field is mapped on to a $512\times 512$ grid from which we compute the 2D Fourier transform. Prior to the transformation the signal is zero-padded. The spherically averaged power spectrum is
\begin{align*}
P(k) = \frac{2\pi k}{N} \left\langle \left| \hat{\rho} \right|^2 \right\rangle_k,
\end{align*}
where $\hat{\rho}$ is the Fourier transform of $\rho$, $k$ is the wave number, and $N$ is the number of mesh cells. $P(k)$ is then calculated in 24 bins distributed evenly in $\log k$. In the lower panels of Fig.~\ref{Fig6PowerSpectrum} the binned power spectra are shown together with the 16-84\% percentiles demonstrating the scatter in each $\log k$-bin. In the range $ -0.3 \leq \log k R_\text{cloud}\leq 1$, we perform a power law fit to the power spectrum. In the figure the fit is shown as a thick-dashed line and the slope, $\text{d}\log P(k)/\text{d}\log k$, is shown in the lower-left corner. Following this procedure the spectral slope, $\alpha \equiv \text{d} \log P(k)/ \text{d}\log k$, is determined for all the snapshots of the simulation.

\subsubsection{Calculation of the 3D power spectrum}

For the 3D simulation a 3D power spectrum is calculated in a similar fashion. $y_\text{head}$ is determined exactly as in 2D, and $x_\text{centre}$ and $z_\text{centre}$ are fixed to be in the centre of the box. The main cloud region is chosen to be $|x-x_\text{centre}|\leq 2 R_\text{cloud}$, $|z-z_\text{centre}|\leq 2 R_\text{cloud}$ and $y_\text{head}\leq y \leq y_\text{head}+4 R_\text{cloud}$. For the downstream regions the $y$-coordinates are shifted to $y_\text{head}+4 R_\text{cloud}\leq y \leq y_\text{head}+8 R_\text{cloud}$. The Fourier transform is then calculated on a $128^3$ grid for each of the two regions, and the 3D power spectrum is computed as
\begin{align*}
P(k) = \frac{4\pi k^2}{N} \left\langle \left| \hat{\rho} \right|^2 \right\rangle_k,
\end{align*}
and fitted in the same way as in 2D.

\subsubsection{Evolution of the power spectral slope}

The evolution of the power spectral slope, $\alpha$, which we determine based on the above fits, is plotted in Fig.~\ref{Fig7SpectralSlope}. We remind the reader that a large value of the power spectral slope (i.e., a shallower power spectrum) corresponds to more structure on sub-$R_\text{cloud}$ scales. Thus, $\alpha$ serves as a diagnostic of shattering. The simulations with radiative cooling behave remarkably similar in 2D and 3D. For the 100 pc simulations the spectral slope increases monotonically throughout the simulation (in both 2D and 3D), and for the 1 pc simulations the slope does not change much after $4 t_\text{cc}$. For the 10 pc simulations the slope increases monotonically in the main-cloud regions, but it stagnates in the downstream regions after $5 t_\text{cc}$ in 2D and after $4 t_\text{cc}$ in 3D. Near the end of the simulation the power spectra are more shallow in the 100 pc simulations compared to any of the other clouds. This is consistent with the FoF-analysis, which reveals more sub-$R_\text{cloud}$ cloudlets in the 100 pc cloud simulations.

For the robustness of our results it is reassuring that the slope evolves almost identically in the downstream- and main-cloud-regions for the 1 pc and 100 pc simulations with radiative cooling (in both 2D and 3D). Based on these simulations we are able to conclude that radiative cooling causes a shallower power spectrum in clouds with $R_\text{cloud}\gg l_\text{cool}$.

\subsection{Numerical convergence}

When studying fragmentation in simulations it is important to check for numerical convergence. In Appendix~\ref{Sec:ShatteringAndConvergence} we perform a comprehensive set of resolution tests. In 2D simulations we find that a resolution of $R_\text{cloud}/\Delta x\geq 152$ is required to obtain an increased number of FoF groups of 100 pc clouds in comparison to 10 pc clouds. An identical result is found in 3D.

For the 3D simulations we also study the convergence of the power spectral slope, and find that a 100 pc cloud is more fragmented in comparison to a 10 pc cloud only for a resolution of $R_\text{cloud}/\Delta x\geq 160$. The conclusion that a 10 and 100 pc cloud is more fragmented than a 1 pc simulation is, however, also obtained for a lower resolution of $R_\text{cloud}/\Delta x= 80$. Our main conclusion that clouds with $R_\text{cloud}\gg l_\text{cool}$ undergo excessive fragmentation in comparison to clouds with $R_\text{cloud}\lesssim l_\text{cool}$ is thus robust.

\section{Discussion}\label{Sec:Discussion}

\begin{figure}
\centering
\includegraphics[width = 1.0\linewidth]{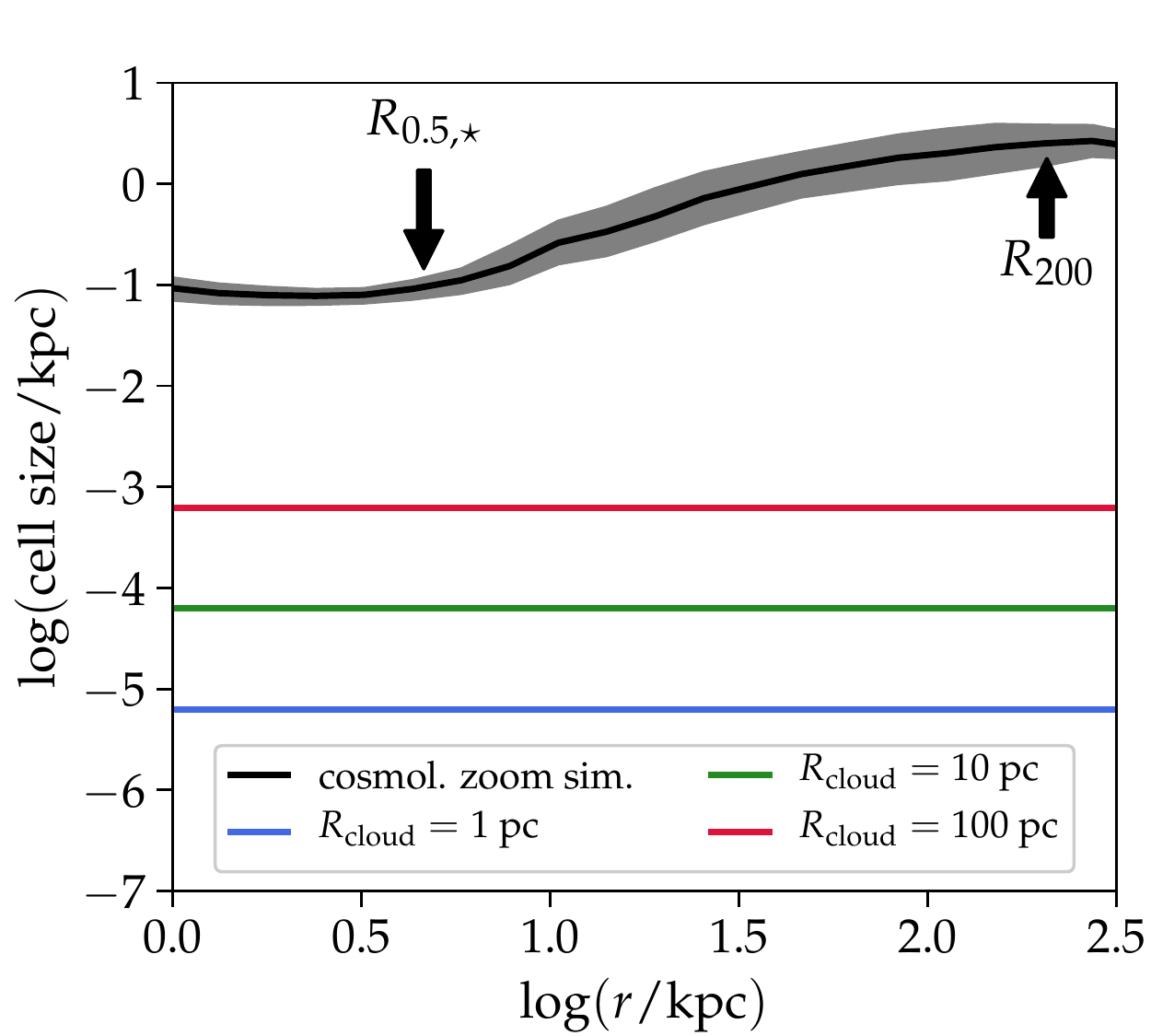}
\caption{The linear cell sizes of the dense gas in our 3D high-resolution simulations are compared to a cosmological zoom simulation from \citet{2016MNRAS.462.2418S}. $R_{200}$ and the stellar half-mass-radius ($R_{0.5,\star}$) are marked by arrows. The black line shows the mean cell size for the cosmological zoom simulation and the grey shaded area shows the 16--84\% interval. The fragmentation processes studied in this paper are completely unresolved in the current generation of cosmological galaxy formation simulations.}
\label{PlotCellSize_Radius}
\centering
\includegraphics[width = 1.0\linewidth]{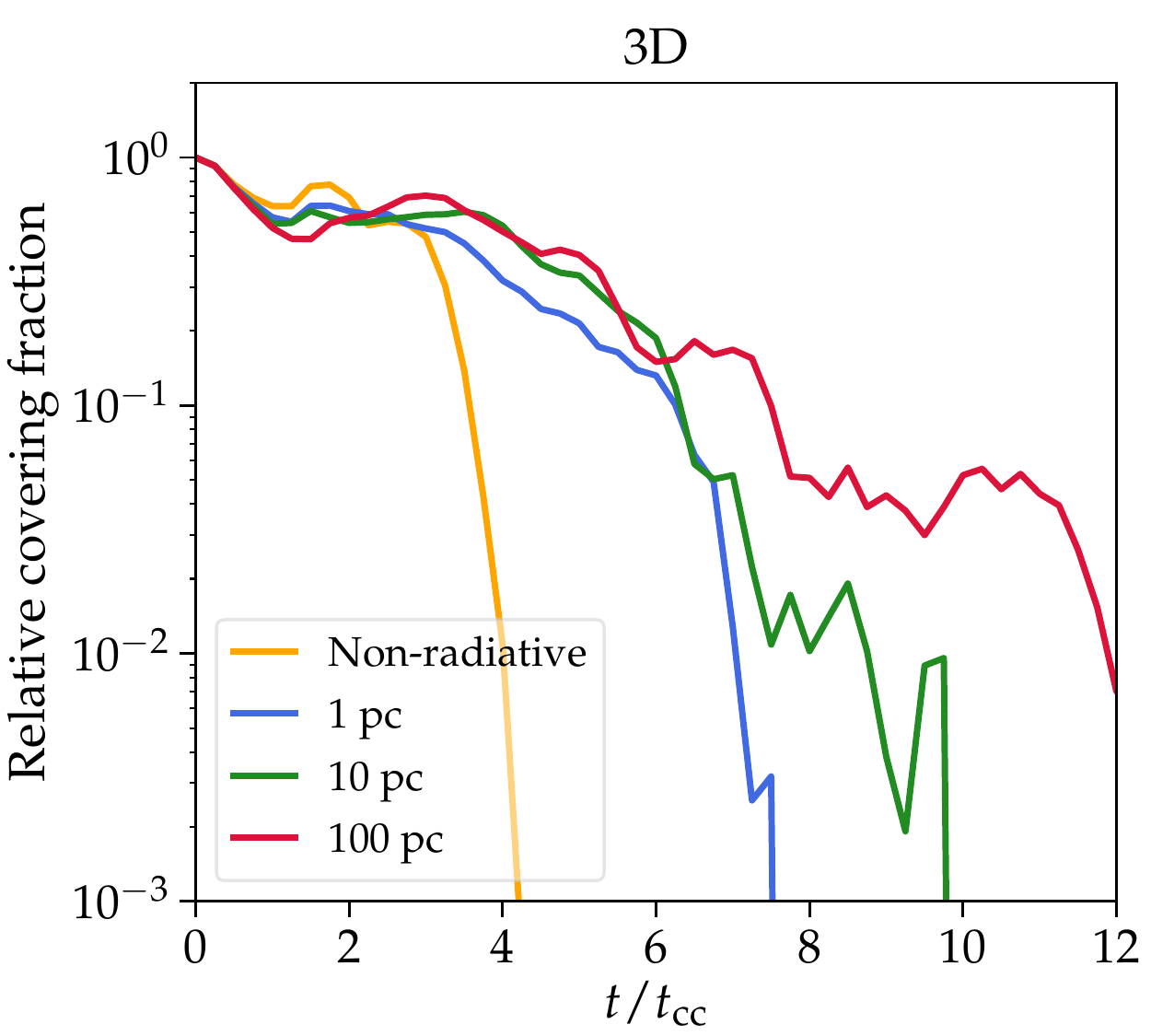}
\caption{For gas with $n\geq \frac{1}{3}n_\text{cloud}$ we compute the column number density, $\Sigma$, along the $z$-axis and the covering fraction of sight-lines with $\Sigma\geq 2 R_\text{cloud}n_\text{cloud}$. The covering fraction is normalised by the $t=0$ value. As previously established (Figure~\ref{Fig1CloudSurvival}) the dense gas survives longest in the 100 pc cloud simulation, and this leads to an enhanced covering fraction at late times ($t\gtrsim 7 t_\text{cc}$). Even though the covering fraction is enhanced at late times, the covering fraction of dense gas is much smaller than predicted by the shattering hypothesis of \citet{2018MNRAS.473.5407M}, who suggests that all gas resides in long-lived cloudlets with a size comparable to the cooling length.}
\label{Fig8CoveringFraction}
\end{figure}

\subsection{Cosmological galaxy simulations}

The idealised simulations presented in this paper have a spatial resolution much higher than in cosmological simulations. The dense clouds in the high-resolution 3D simulations of 1 pc, 10 pc and 100 pc clouds are resolved by cells with a linear size of 0.00625 pc, 0.0625 pc and 0.625 pc, respectively. In Fig.~\ref{PlotCellSize_Radius} the resolution of our idealised simulations is compared to the radial dependence of the mean cell size in a cosmological zoom simulation of one of the galaxies from \citet{2016MNRAS.462.2418S}. (We select the 1605-3 galaxy at $z=0$ with $M_{200}\simeq 10^{12}\msun$ and $M_*\simeq 10^{10.89}\msun$.) This simulation uses the Auriga galaxy formation model \citep{2017MNRAS.467..179G} with a resolution comparable to the fiducial (\emph{level 4}) Auriga simulations. This simulation is representative of state-of-the-art cosmological zoom simulations of Milky-Way-like galaxies. In the region defining the CGM -- around $(0.5$-$1.0) R_{200}$ -- the cosmological zoom simulation has mean cell sizes of around 1-2 kpc, which is orders of magnitudes larger than the resolution of our cold cloud and the hot wind. Even in the inner parts of the galaxy we are far from resolving the cooling length of $\simeq 1\;$pc for dense gas with $n\simeq 0.1\;$cm$^{-3}$. Cosmological simulations are therefore not able to capture the cooling-induced fragmentation, which we have studied in this paper.

Several simulation groups are currently developing new techniques to refine the CGM of cosmological galaxy formation simulations. This may enable resolving the small-scale-structure, but it is currently unclear whether we will be able to increase the spatial resolution in the CGM by the $\sim$3 orders of magnitudes (this is required at $R_{200}$ to reach the same resolution as the $R_\text{cloud}=100$ pc simulation), which will be necessary to see the effect of fragmentation studied in this paper. Irrespective of whether cosmological simulations will in the near future be able to resolve the sub-parsec-scale structure of gas, here we have demonstrated how idealised simulations can be used as a powerful tool to study the behaviour of galaxy formation physics at extremely high resolution.

\subsection{Shattering}

The shattering hypothesis of \citet{2018MNRAS.473.5407M} predicts galaxy haloes to be filled with high-density gas cloudlets with a size comparable to $l_\text{cool}$, and therefore also a high covering fraction of sight-lines with high column-densities. Thus, the covering fraction of gas can be increased by several orders of magnitudes compared to what is expected from a uniform gas distribution. Shattering has the potential to explain the large extent of observed Lyman-$\alpha$ haloes \citep{Wisotzki:2016hw,2018Natur.562..229W}, the characteristics of high-velocity clouds \citep{2005A&A...442L..49R,2005A&A...436L..53B}, and the broad-line regions of AGNs \citep{2018MNRAS.473.5407M}.

In Figure~\ref{Fig8CoveringFraction} we test how the covering fraction of dense gas with $n\geq n_\text{cloud}/3$ behaves in our high-resolution 3D simulations. By comparing the 100 pc and 1 pc simulations with radiative cooling, we see that the former exhibits covering fractions above 1 per cent of the initial value for twice as long as the latter. This factor of two is much smaller than the order-of-magnitude-increase expected by the original formulation of the shattering hypothesis. It is, however, possible that alternative physics, or variations of the initial conditions (e.g., to account for thermal instability or coagulation) can change this result.

A visual inspection of the 10 and 100 pc cloud crushing simulations of \citealt{2018MNRAS.473.5407M} (see their fig. 5) also reveals that small cloudlets are not contributing substantially to the dense gas covering fraction. The results of our simulations are therefore completely consistent with the simulations of \citet{2018MNRAS.473.5407M}. The strongest effect of shattering is indeed not seen in their cloud crushing simulations, but rather in their \emph{thermal instability simulation} \citep[Fig. 4 from][]{2018MNRAS.473.5407M}, where cloudlets precipitate out of a over-dense medium.

It is also possible that the recently proposed process by \citet{2018arXiv180602728G} is more efficient in producing high covering fractions in comparison to our cloud crushing simulations. They demonstrate that gas far downstream from the main cloud, which consists of a well-mixed phase of cold and hot gaseous phases, can cool efficiently if the mixed gas temperature is near the peak of the cooling curve, and if the cloud radius is sufficiently large. In our simulations, the wind is sufficiently hot so that the mixed gas cannot cool efficiently, and we do not observe the condensation of dense gas from the mixed downstream material. For our setup it would require a cloud radius of $R_\text{cloud}\gtrsim 160$ pc to see condensation of the mixed gas, following the equations in \citet{2018arXiv180602728G}. Here, we have not studied such large clouds because we focused on resolving the cooling length of the cold phase, which is difficult for such large clouds.

It is not surprising that not all types of simulation lead to the formation of cloudlets with sizes comparable to the cooling length. \citet{2018MNRAS.473.5407M} themselves conclude that the observations of the CGM of low-redshift haloes with $M_\text{halo}\simeq 10^{12}\msun$ \citep[presented in][]{2014ApJ...792....8W} are inconsistent with shattering. Even though we conclude that shattering into cloudlets is not always important for increasing the covering fraction of clouds, we find that there is solid evidence for additional fragmentation when $R_\text{cloud}\gg l_\text{cool}$. For some applications it will be of particular importance to take this excessive amount of fragmentation into account. The mock absorption spectra of gas clouds are for example sensitive to the detailed velocity- and density-distribution of the gas.

In future work we will estimate the role of fragmentation and shattering in further simulations with additional physics including magnetohydrodynamics, anisotropic thermal conduction and self-gravity. These processes will introduce additional length scales into the simulations; self-gravity for example introduces the Jeans length and thermal conduction introduces the Field length \citep{1965ApJ...142..531F}. Various works have studied cloud-crushing simulations with magnetic fields \citep{2008ApJ...677..993D,2015MNRAS.449....2M} and thermal conduction \citep{2016MNRAS.462.4157A,2016ApJ...822...31B}, but their main focus has been on the gas survival rather than on shattering. Recently, \citet{2018arXiv180610688L} found shattering to also occur in 2D cloud-crushing simulations with thermal conduction and magnetic fields, but a study of shattering in 3D simulations including these processes still remains to be done.

\section{Conclusion}\label{Sec:Conclusion}

In this paper we have performed \emph{cloud crushing simulations} to study the behaviour of multiphase gas. We have examined how cooling introduces a characteristic scale, the cooling length ($l_\text{cool}$), which breaks the self-similarity of non-radiative cloud crushing simulations. In the following we summarise our results:
\begin{itemize}
\item In 2D and 3D simulations, clouds with radii $R_\text{cloud}\gg l_\text{cool}$ survive longer and undergo excessive fragmentation compared to smaller clouds with $R_\text{cloud}\simeq l_\text{cool}$. We have determined the presence of fragmentation based on visual inspection of the density field, density power spectra and a friends-of-friends cloudlet analysis.
\item The density power spectrum analysis reveals that clouds with $R_\text{cloud}\gg l_\text{cool}$ have a shallow spectral index near the end of a clouds lifetime in comparison to smaller clouds. This occurs in 2D and 3D, and is a signature of shattering as predicted by \citet{2018MNRAS.473.5407M}. For the first time we have demonstrated that effects of shattering occur in 3D simulations.
\item The increase in covering fraction for large clouds is less than expected by the shattering hypothesis of \citet{2018MNRAS.473.5407M}. Even though shattering plays a less important role in shaping cold clouds accelerated by a hot wind, it remains to be seen whether shattering is important for other 3D simulation setups and with additional physics. The effect of processes such as self-gravity, magnetohydrodynamics and thermal conduction also remain to be studied in future work.
\item The growth of instabilities in 2D and 3D is different, because a 2D simulation technically corresponds to a 3D simulation, where symmetry is strictly enforced along the $z$-axis.  In 2D instabilities can therefore only grow in the $x$--$y$-plane. Furthermore, we have demonstrated that (at least in some cases) instabilities working in the $x$--$y$-plane have a larger effect in 2D than in 3D, because a 3D flow has the freedom to use the $z$-direction to go around dense clouds rather than penetrating them. This causes vigorous fragmentation of 2D clouds in comparison to 3D analogues, because the Richtmyer--Meshkov instability works more efficiently in the former. Because of these fundamental differences between 2D and 3D simulations we discourage the use of 2D simulations for providing predictions for observations.
\item A spatial resolution many times higher than state-of-the-art cosmological simulations is required to resolve the fragmentation processes studied in this paper. Cosmological simulations therefore underestimate the clumpiness of the gas in the outskirts of galaxy haloes.
\end{itemize}

\section*{Acknowledgements}
We thank Federico Marinacci, Peng Oh, Philipp Girichidis and Volker Springel for useful comments and discussions. CP and MS acknowledges support by the European Research Council under ERC-CoG grant CRAGSMAN-646955. MV acknowledges support through an MIT RSC award, the Alfred P. Sloan Foundation, NASA ATP grant NNX17AG29G, and a Kavli Research Investment Fund.

\footnotesize{
\bibliographystyle{mnras}
\bibliography{ref}
}


\appendix

\section{Convergence study}\label{Sec:ShatteringAndConvergence}

In this section we perform several convergence tests of our simulations.

\begin{figure*}
\centering
\includegraphics[width = 1.0\linewidth]{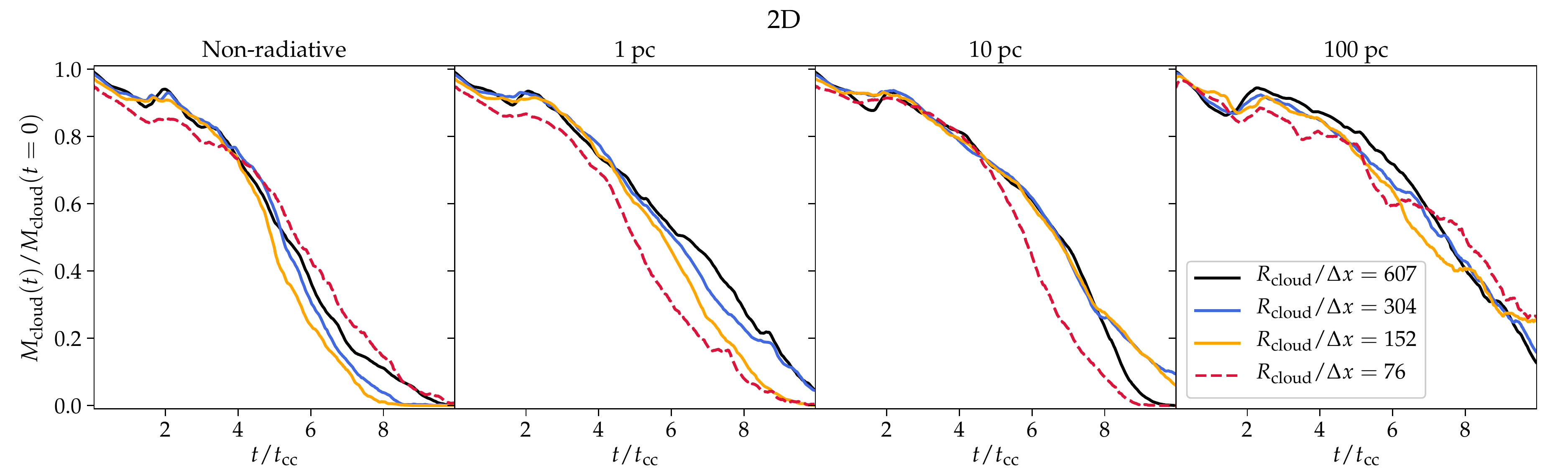}
\caption{A convergence study of the evolution of the mass of dense gas with $n\geq n_\text{cloud}/3$. For all cloud sizes the three highest resolution levels show good agreement, especially at early times $t\lesssim 5 t_\text{cc}$, implying that these simulations are well converged. At the same time, the low-resolution simulations with $R_\text{cloud}/\Delta x = 76$ (dashed lines) systematically differ from the other simulations. Thus, only the dense gas survival mass fraction in the 2D simulations with $R_\text{cloud}/\Delta x \geq 152$ are converged.}
\label{Fig030_CloudSurvival2D_2D}
\end{figure*}
\begin{figure*}
\centering
\includegraphics[width = 0.9\linewidth]{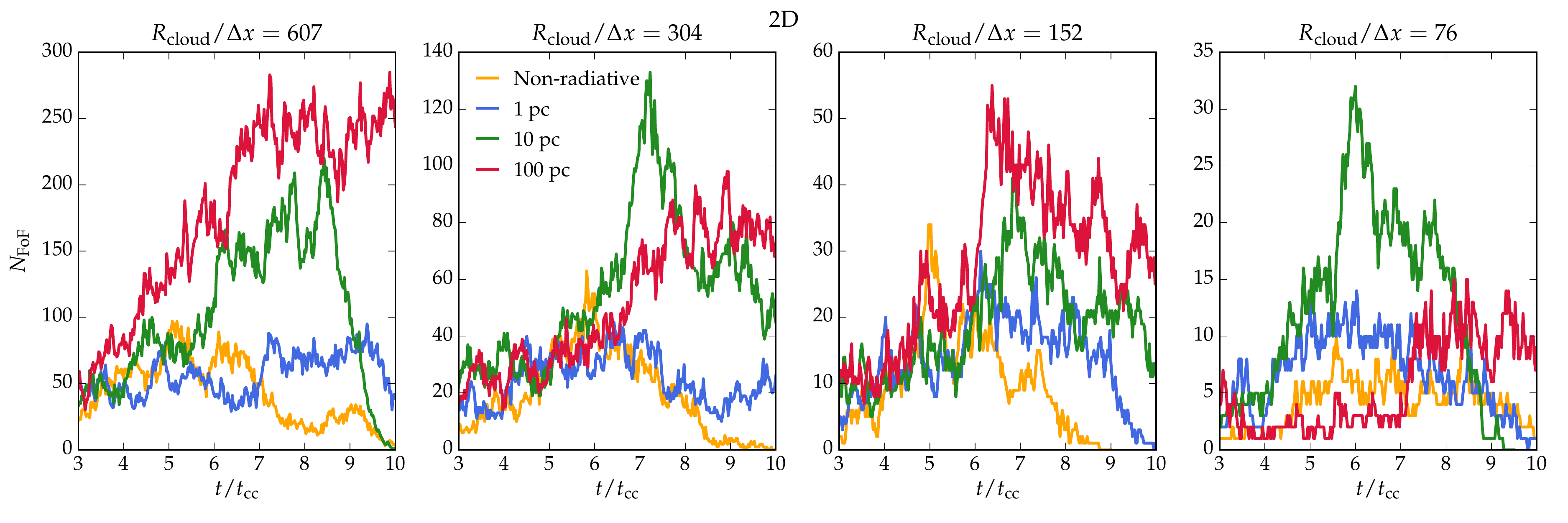}
\caption{The number of FoF groups in 2D simulations with a resolution of 607, 304, 152 and 76 cells per cloud radius (from left to right). For $R_\text{cloud}/\Delta x\geq 152$ the 10 and 100 pc cloud simulations have a higher peak number of FoF groups in comparison to the non-radiative and 1 pc cloud simulations. A resolution of $R_\text{cloud}/\Delta x = 76$ is evidently not enough to cause fragmentation of a 100 pc cloud since the peak $N_\text{FoF}$-value is comparable to the 1 pc simulation with radiative cooling. We conclude that a resolution of $R_\text{cloud}/\Delta x\geq 152$ for our 2D simulations is sufficient to capture fragmentation in 1, 10 and 100 pc simulations with cooling.}
\label{Fig276_FoF_EvolutionConverge_2D}
\end{figure*}
\begin{figure}
\centering
\includegraphics[width = 0.99\linewidth]{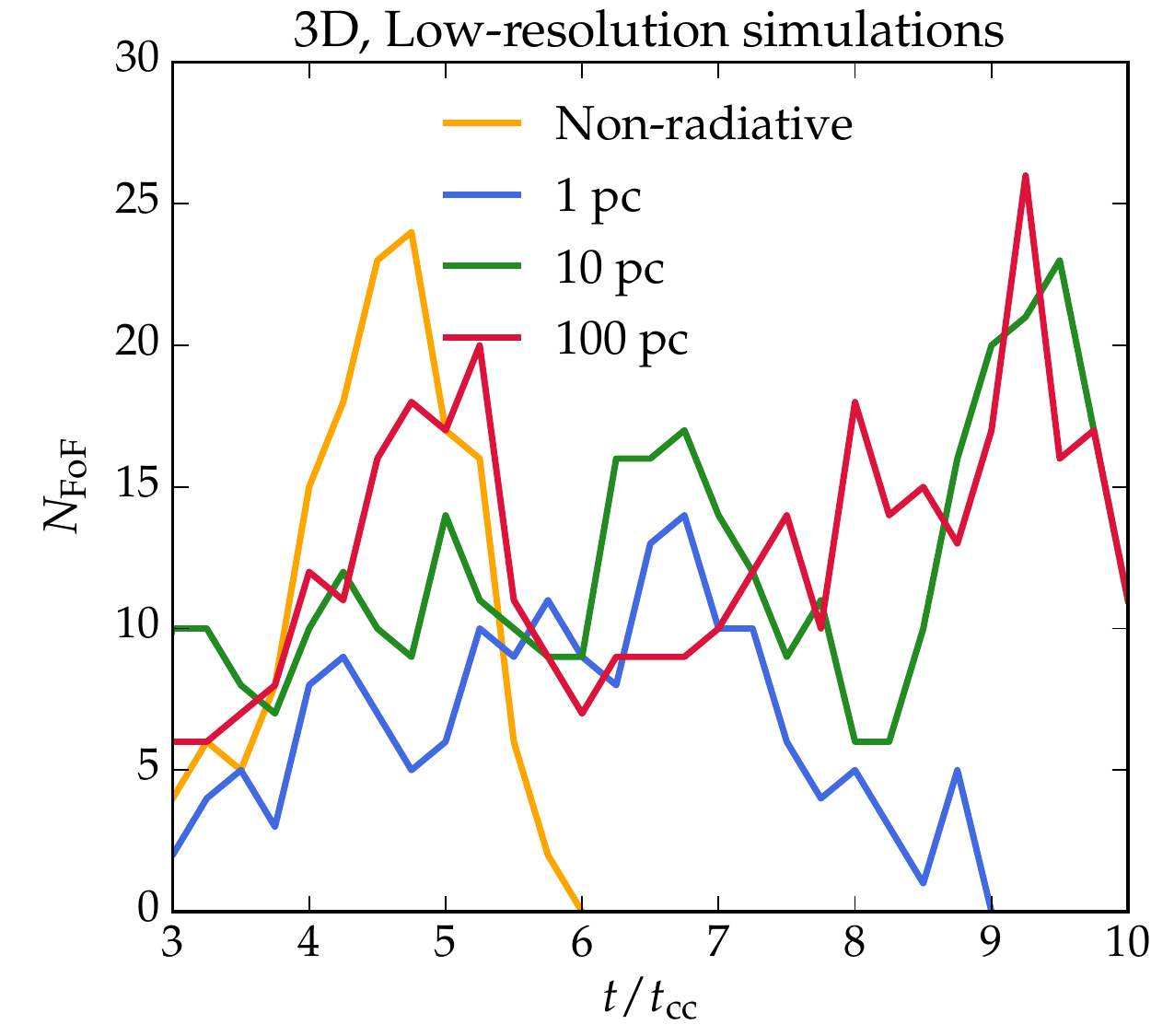}
\caption{$N_\text{FoF}$ as a function of time for the low-resolution 3D simulations (with $R_\text{cloud}/\Delta x= 80$). The relative differences between the 1, 10 and 100 pc simulations with radiative cooling are smaller in comparison to the high-resolution simulations (Fig.~\ref{Fig5FoF}). A resolution of $R_\text{cloud}/\Delta x= 160$ is required to capture excessive fragmentation of 100 pc clouds in comparison to a 10 pc cloud. This is, however, not surprising, since we concluded in Fig.~\ref{Fig1CloudSurvival} that the high resolution simulation of the 100 pc cloud is not necessarily converged. It is nevertheless reassuring that the low-resolution 3D runs of the 10 and 100 pc clouds show more fragmentation than the corresponding 1 pc simulation.}
\label{Fig371_FoF_Evolution_3D_LowRes}
\end{figure}
\begin{figure*}
\centering
\includegraphics[width = 0.99\linewidth]{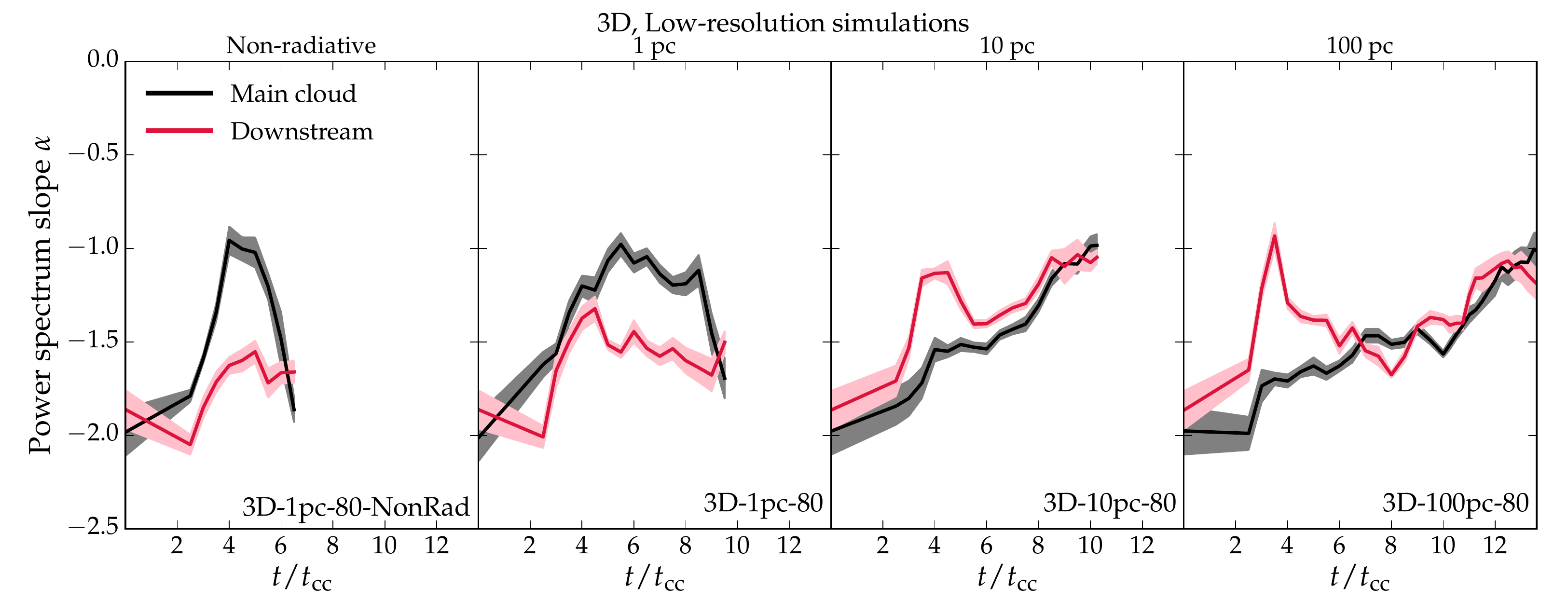}
\caption{The evolution of the power spectral slope for the low-resolution 3D simulations. The peak value of the {\tt 3D-100pc-80} simulation is smaller than for the corresponding high-resolution simulation ({\tt 3D-100pc-160} in Fig.~\ref{Fig7SpectralSlope}). Indeed the peak value at the end of the 10 and 100 pc simulations have similar values. This is consistent with the peak value of $N_\text{FoF}$ being similar for these two simulations, and it strengthens our conclusion that the simulations with $R_\text{cloud}/\Delta x= 80$ do not capture the excessive fragmentation of a 100 pc cloud, which is seen at high-resolution. Finally, we note that it is reassuring that the spectral slope of the {\tt 2D-100pc-80} simulation gradually becomes shallower, which is also seen in the high-resolution simulation ({\tt 3D-100pc-160}).}
\label{Fig860_PowerSpecSlopeEvolution3D_LowRes_3D}
\end{figure*}

\subsection{2D simulations}

Figure~\ref{Fig030_CloudSurvival2D_2D} shows a convergence test of the dense gas survival fraction for the 2D simulations. All simulations with $R_\text{cloud}/\Delta x=76$ show a significantly different behaviour than the simulations at higher resolution. The $R_\text{cloud}=1$ pc cloud with this resolution for example has a lower cloud survival fraction throughout the simulation in comparison to the simulations with $R_\text{cloud}/\Delta x \geq 152$. The $R_\text{cloud}/\Delta x=76$ simulation yields a similar offset at $t\lesssim 4 t_\text{cc}$ in the panel showing the non-radiative simulations, and at late times ($t\gtrsim 5 t_\text{cc}$) for the 10 pc simulations. We therefore conclude that only the simulations with $R_\text{cloud}/\Delta x\geq 152$ are resolved in 2D. At first sight, this is somewhat surprising, because the 3D simulations with a very similar resolution of $R_\text{cloud}/\Delta x=80$ give converged results according to Fig.~\ref{Fig1CloudSurvival}. The natural explanation is that the ratio of the resolution inside and outside the cold cloud is different in 2D and 3D. Even though the lowest-resolution simulations in 2D and 3D have almost identical resolutions of $R_\text{cloud}/\Delta x=76$ and 80 inside the cloud, the resolutions outside are different. The hot wind of a 2D simulation is indeed represented by a factor of $\simeq 3.16$ larger gas cells in comparison to 3D, if the 2D and 3D simulations have a matched resolution inside the cloud (see Sec.~\ref{Refinement} for an explanation).

The evolution of the number of FoF-groups for the 2D simulations is shown in Fig.~\ref{Fig276_FoF_EvolutionConverge_2D}. We confirm the conclusion that clouds with $R_\text{cloud}\gg l_\text{cool}$ undergo excessive fragmentation for a resolution of $R_\text{cloud}/\Delta x\geq 152$. Only the low-resolution simulations with $R_\text{cloud}/\Delta x=76$ are not capturing this result in 2D.

\subsection{3D simulations}\label{3DConvergence}

Because 2D and 3D clouds may be affected differently by instabilities, we carry out the same resolution studies for the 3D simulations.

First, we remind the reader that the dense gas mass survival fraction studied in Fig.~\ref{Fig1CloudSurvival} behaves very similar for our low- and high-resolution 3D simulations, with the only notable difference being that the dense gas in the 100 pc simulation at low resolution survives longer in comparison to the high-resolution simulation.

To further study convergence properties we show the evolution of $N_\text{FoF}$ in Fig.~\ref{Fig371_FoF_Evolution_3D_LowRes}. The figure confirms that the 10 and 100 pc simulations are more fragmented than the non-radiative simulation and the 1 pc simulation with radiative cooling. Overall, this confirms our result that clouds larger than the cooling length shatter to smaller cloudlets. However, the figure also shows that fragmentation of the 100 pc cloud in 3D is not fully captured, because the 10 and 100 pc clouds here have a similar number of FoF-groups throughout the simulations. This conclusion is further supported by the evolution of the power spectral slope of the low-resolution simulations shown in Figure~\ref{Fig860_PowerSpecSlopeEvolution3D_LowRes_3D} because the peak value of the slope is not larger in the 100 pc simulation in comparison to the 10 pc simulation. Excessive fragmentation of the 100 pc cloud in comparison to the 10 pc cloud is thus only seen in our high-resolution 3D simulations.

An implication is that we have not demonstrated convergence for our 100 pc cloud simulation in 3D. Overall, this lack of convergence for the largest clouds demonstrates the importance of resolving the cooling scale in order to achieve convergence.

\section{The stand-off distance}\label{standoffdistance}

\begin{figure*}
    \centering
    \begin{minipage}{.33\textwidth}
        \centering
        \includegraphics[width=\linewidth]{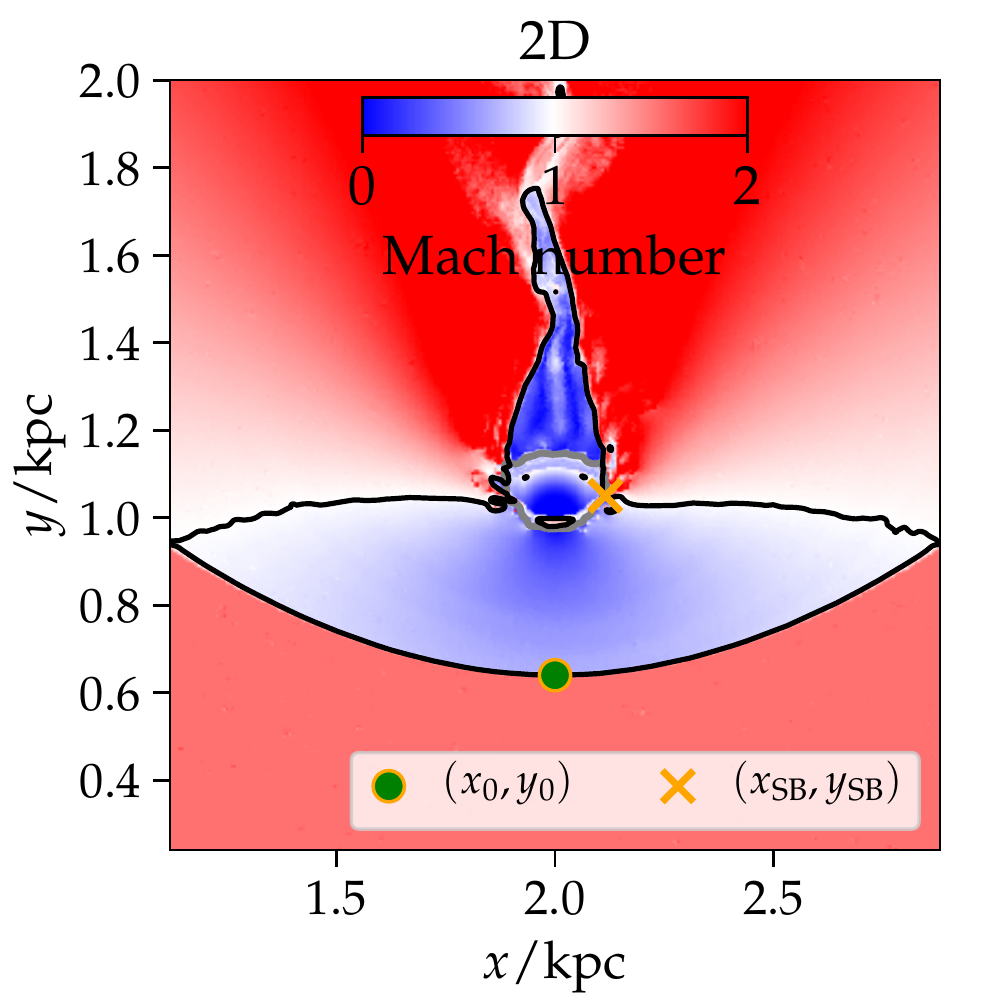}
    \end{minipage}%
    \begin{minipage}{.33\textwidth}
        \centering
        \includegraphics[width=\linewidth]{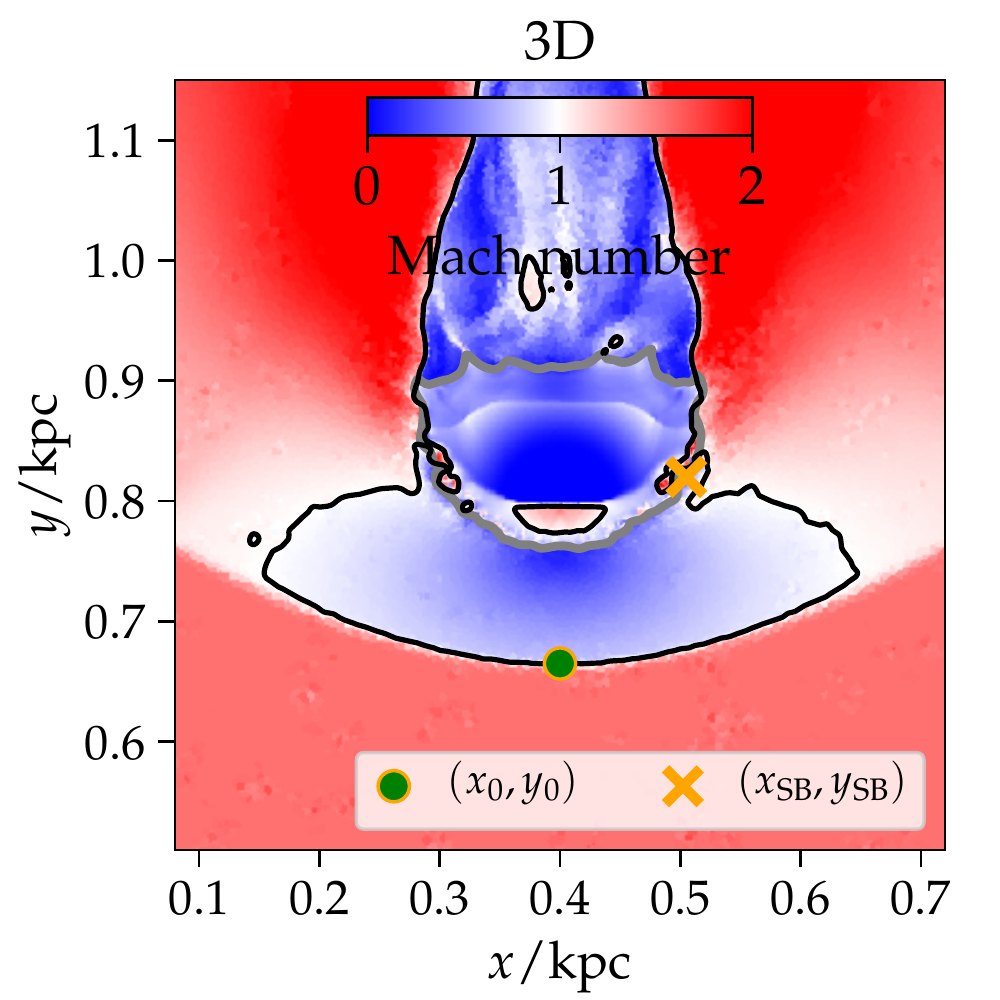}
    \end{minipage}%
    \begin{minipage}{0.33\textwidth}
        \centering
        \includegraphics[width=\linewidth]{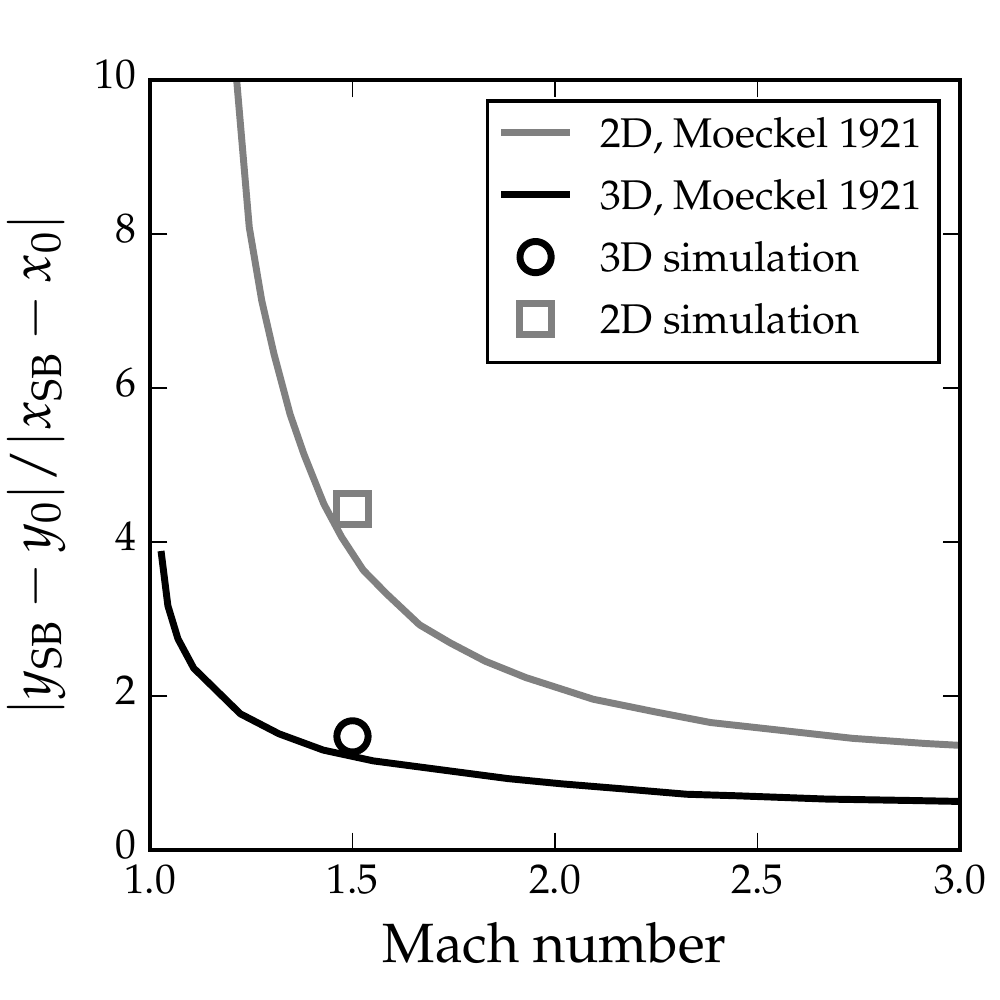}
    \end{minipage}
        \caption{The Mach number of the 2D and 3D simulations from Fig.~\ref{Fig03EarlyEvolution} at $t=t_\text{cc}$. The width of the left and central panel are adjusted, so that the entire subsonic region is included in each panel. $(x_0,y_0)$ marks the head of the bow shock and $(x_\text{SB},y_\text{SB})$ marks the sonic body point. The black contour (with $\M=1$) marks the transition from the subsonic to the supersonic region. The grey contour ($n= 0.01$ cm$^{-3}$) estimates the surface of the cold cloud. The right panel reveals good agreement between the bow shock stand-off distance in our simulations and the theoretical \emph{continuity model} of \citet{Moeckel49approximatemethod}.}
        \label{FigMoeckelComparison}
\end{figure*}
\begin{figure*}
    \centering
    \begin{minipage}{.33\textwidth}
        \centering
        \includegraphics[width=\linewidth]{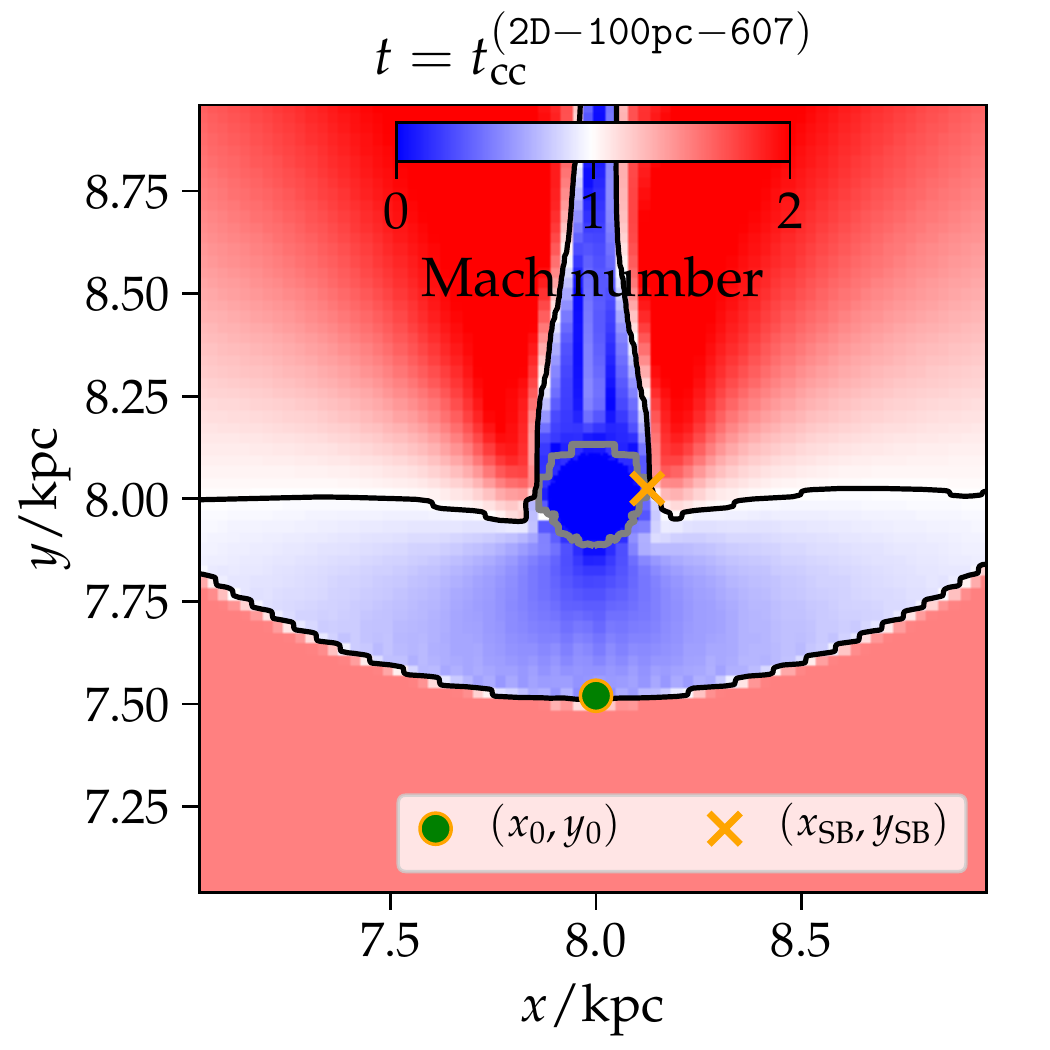}
    \end{minipage}%
    \begin{minipage}{.33\textwidth}
        \centering
        \includegraphics[width=\linewidth]{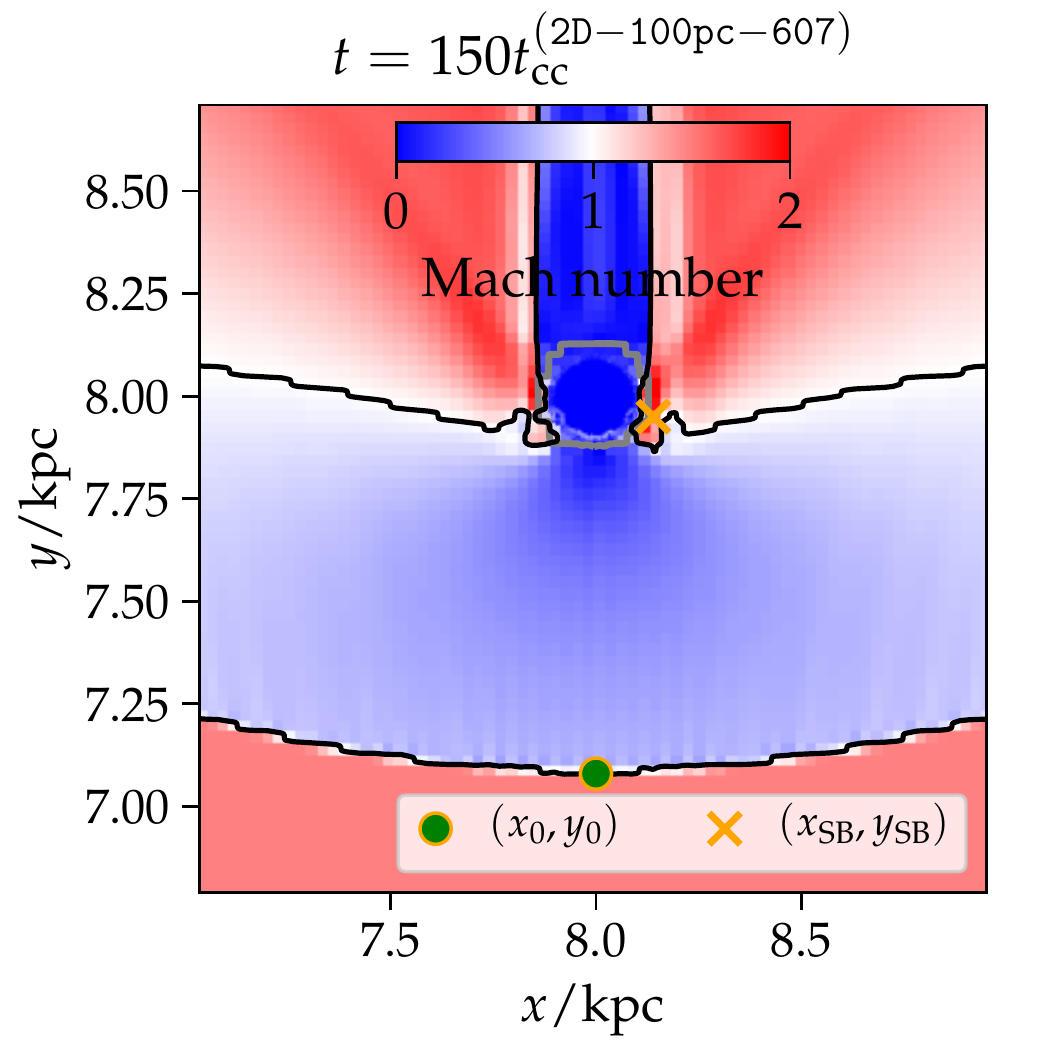}
   \end{minipage}%
    \begin{minipage}{0.33\textwidth}
       \centering
        \includegraphics[width=\linewidth]{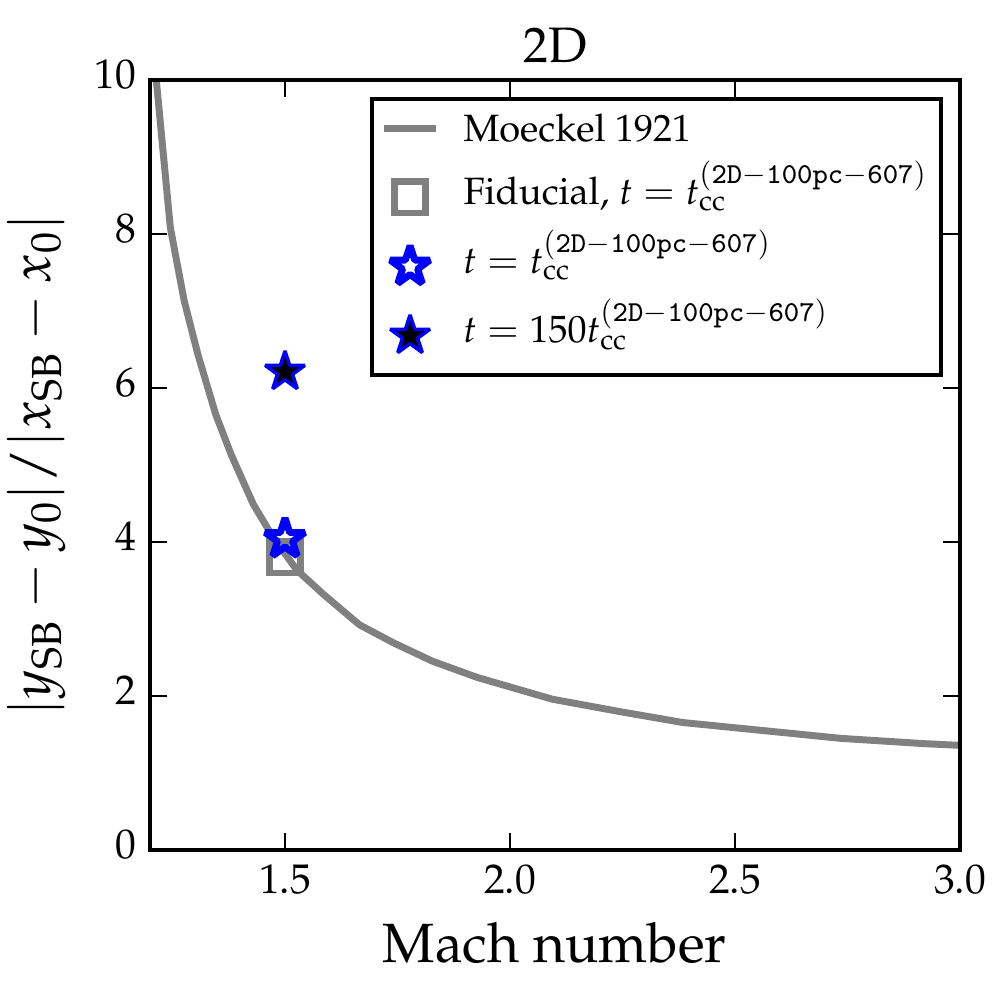}
    \end{minipage}
        \caption{A 2D simulation with a $10^4$ times higher density in the cold cloud than for {\tt 2D-100pc-607}. A higher density contrast enables us to study the evolution of the bow shock for a longer time, because the cloud survives longer. The Mach numbers and subsonic regions are shown at $t=t_\text{cc}^\text{({\tt 2D-100pc-607})}$ (left panel) and $t=150 t_\text{cc}^\text{({\tt 2D-100pc-607})}$ (central panel). We plot the stand-off distance at these two times in the right panel (blue symbols), as well as the value for {\tt 2D-100pc-607} (grey square). At $t=150 t_\text{cc}^\text{({\tt 2D-100pc-607})}$, where an equilibrium configuration is reached, the stand-off distance is 1.5 times larger than at $t=t_\text{cc}^\text{({\tt 2D-100pc-607})}$, implying that we have slightly underestimated the equilibrium stand-off distance in the analysis of the 2D simulation in Fig.~\ref{FigMoeckelComparison}. Our equilibrium configuration in 2D gives a slightly larger stand-off distance than Moeckel 1921, but it is well within the expected scatter around their solution (see text for more discussion).}
        \label{FigMoeckelComparisonIdealised}
\end{figure*}

Figure~\ref{Fig03EarlyEvolution} shows that the distance from the front of the bow-shock to the cloud is larger in 2D in comparison to 3D. A similar qualitative result is also found in \citet{2013ApJ...776..101B}, who analysed simulations with bow-shocks in a different astrophysical context. A general analytical derivation of the stand-off distance of a bow shock has not yet been done. \citet{Moeckel49approximatemethod} derived an analytical expression of the stand-off distance by assuming a particular shape of the bow shock, and a straight \emph{sonic line}, which connects the sonic points on the bow shock and the body. Several important elements of Moeckel's method are summarised in appendix~B of \citet{2001ApJ...551..160V}.

In Fig.~\ref{FigMoeckelComparison} we plot the distribution of Mach numbers, $\M$, of the {\tt 2D-100pc-607} and {\tt 3D-100pc-160} simulations at $t=t_\text{cc}$. We have chosen the simulations with a radius of 100 pc, because they have the longest survival time, and the flow around them thus mimics the flow around a solid sphere as much as possible.

The Mach number distribution allows an easy identification of the subsonic region and also the sonic body point, $(x_\text{SB},y_\text{SB})$. This point is marked together with the head of the bow shock, $(x_0,y_0)$. An estimate of the stand-off distance can be computed as $|y_\text{SB}-y_0|/|x_\text{SB}-x_0|$. In the right panel we compare this to the analytical theory of \citet{Moeckel49approximatemethod}, and find good agreement.

By inspecting the time evolution of the bow shock a subtle issue is, however, revealed. Unlike the 3D simulation, the stand-off distance in the 2D case has not yet reached an equilibrium at the time we have performed the analysis. At later times the cloud starts to disrupt, so it is not possible to determine an equilibrium stand-off-distance for the 2D simulations shown. To investigate this issue further, we run an additional 2D simulation with a $10^4$ higher cold cloud density than our fiducial setup. Following Eq.~\ref{tcc} this extends the cloud lifetime by a factor of 100, increasing the chance that the bow shock can settle into an equilibrium, before the cloud is evaporated. In our analysis of this simulation we report the time in units of cloud crushing time-scales of the {\tt 2D-100pc-607} simulation ($t_\text{cc}^\text{(2D-100pc-607)}$) to ease comparison to our previous simulations. The simulation is run on a static rectangular mesh with $L_x\times L_y =1.6 \text{ kpc} \times 3.2 \text{ kpc}$ and $N_x \times N_y = 600\times 1200$. With this static mesh setup the spatial resolution of the cold phase is much worse in comparison to our moving-mesh refinement scheme, but this is not a problem here, because we are mainly analysing the location of bow shock.

Figure~\ref{FigMoeckelComparisonIdealised} shows this simulation at $t=t_\text{cc}^\text{({\tt 2D-100pc-607})}$ and $t=150 t_\text{cc}^\text{({\tt 2D-100pc-607})}$. The stand-off-distance keeps expanding after $t_\text{cc}^\text{({\tt 2D-100pc-607})}$, but it eventually reaches an equilibrium configuration at $150t_\text{cc}^\text{({\tt 2D-100pc-607})}$. The value at the latter time, is, however  $\simeq$50\% larger than the analytical solution of \citet{Moeckel49approximatemethod}. The most likely reason is that the assumptions of \citet{Moeckel49approximatemethod} are not fully valid in our case. Figures~\ref{FigMoeckelComparison} and \ref{FigMoeckelComparisonIdealised} indeed reveal that the sonic line is not straight. Experiments also reveal some scatter around the Moeckel solution (see Moeckel's figure~7), and taking this into account the tension between our simulation and the model weakens. Overall, we regard our simulations as being in broad agreement with the model of \citet{Moeckel49approximatemethod}, and most importantly the model and simulation agree that the stand-off distance is significantly larger in 2D than in 3D.

\section{Potential flow solutions}\label{Sec:PotentialFlow}

In this section we derive the 3D incompressible potential flow solutions, which obey $\bs{\nabla} \bs{\cdot} \bs{\varv} = 0$ and $\bs{\nabla \times \varv} = \bs{0}$, around a sphere and a cylinder. The latter corresponds to the flow around a 2D sphere.

\begin{figure*}
\centering
\includegraphics[width = 0.98\linewidth]{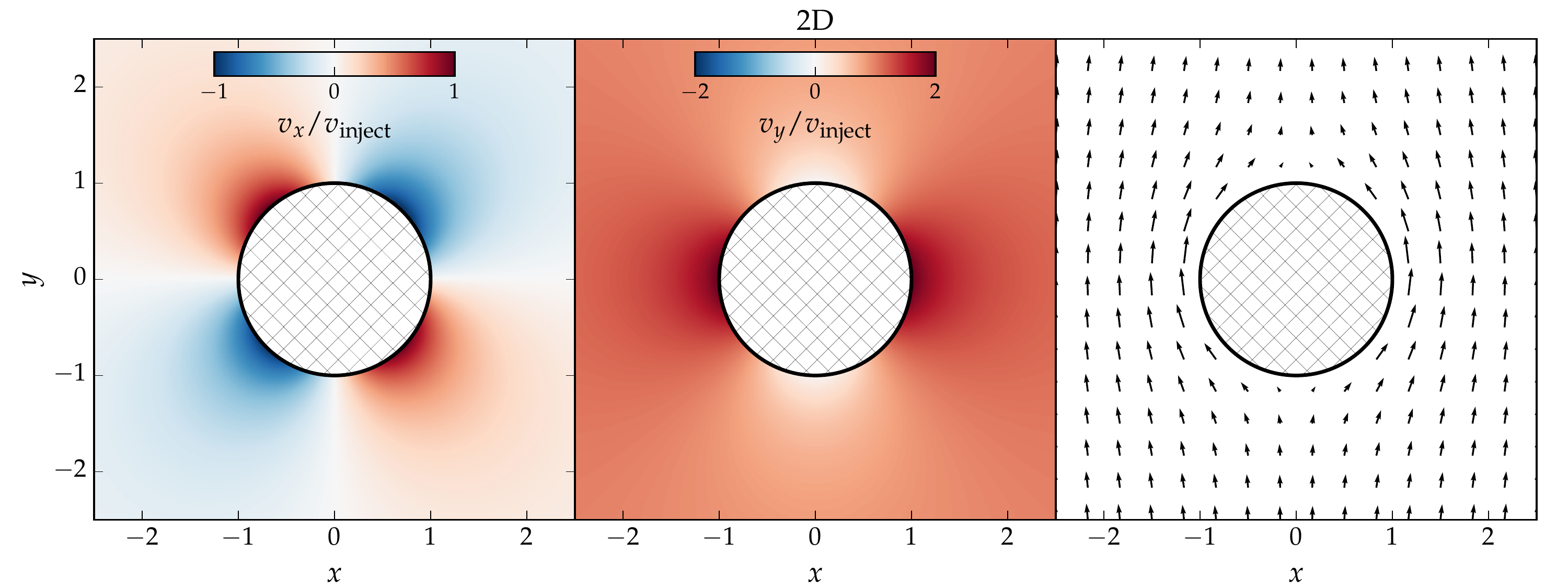}
\includegraphics[width = 0.98\linewidth]{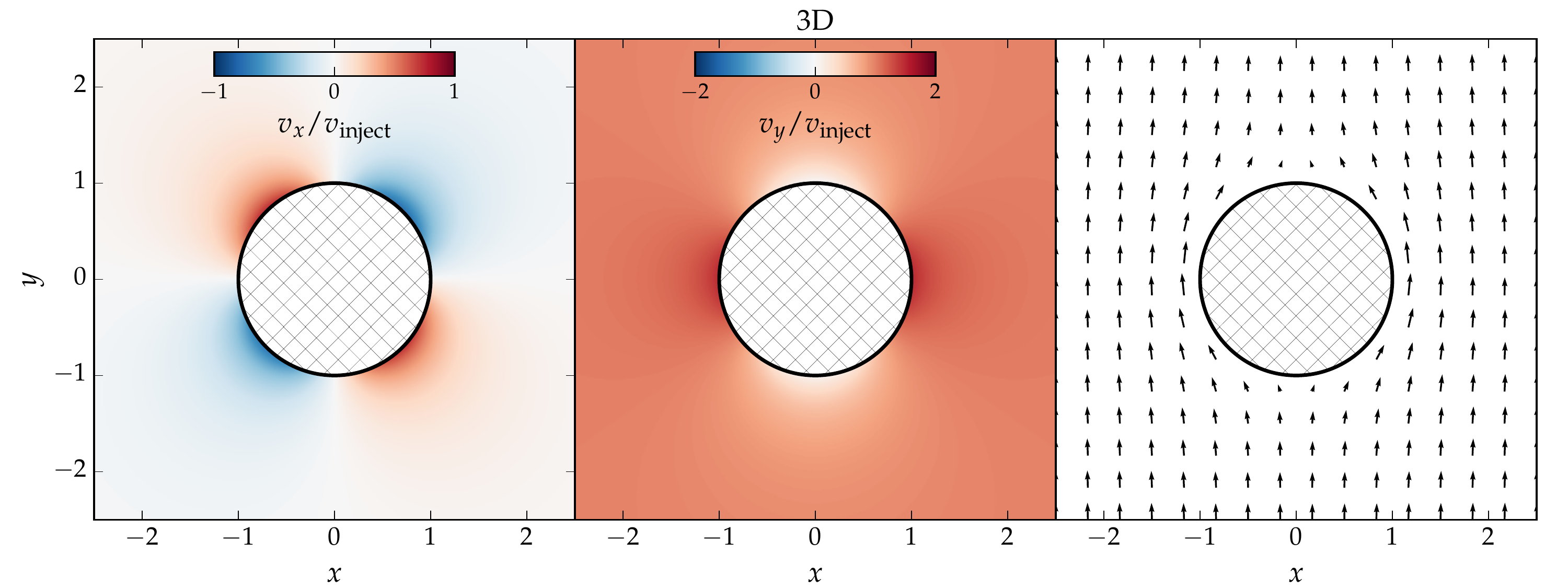}
\caption{A slice in the $z=0$ plane of the potential flow solution around a 2D and a 3D sphere (upper and lower panels, respectively). The flow is oriented so the wind is blowing in the $y$-direction with an initial speed, $\varv_\text{inject}$. The left panels show $\varv_x/\varv_\text{inject}$, the central panels show $\varv_y/\varv_\text{inject}$ and the right panels show the orientation of the velocity field. Near the surface of the sphere there are larger velocities visible in 2D in comparison to the 3D case. To obtain a realistic velocity field in a cloud crushing simulation it is therefore not sufficient to carry out 2D simulations.}
\label{PotentialFlow_3D}
\end{figure*}

\subsection{Potential flow around a 3D cylinder}

The flow around a cylinder is most easily derived in cylindrical coordinates, $(\rho,\phi,z)$, where $x = \rho \cos \phi$ and $y = \rho \sin \phi$. Here the angle is defined in the interval $0\leq\phi<2\pi$. An infinite cylinder, with radius $R$, is placed along the $z$-axis and $\hat{x}$ is the direction of the flow. It is a requirement that no flow goes through the body, hence $\varv_\rho =0$ at $\rho=R$. We seek a solution which asymptotically approaches the injection velocity $\bs{\varv}=\varv_\text{inject} \bs{\hat{x}}$ at large distances. Accounting for the symmetry of the problem, we adopt the ansatz
\begin{align}
\varv_\rho &= \varv_\text{inject} \left (1-\frac{R^n}{\rho^n}\right) \cos \phi, \label{flowA}\\
\varv_\phi &= - \varv_\text{inject} \left (1+\frac{R^n}{b\rho^n}\right) \sin \phi,\label{flowB}\\
\varv_z &= 0.
\end{align}
In cylindrical coordinates the divergence and curl of the velocity are given by:
\begin{align}
\bs{\nabla \cdot \varv} &= \frac{1}{\rho} \frac{\partial \left( \rho \varv_\rho\right)}{\partial \rho} + \frac{1}{\rho} \frac{\partial \varv_\phi }{\partial\phi },\\
\bs{\nabla \times \varv} &= \frac{1}{\rho} \left( \frac{\partial\left( \rho \varv_\phi \right)}{\partial \rho} - \frac{\partial \varv_\rho}{\partial \phi}\right) \bs{\hat{z}},
\end{align}
where we have imposed that the $\partial /\partial z$-terms must vanish for symmetry reasons. Plugging \eqref{flowA} and \eqref{flowB} into these equations, and imposing the conditions $\bs{\nabla \cdot \varv} = 0$ and $\nabla \bs{\times} \bs{\varv} = \bs{0}$, then yields $b=1$ and $n=2$. We have discarded the trivial solution $\bs{\varv}=\bs{0}$, because the velocity field does not have the desired behaviour at large distances from the cylinder.

\subsection{Potential flow around a 3D sphere}

To derive the flow around a sphere we use spherical coordinates, $(r,\theta,\varphi)$, with $x = r \sin \theta \cos \varphi$, $y = r \sin \theta \sin \varphi$ and $z = r \cos \theta$. The domains of the angles are $0\leq \theta\leq \pi$ and $0\leq \varphi < 2\pi$. The solution around a solid sphere of radius $R$ must furthermore ensure that no gas is flowing through the surface of the sphere, implying $\varv_r=0$ at $r=R$. At large distance, the velocity field should asymptotically approach the injection velocity, which we choose as $\varv_\text{inject}\bs{\hat{z}}$. We adopt the ansatz
\begin{align}
\varv_r &= \varv_\text{inject} \left (1-\frac{R^n}{r^n}\right) \cos \theta,\\
\varv_\theta &= -\varv_\text{inject} \left (1+\frac{R^n}{br^n}\right) \sin \theta,\\
\varv_\varphi &= 0.
\end{align}
In spherical coordinates the divergence and curl of the velocity are given by:
\begin{align}
\bs{\nabla \cdot \varv} &= \frac{1}{r^2} \frac{\partial \left( r^2 \varv_r\right)}{\partial r} + \frac{1}{r\sin\theta} \frac{\partial \left( \varv_\theta \sin\theta \right)}{\partial\theta },\\
\bs{\nabla \times \varv} &= \frac{1}{r} \left( \frac{\partial\left( r \varv_\theta \right)}{\partial r} - \frac{\partial \varv_r}{\partial \theta}\right) \bs{\hat{\varphi}},
\end{align}
where we have used that the $\partial / \partial \varphi$-terms must vanish for symmetry reasons. We then obtain $b=2$ and $n=3$. We have again discarded the solution $\bs{\varv}=\bs{0}$.

The potential flow solutions are shown in Fig.~\ref{PotentialFlow_3D}. In this figure a coordinate rotation is performed, so the flow is directed in the $y$-direction (to match our simulation coordinate system).

\section{Friends-of-friends analysis}\label{Sec:FoFMass}

\begin{figure}
\centering
\includegraphics[width = 0.99\linewidth]{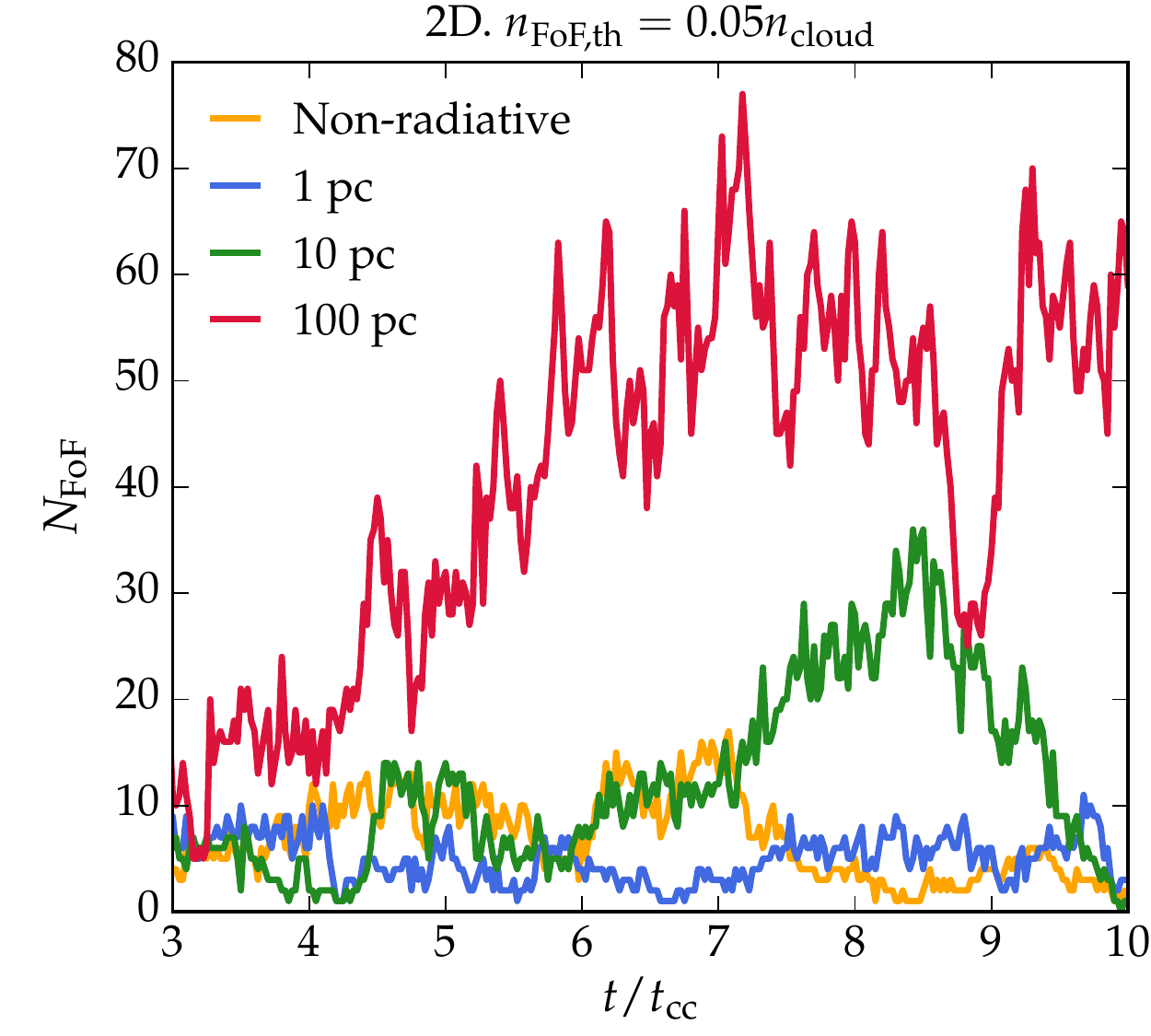}
\caption{The 2D simulations analysed as in Fig.~\ref{Fig5FoF}, but with a lower value of the FoF-density threshold of $n_{\rm FoF,th}=0.05 n_{\rm cloud}$. The number of FoF-groups is significantly lower than for the fiducial value of $n_{\rm FoF,th}=0.25 n_{\rm cloud}$, but the relative differences between the simulations are similar. We again see that clouds with $R_\text{cloud}\gg l_\text{cool}$ experience more fragmentation in comparison to the smaller clouds. The main result of our FoF-analysis is hence insensitive to the actual value of $n_{\rm FoF,th}$.}
\label{Fig277_FoF_EvolutionConverge_2D}
\end{figure}

In this section we study the robustness of the results of our friends-of-friends cloudlet finding algorithm. The most important free parameter that enters our FoF-method is the density threshold, $n_\text{FoF,th}$. To examine the sensitivity of our results to this free parameter, we have created a version of Fig.~\ref{Fig5FoF} that shows the evolution of $N_\text{FoF}$ for $n_\text{FoF,th}=0.05 n_\text{cloud}$ instead of our fiducial value of $0.25 n_\text{cloud}$. This is shown in Fig.~\ref{Fig277_FoF_EvolutionConverge_2D}.

By comparing Fig.~\ref{Fig5FoF} (left panel) and Fig.~\ref{Fig277_FoF_EvolutionConverge_2D} we find that the actual number of FoF-groups is overall smaller for the smaller choice of $n_\text{FoF,th}$, which is expected because a lower density threshold corresponds to a larger linking length. The relative behaviour of the different simulations is, however, remarkably similar in the two figures. Fragmentation becomes gradually more important as $R_\text{cloud}$ becomes larger than $l_\text{cool}$, independent of the chosen value of $n_\text{FoF,th}$.

\bsp	
\label{lastpage}
\end{document}